\documentclass[preprints,article,accept,moreauthors,10pt,a4paper]{mdpi}

\firstpage{1}
\pubvolume{xx}
\issuenum{1}
\articlenumber{1}
\pubyear{2020}
\copyrightyear{2020}
\externaleditor{Academic Editor: name}
\history{Received: date; Accepted: date; Published: date}

\usepackage{amssymb}
\usepackage{hyperref}
\usepackage[hyphenbreaks]{breakurl}

\def\bal#1\eal{\begin{align}#1\end{align}}
\newcommand\beq{\begin{equation}}
\newcommand\eeq{\end{equation}}
\newcommand\beqa{\begin{eqnarray}}
\newcommand\eeqa{\end{eqnarray}}
\newcommand{\nn}{\nonumber\\}
\newcommand{\ep}{\={e}}

\newcommand{\GG}{\mathcal{G}}
\newcommand{\ex}{\text{ex}}
\newcommand{\dd}{\text{d}}
\newcommand{\eff}{\eta_{\text{eff}}}

\newcommand{\Sn}{\Omega}
\newcommand{\NN}{s}
\newcommand{\muM}{M}

\newcommand{\mn}{\muM_n}
\newcommand{\pure}{{\text{s}}}

\newcommand{\eed}{\bibliography{D:/Dropbox/Mis_Dropcumentos/bib_files/liquid}\end{document}}

\Title{Equation of State of Four- and Five-Dimensional Hard-Hypersphere Mixtures}

\Author{{Mariano} L\'{o}pez de Haro $^{1}$\orcidA{}, Andr\'es Santos $^{2,}$*\orcidB{} and Santos B. Yuste $^{2}$\orcidC{}}

\AuthorNames{Mariano L\'{o}pez de Haro, Andr\'es Santos  and Santos B. Yuste}

\address{%
$^{1}$ \quad Instituto de Energ\'{\i}as Renovables, Universidad Nacional Aut\'onoma de M\'exico (U.N.A.M.),
Temixco, Morelos 62580, Mexico; malopez@unam.mx\\
$^{2}$ \quad Departamento de F\'{\i}sica and Instituto de Computaci\'on Cient\'ifica Avanzada (ICCAEx), Universidad de Extremadura,
E-06006 Badajoz, Spain; santos@unex.es}

\corres{Correspondence: andres@unex.es; Tel.: +34-924-289-651}

\abstract{New proposals for the equation of state of four- and five-dimensional hard-hypersphere mixtures in terms of the equation of state of the corresponding monocomponent hard-hypersphere fluid are introduced. Such proposals (which are constructed in such a way so as to yield the exact third virial coefficient) extend, on the one hand, recent similar formulations for hard-disk and (three-dimensional) hard-sphere mixtures and, on the other hand, two of our previous proposals also linking the mixture equation of state and the one of the monocomponent fluid but unable to reproduce the exact third virial coefficient. The old and new proposals are tested by comparison with published molecular dynamics and Monte Carlo simulation results and their relative merit is~evaluated.}


\begin{document} 

\section{Introduction}
The interest in studying systems of $d$-dimensional hard spheres has been present for many decades and still continues to stimulate intensive research~\cite{FRW85,L86,FRW86,KF86,WRF87,BR88,EF88,CFP91,FP99,PS00,YFP00,CKPUZ16,SH05,FI81,L84,MT84,BC86,BC87,R87,R88,ASV89,SMS89,SM90,GGS90,LM90,MSAV91,GGS91,GGS92,VMN99,BMC99,FSL01,SYH99,MP99,S00,YSH00,GAH01,
SYH02,RHS04,SHY05,BWK05,BW05,LB06,HYS06,BW07,RHS07,WBT07,RS07,HYS08,BCW08,AKV08,RRHS08,vMCFC09,LBW10,RS11,LS11,ER11,BW12,BW13,BW16,AR13,A14,HHL15,S16,SYHO17,A17,I18,SYH01,RH64a,LB82,J82,LZKH91,EAGB02,
BMV04,CM04a,CM04b,BMV05,CM05,L05,CM06,ZM16b,SDST06,TS06,TS06a,TUS06,PZ06,SST08,vMFC09,AW10,TS10,ZT13,KBS19,BCK19,SYHOO14,BMS85,CB86,L05b}.
This interest is based on the versatility of such systems that allows one to gain insight into, among~other things, the~equilibrium and dynamical properties of simple fluids, colloids, granular matter, and~glasses with which they share similar phenomenology. For~instance, it is well known that all $d$-dimensional hard-sphere systems undergo a fluid-solid phase transition which occurs at smaller packing fractions as the spatial dimension is increased. This implies that mean-field-like descriptions of this transition  become mathematically simpler and more accurate as one increases the number of dimensions. Additionally, in~the limit of infinite dimension one may even derive analytical results for the thermodynamics, structure, and~phase transitions of such hypersphere fluids~\cite{FRW85,L86,FRW86,KF86,WRF87,BR88,EF88,CFP91,FP99,PS00,YFP00,CKPUZ16,SH05}. In~particular, the~equation of state (EOS)  truncated  at the level of the second virial coefficient becomes exact in this limit~\cite{CFP91}.

While of course real experiments cannot be performed in these systems, they are amenable to computer simulations and theoretical developments. Many aspects concerning hard hyperspheres have been already dealt with, such as thermodynamic and structural properties~\cite{FI81,L84,MT84,BC86,BC87,R87,R88,ASV89,SMS89,SM90,GGS90,LM90,MSAV91,GGS91,GGS92,VMN99,BMC99,FSL01,SYH99,MP99,S00,YSH00,GAH01,SYH01,SYH02,RHS04,SHY05,SH05,BWK05,BW05,
LB06,HYS06,BW07,RHS07,WBT07,RS07,HYS08,BCW08,AKV08,RRHS08,vMCFC09,LBW10,RS11,LS11,ER11,BW12,BW13,AR13,A14,HHL15,S16,BW16,A17,SYHO17,I18}, virial \mbox{coefficients~\cite{RH64a,LB82,J82,LZKH91,SYH01,EAGB02,BMV04,CM04a,CM04b,BMV05,CM05,L05,CM06,ZM16b}}, and~disordered packings~\cite{SDST06,TS06a,TS06,TUS06,PZ06,SST08,vMFC09,vMCFC09,AW10,TS10,ZT13,KBS19} or glassy behavior~\cite{SDST06,TS06,CKPUZ16,BCK19}. Nevertheless, due to the fact that (except in the infinite dimensional case) no exact analytical results are available, efforts to clarify or reinforce theoretical developments are worth pursuing. In~the case of mixtures of hard hyperspheres this is particularly important since, comparatively speaking, the~literature pertaining to them is not very abundant. To~the best of our knowledge, the~first paper reporting an  (approximate) EOS for  additive binary hard-hypersphere fluid mixtures is the one by \mbox{Gonz\'alez {et al.}}~\cite{GGS92}, in~which they used the overlap volume approach. What they did was to compute the partial direct correlation functions through an interpolation between the exact low-density and the Percus--Yevick high-density behavior of such functions to produce a Carnahan--Starling-like EOS which they subsequently compared with the (very few then) available simulation data for additive hard-disk mixtures. A~few years later, we~\cite{SYH99,HYS08} proposed an ansatz for the contact values of the partial radial distribution functions complying with some exact limiting conditions to derive an EOS (henceforth denoted with the label ``e1'') of a multicomponent $d$-dimensional hard-sphere fluid  in terms of the one of the single monocomponent system. To~our knowledge, the~first simulation results for the structural and thermodynamic properties of additive hard-hypersphere mixtures were obtained via molecular dynamics (MD) for a few binary mixtures in four and five spatial dimensions by Gonz\'alez-Melchor {et al.}~\cite{GAH01}, later confirmed by Monte Carlo (MC) computations by Bishop and Whitlock~\cite{BW05}. The~comparison between such simulation results and our e1 EOS~\cite{SYH99} led to very reasonable agreement.
Later, we proposed a closely related EOS (henceforth denoted with the label ``e2'') stemming from additional exact limiting conditions applied to the contact values of the partial radial distribution functions~\cite{HYS08,SYH02}. {A limitation of these proposals is that, except~in the three-dimensional case, they are unable to yield the exact third virial coefficient. As~shown below, extensions of these EOS (denoted as ``\ep1'' and ``\ep2'') complying with the requirement that the third virial coefficient computed from them is the exact one, may be introduced with little difficulty.}
More recently, we have developed yet another approximate EOS (henceforth denoted with the label ``sp'') for $d$-dimensional hard-sphere fluid mixtures~\cite{SYHOO14,SYHO17,S16}, and~newer simulation results for hard hypersphere mixtures have also been obtained~\cite{BW12,BW13,BW16}. It is the aim of this paper to carry out a comparison between  available simulation data for binary additive four- and five-dimensional hypersphere fluid mixtures and our theoretical~proposals.

The paper is organized as follows. In~order to make it self-contained, in~Section~\ref{sec2} we provide a brief outline of the approaches we have followed to link the EOS of a polydisperse $d$-dimensional hard-sphere mixture and that of the corresponding monocomponent system. Section~\ref{sec3} presents the specific cases of four and five spatial dimensions, the~choice of the EOS of the monocomponent system to complete the mapping, and~the comparison with the simulation data. We close the paper in Section~\ref{sec4} with a discussion of the results and some concluding~remarks.

\section{Mappings Between the Equation of State of the Polydisperse Mixture and that of the Monocomponent~System}
\label{sec2}
Let us  begin by considering a mixture of additive hard spheres in $d$ dimensions with an arbitrary number $\NN$ of components.
This number $\NN$ may even be infinite, {i.e.}, the~system may also be a polydisperse mixture with a continuous size distribution.
The {additive}  hard core of the interaction between a sphere of species $i$ and a sphere of species $j$ is $\sigma_{ij}=\frac{1}{2}(\sigma _{i}+\sigma _{j})$,
where the diameter of a sphere of species $i$ is $\sigma _{ii}=\sigma _{i}$. Let the number density of the mixture be $\rho $ and the mole fraction of species $i$ be
$x_{i}=\rho _{i}/\rho $, where $\rho_i$ is the number density of species $i$. In~terms of these quantities, the~packing fraction
is given by  $\eta =v_{d}\rho \muM_d $, where $v_{d}=(\pi/4)^{d/2}/\Gamma (1+d/2)$ is the volume of a $d$-dimensional sphere of unit diameter, {$\Gamma (\cdot)$} is the Gamma function, and~
$\muM_n \equiv \langle \sigma^n\rangle=\sum_{i=1}^\NN x_{i}\sigma_{i}^{n}$
denotes the $n$th moment of the diameter~distribution.

Unfortunately, no exact explicit EOS for a fluid mixture of $d$-dimensional hard spheres is available. The~ (formal) virial expression for  such EOS involves only the contact values
$g_{ij}(\sigma_{ij}^+)$ of the radial distribution functions $g_{ij}(r)$, where $r$ is the
distance, namely
\begin{equation}
Z(\eta )=1+\frac{2^{d-1}}{\muM_d} \eta\sum_{i,j=1}^\NN
x_{i}x_{j}{\sigma _{ij}^{d} } g_{ij}(\sigma_{ij}^+),
\label{1}
\end{equation}
where $Z=p/\rho k_{B}T$ is the compressibility factor of the mixture, $p$ being the pressure, $k_{B}$ the Boltzmann constant, and~$T$ the absolute temperature. Hence, a~useful way to obtain approximate expressions for the EOS of the mixture is to
propose or derive approximate expressions for the contact values $g_{ij}(\sigma_{ij}^+)$. We have already followed this route and the outcome is briefly described in Sections \ref{sec:2.1} and \ref{sec:2.2}. More details may be found in Ref.\ \cite{HYS08} and references~therein.

\subsection{The e1~Approximation}
\label{sec:2.1}

The basic assumption is  that, at~a given packing fraction $\eta$, the~dependence of $g_{ij}(\sigma_{ij}^+)$ on
the sets of $\{\sigma _{k}\}$ and $\{x_{k}\}$ takes place \textit{only} through the scaled quantity
\begin{equation}
z_{ij}\equiv \frac{\sigma _{i}\sigma_{j}}{\sigma
_{ij}}\frac{\muM_{d-1}}{\muM_d},
\label{zij}
\end{equation}
which we express as
\begin{equation}
g_{ij}(\sigma_{ij}^+)=\GG(\eta,z_{ij}),
\label{5}
\end{equation}
where the function $\GG(\eta,z)$ is \textit{universal}, {i.e.}, it is a common function for all the pairs $(i,j)$, regardless
of the {composition and} number of components of the mixture. Next, making use of some consistency conditions, we have derived two approximate
expressions for the EOS of the mixture. The~first one, labeled ``e1,'' indicating  that (i) the contact
values $g_{ij}(\sigma_{ij}^+)$ used are an \emph{extension} of the monocomponent fluid contact value $g_\pure\equiv g(\sigma^+)$ and that (ii) $\GG(\eta,z)$ is
a \emph{linear} polynomial in $z$, leads to an EOS that exhibits an excellent agreement with simulations in $2$, $3$, $4$, and~$5$ dimensions, provided
that an accurate $g_\pure$ is used as input~\cite{SYH99,SYH01,GAH01,BW12,BW16}. This EOS may be written as
\begin{equation}
Z_\text{e1}(\eta
)=1+\frac{\eta}{1-\eta}2^{d-1}(\Sn_0-\Sn_1)+\left[Z_\pure(\eta
)-1\right]\Sn_1,
 \label{Ze1}
\end{equation}
where   the coefficients $\Sn_{m}$ depend only on the composition of
the mixture and are defined by
\begin{equation}
\Sn_{m}=2^{-(d-m)}\frac{\muM_{d-1}^{m}}{\muM_{d}^{m+1}}\sum_{n=0}^{d-m}\binom{d-m}{n}
{\muM_{n+m} }{\muM_{d-n}}.
\label{Omega}
\end{equation}

It is interesting to point out that from Equation~(\ref{Ze1}) one may write the virial
coefficients of the mixture $B_n$, defined by
\beq
Z(\rho)=1+\sum_{n=1}^\infty B_{n+1} \rho^{n},
\label{virial}
\eeq
in terms of the (reduced) virial coefficients of the single component fluid $b_n$
defined by
\beq
Z_\pure(\eta)=1+\sum_{n=1}^\infty b_{n+1} \eta^{n}.
\label{virial_s}
\eeq
The result is
\begin{equation}
\bar{B}_n^{\text{e1}}=\Sn_1
b_n+2^{d-1}(\Sn_0-\Sn_1),
\label{Virial}
\end{equation}
where $\bar{B}_n\equiv B_n/(v_d M_d)^{n-1}$ are {reduced} virial coefficients. Since $b_2=2^{d-1}$, Equation \eqref{Virial} yields the \emph{exact} second virial coefficient~\cite{S16}
\beq
\label{B2}
\bar{B}_2= 2^{d-1}\Sn_0.
\eeq
In general, however, $\bar{B}_n^{\text{e1}}$ with $n\geq 3$ are only approximate.
In particular,
\begin{subequations}
\label{B3e14D&5D}
\beq
\label{B3e14D}
\bar{B}_3^{\text{e1}}=1+\left(\frac{b_3}{4}+2\right)\frac{\muM_1\muM_3}{\muM_4}+3\frac{\muM_2^2}{\muM_4}+\left(\frac{3b_3}{4}-6\right)\frac{\muM_2\muM_3^2}{\muM_4^2},\quad (d=4),
\eeq
\beq
\label{B3e15D}
\bar{B}_3^{\text{e1}}=1+\frac{65}{4}\frac{\muM_1\muM_4}{\muM_5}+10\frac{\muM_2\muM_3}{\muM_5}
+45\frac{\muM_2\muM_4^2}{\muM_5^2}+\frac{135}{4}\frac{\muM_3^2\muM_4}{\muM_5^2},\quad (d=5).
\eeq
\end{subequations}
In Equation \eqref{B3e14D},
\beq
\label{b34D}
b_3=64\left(\frac{4}{3}-\frac{3\sqrt{3}}{2\pi}\right),\quad (d=4),
\eeq
is the reduced third virial coefficient of a monocomponent four-dimensional fluid, while in Equation~\eqref{B3e15D} we have taken into account that $b_3=106$ if $d=5$.

It is interesting to note that, by~eliminating $\Sn_0$ and $\Sn_1$ in favor of $\bar{B}_2$ and  $\bar{B}_3^{\text{e1}}$, Equation \eqref{Ze1} can be rewritten as
\begin{equation}
Z_\text{e1}(\eta)=1+\frac{\eta}{1-\eta}\frac{b_3 \bar{B}_2- b_2 \bar{B}_3^{\text{e1}}}{b_3-b_2}+\left[Z_\pure(\eta)-1\right]\frac{\bar{B}_3^{\text{e1}}-\bar{B}_2}{b_3-b_2}.
\label{Ze1bis}
\end{equation}

\subsection{The e2~Approximation}
\label{sec2.1}

The second approximation, labeled ``e2,'' similarly indicates that (i) the resulting contact values represent an \emph{extension} of the single
component contact value $g_\pure$ and that (ii) ${\GG}(\eta, z)$ is a \emph{quadratic} polynomial in $z$. In~this case, one also gets a closed expression
for the compressibility factor in terms of the packing fraction $\eta $ and the first few moments $\mn$, $n\leq d$.  Such an expression is
\beq
Z_{\text{e2}}(\eta) =Z_{\text{e1}}(\eta)-(\Sn_{2}-\Sn_{1})\left[Z_\pure(\eta)\left(1-2^{d-2}\eta\right)
-1-2^{d-2}\frac{\eta}{1-\eta}\right].
\label{Ze2}
\eeq
The associated (reduced) virial coefficients are
\beq
\label{Virial_e2}
\bar{B}_n^{\text{e2}}=\bar{B}_n^{\text{e1}}-(\Sn_{2}-\Sn_{1})\left[b_n-2^{d-2}\left(1+b_{n-1}\right)\right].
\eeq
Again, since $b_1=1$ and $b_2=2^{d-1}$, the~exact second virial coefficient, Equation \eqref{B2}, is recovered for any dimensionality. Additionally, in~the case of spheres ($d=3$), $b_3=10$ and thus $\bar{B}_3^{\text{e1}}=\bar{B}_3^{\text{e2}}=4\Sn_0+6\Sn_1$, which is the exact result for that dimensionality.
In the cases of $d=4$ and $d=5$, one has
\begin{subequations}
\label{B3e24D&5D}
\beq
\label{B3e24D}
\bar{B}_3^{\text{e2}}=1+\left(\frac{b_3}{2}-7\right)\frac{\muM_1\muM_3}{\muM_4}+3\frac{\muM_2^2}{\muM_4}+\left(b_3-15\right)\frac{\muM_2\muM_3^2}{\muM_4^2}+\left(18-\frac{b_3}{2}\right)\frac{\muM_3^4}{\muM_4^3},
\quad (d=4),
\eeq
\beq
\label{B3e25D}
\bar{B}_3^{\text{e2}}=1+\frac{25}{2}\frac{\muM_1\muM_4}{\muM_5}+10\frac{\muM_2\muM_3}{\muM_5}+\frac{75}{2}\frac{\muM_2\muM_4^2}{\muM_5^2}+\frac{45}{2}\frac{\muM_3^2\muM_4}{\muM_5^2}+\frac{45}{2}\frac{\muM_3\muM_4^3}{\muM_5^3},
\quad (d=5).
\eeq
\end{subequations}

It is also worthwhile noting that $\Sn_1=\Sn_2$ in the case of disks ($d=2$) and thus $Z_{\text{e1}}(\eta)=Z_{\text{e2}}(\eta)$ for those~systems.

\subsection{Exact Third Virial Coefficient. Modified Versions of the e1 and e2~Approximations}
As said above, both $\bar{B}_3^{\text{e1}}$ and $\bar{B}_3^{\text{e2}}$ differ from the exact third virial coefficient, except~in the three-dimensional case ($d=3$). The~exact expression is~\cite{S16}
\begin{subequations}
\label{B3exact}
\begin{equation}
\bar{B}_{3}=\frac{1}{\muM_d^2}\sum_{i,j,k=1}^\NN x_{i}x_{j}x_{k}\widehat{B}_{ijk},
\label{Thirdvir}
\end{equation}
\beq
\widehat{B}_{ijk}=\frac{d^2 }{3} 2^{5d/2-1} \Gamma (d/2)\left(\sigma_{ij}\sigma_{ik}\sigma_{jk}\right)^{d/2} \int_{0}^{\infty} \frac{\dd\kappa}{\kappa^{1+d/2}} J_{d/2}(\kappa\sigma_{ij}) J_{d/2}(\kappa\sigma_{ik})J_{d/2}(\kappa\sigma_{jk}),
\label{PartThirdvir}
\eeq
\end{subequations}
where
$J_{n} (\cdot)$ is the Bessel function of the first kind of order $n$.

For odd dimensionality, it turns out that the composition-independent
coefficients $\widehat{B}_{ijk}$ have a polynomial dependence on $\sigma_i$, $\sigma_j$, and~$\sigma_k$. As~a consequence, the~third virial coefficient $\bar{B}_{3}$ can be expressed in terms of moments $\muM_n$ with $1\leq n\leq d$. In~particular~\cite{S16},
\beq
\bar{B}_{3}=1+10\frac{\muM_1\muM_4}{\muM_5}+20\frac{\muM_2\muM_3}{\muM_5}+25\frac{\muM_2\muM_4^2}{\muM_5^2}+50\frac{\muM_3^2\muM_4}{\muM_5^2},\quad (d=5).
\eeq
On the other hand, for~even dimensionality the dependence of $\widehat{B}_{ijk}$ on $\sigma_i$, $\sigma_j$, and~$\sigma_k$ is more complex than polynomial. In~particular, for~a binary mixture ($\NN=2$) with $d=4$ one has
\begin{subequations}
\label{B111&B112}
\beq
\label{B111}
\widehat{B}_{111}=b_3\sigma_1^8,\quad (d=4),
\eeq
\bal
\label{B112}
\widehat{B}_{112}=&\sigma_1^8\frac{16(1+q)^4}{3}\Big[1
-\frac{1}{8\pi}(1-q)(3+q)(5+2q+q^2)
\arcsin\frac{1}{1+q}
-\frac{\sqrt{q(2+q)}}{24\pi(1+q)^4}\left(45+138q\right.\nn
&\left.+113q^2+68q^3+47q^4+18q^5+3q^6\right)\Big],\quad (d=4),
\eal
\end{subequations}
where $q\equiv\sigma_2/\sigma_1$ is the size ratio. The~expressions for $\widehat{B}_{222}$ and $\widehat{B}_{122}$ can be obtained from Equations~\eqref{B111} and \eqref{B112}, respectively, by~the replacements $\sigma_1\to\sigma_2$, $q\to q^{-1}$.

\begin{figure}[H]
\centering
\includegraphics[width=.45\columnwidth]{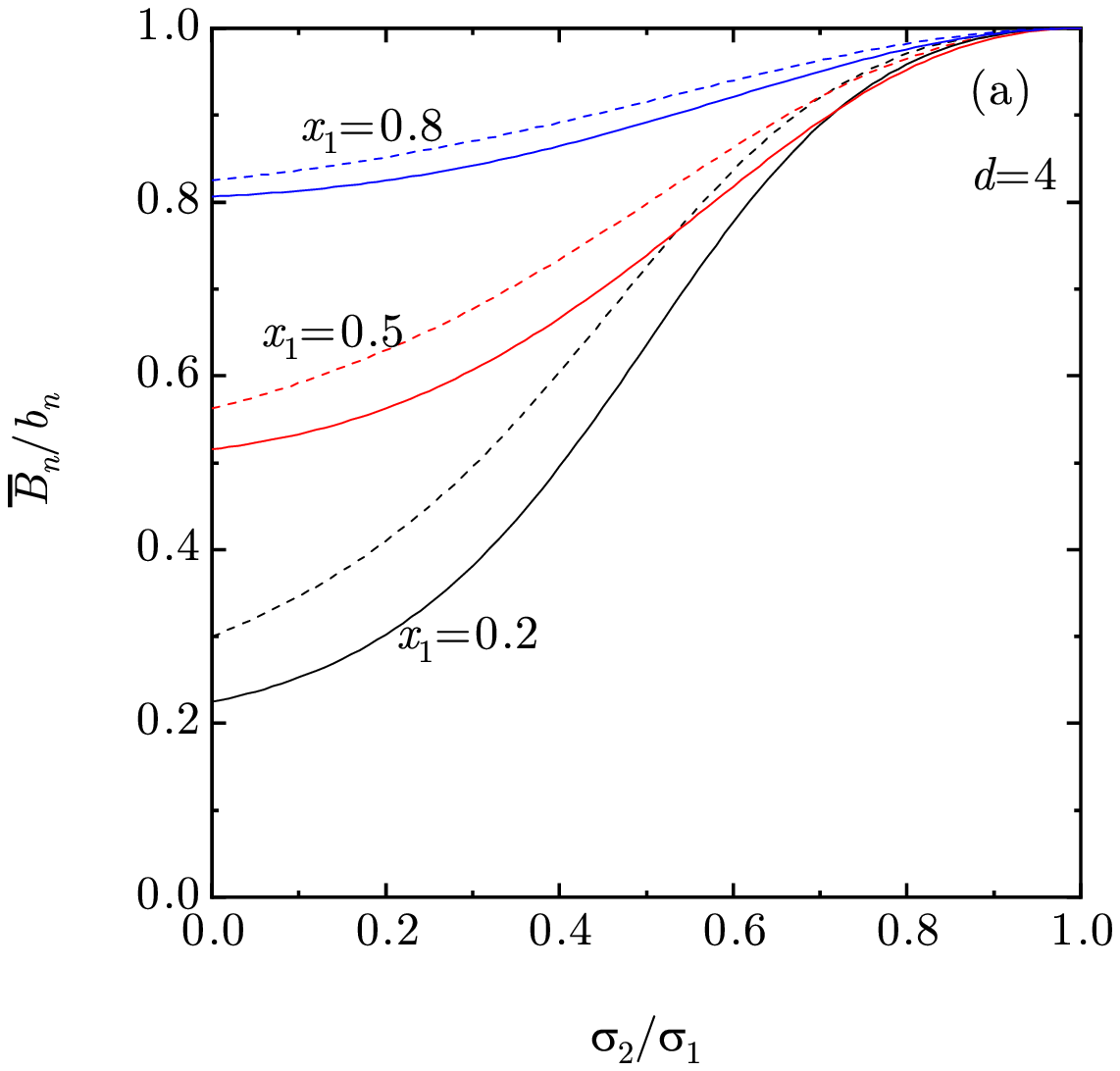}\hspace{1cm}\includegraphics[width=.45\columnwidth]{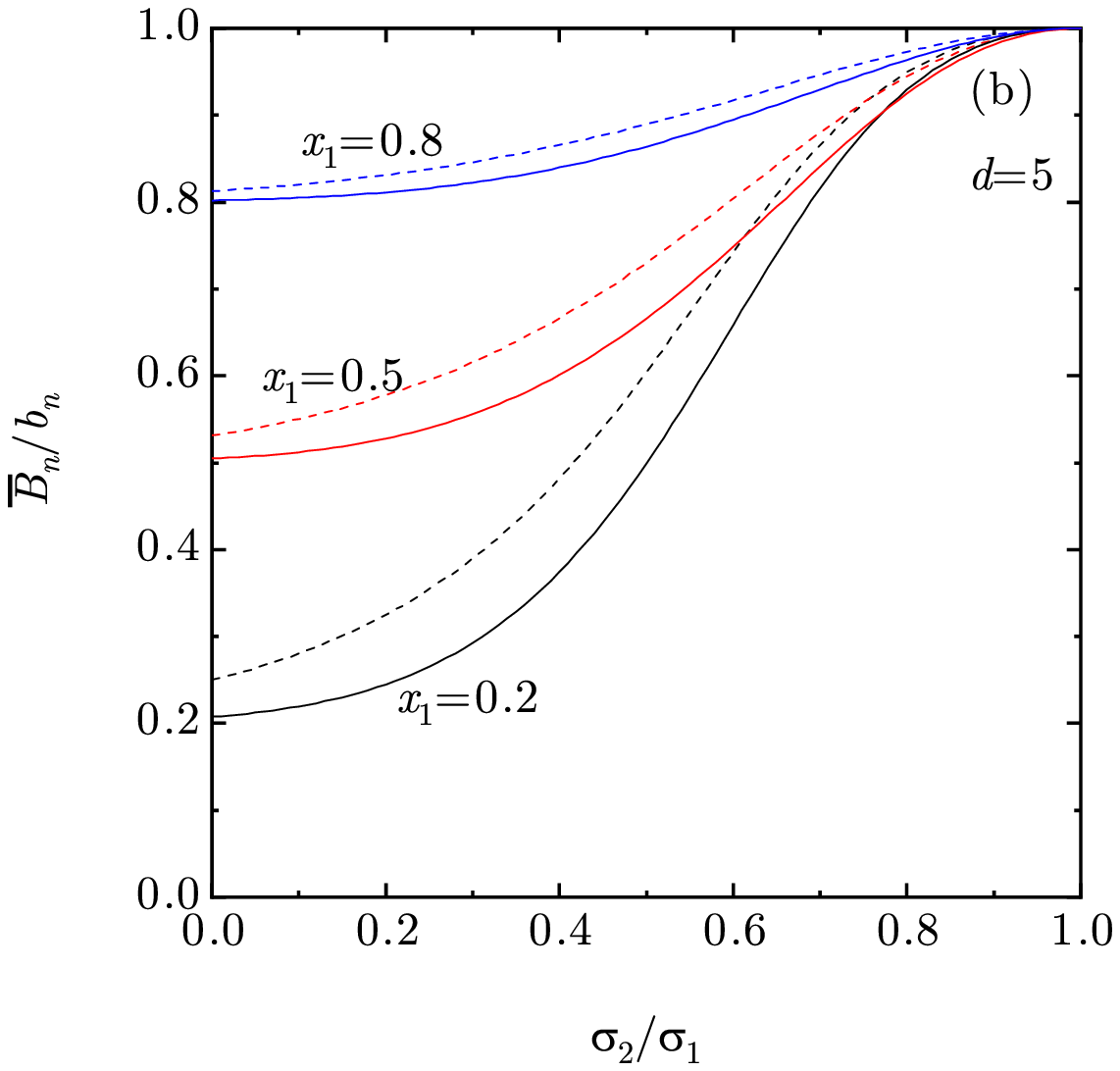}
\caption{Plot of the ratios $\bar{B}_2/b_2$ (dashed lines) and  $\bar{B}_3/b_3$ (solid lines)  vs the size ratio $\sigma_2/\sigma_1$ for binary mixtures with mole fractions $x_1=0.2$, $0.5$, and~$0.8$. Panel (\textbf{a}) corresponds to $d=4$, while panel (\textbf{b}) corresponds to $d=5$.\label{figB2B3}}
\end{figure}

Figure~\ref{figB2B3} displays the size-ratio dependence of the exact second and third virial coefficients for three representative binary compositions of four- and five-dimensional systems. The~degree of bidispersity of a certain binary  mixture can be measured by the distances $1-\bar{B}_2/b_2$ and $1-\bar{B}_3/b_3$. In~this sense, Figure~\ref{figB2B3} shows that, as~expected, the~degree of bidispersity  grows monotonically as the small-to-big size ratio decreases at a given mole fraction. It also increases as the concentration of the big spheres decreases at a given size ratio,
except if the latter ratio is close enough to~unity.

\begin{figure}[H]%
\centering
\includegraphics[width=.45\columnwidth]{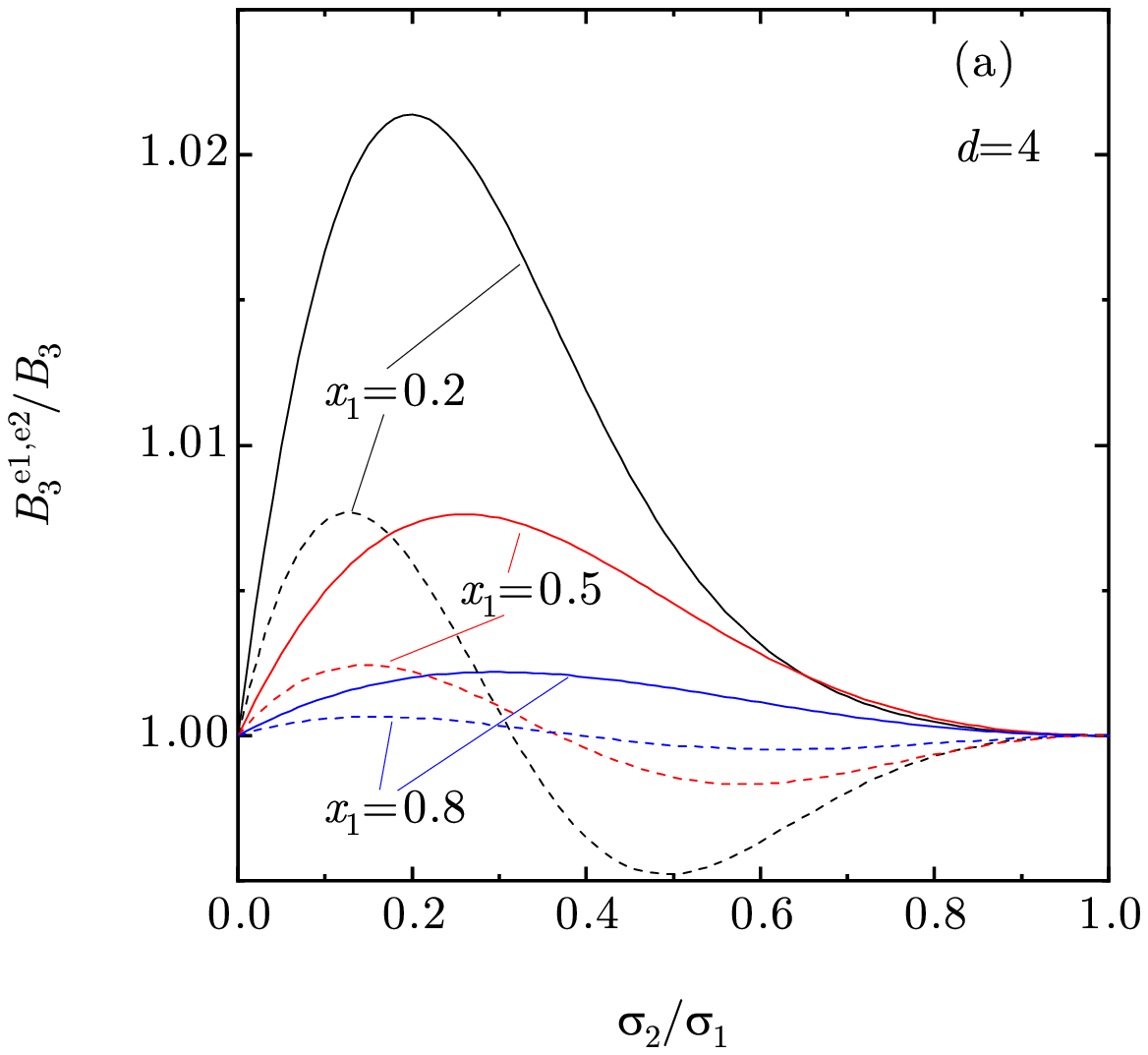}\hspace{1cm}\includegraphics[width=.45\columnwidth]{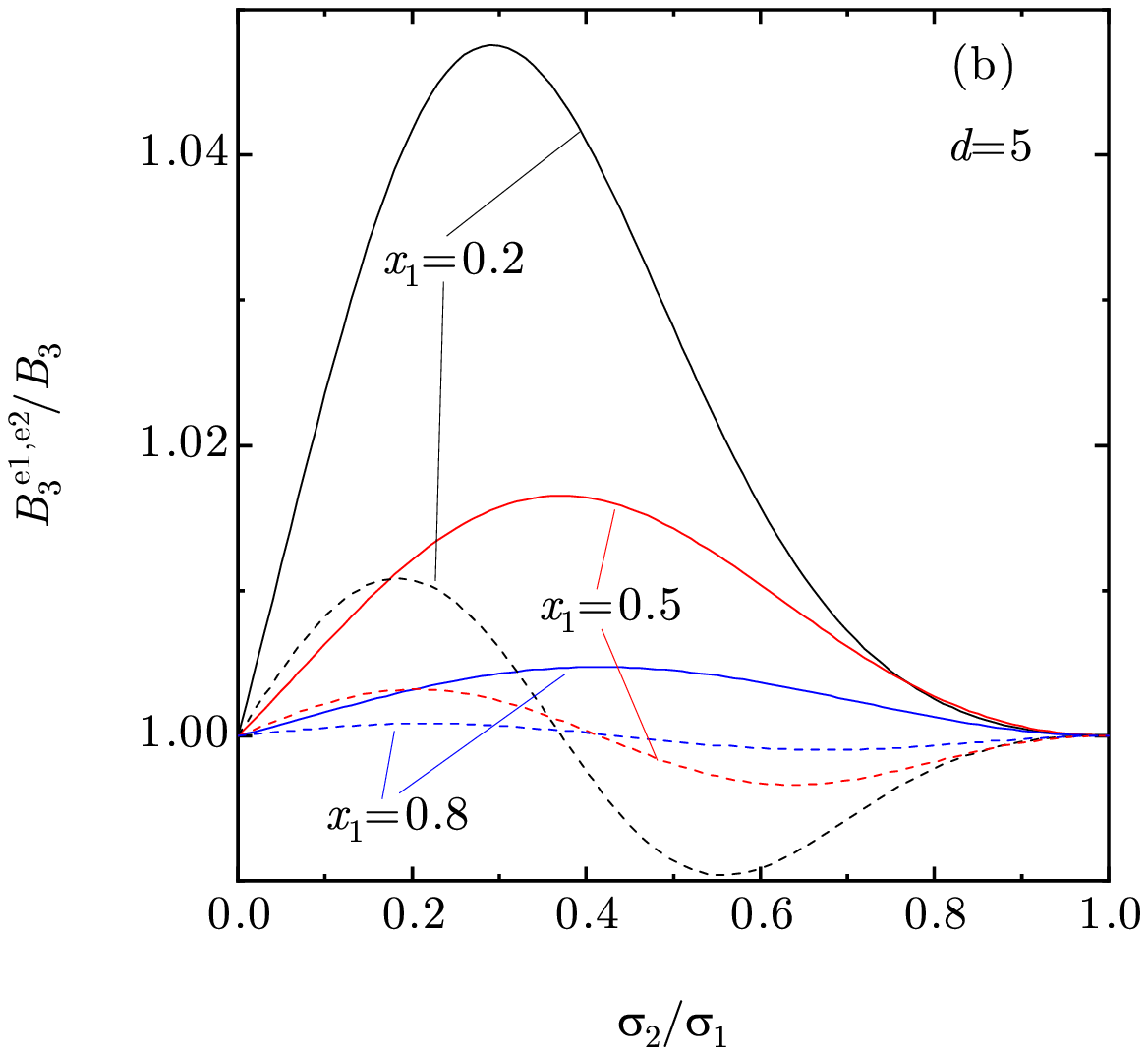}
\caption{Plot of the ratios $B_3^{\text{e1}}/B_3$ (solid lines) and $B_3^{\text{e2}}/B_3$ (dashed lines) vs the size ratio $\sigma_2/\sigma_1$ for binary mixtures with mole fractions $x_1=0.2$, $0.5$, and~$0.8$. Panel (\textbf{a}) corresponds to $d=4$, while panel (\textbf{b}) corresponds to $d=5$.\label{fig0}}
\end{figure}

To assess the quality of the approximate coefficients \eqref{B3e14D&5D} and \eqref{B3e24D&5D}, we plot in Figure~\ref{fig0}  the ratios $B_3^{\text{e1}}/B_3$ and $B_3^{\text{e2}}/B_3$ as functions of the size ratio $\sigma_2/\sigma_1$ for the same three representative binary compositions as in Figure~\ref{figB2B3}. As~we can observe, both the e1 and e2 approximations predict values for the third virial coefficient in overall good agreement with the exact values, especially as the concentration of the big spheres increases. The~e1 approximation overestimates $B_3$ and generally performs worse than the e2 approximation, which tends to overestimate (underestimate) $B_3$ if the concentration of the big spheres is sufficiently small (large). Additionally, the~agreement is better in the four-dimensional case than for five-dimensional hyperspheres. The~latter point is relevant because, as~said before, the~exact expressions of $B_3$ for $d=4$ are relatively involved [see Equations \eqref{B111&B112} in the binary case], whereas $B_3^{\text{e1}}$ and $B_3^{\text{e2}}$ are just simple combinations of moments [see Equations \eqref{B3e14D} and~\eqref{B3e24D}].
The structure of Equation \eqref{Ze1bis}  suggests the introduction of a  \emph{modified}  version (henceforth labeled as ``\ep1'') of the e1 EOS by replacing the approximate third virial coefficient $\bar{B}_3^{\text{e1}}$ by the exact one. More specifically,
\beq
Z_{\text{\ep1}}(\eta)=Z_{\text{e1}}(\eta)+\frac{\bar{B}_3-\bar{B}_3^{\text{e1}}}{b_3-b_2}\left[Z_\pure(\eta)-1-b_2\frac{\eta}{1-\eta}\right].
\label{Ze1'}
\eeq
Analogously, we introduce the modified version (``\ep2'') of the e2 approximation as
\beq
Z_{\text{\ep2}}(\eta)=Z_{\text{e2}}(\eta)+\frac{\bar{B}_3-\bar{B}_3^{\text{e2}}}{b_3-b_2}\left[Z_\pure(\eta)-1-b_2\frac{\eta}{1-\eta}\right].
\label{Ze2'}
\eeq
By construction, both $Z_{\text{\ep1}}(\eta)$ and $Z_{\text{\ep2}}(\eta)$ are consistent with the exact second and third virial coefficients. Moreover, $Z_{\text{\ep1}}(\eta)=Z_{\text{\ep2}}(\eta)$ for $d=2$, while $Z_{\text{\ep1}}(\eta)=Z_{\text{e1}}(\eta)$ and $Z_{\text{\ep2}}(\eta)=Z_{\text{e2}}(\eta)$ for $d=3$.

\subsection{The sp~Approximation}

Additionally, in previous work~\cite{SYHOO14,S16,SYHO17}, we have adopted an  approach to relate the EOS of the polydisperse mixture of $d$-dimensional
hard spheres to the one of the monocomponent fluid which differs from the e1 and e2 approaches in that it does not make use of Equation \eqref{1}. This involves expressing the excess free energy per particle ($a^\ex$) of a polydisperse mixture of packing fraction $\eta$ in terms
of the one of the corresponding monocomponent fluid ($a_\pure^\ex$) of an effective packing fraction $\eta_{\text{eff}}$~as
\beq
\frac{a^\ex(\eta)}{k_BT}+\ln(1-\eta)=\frac{\alpha}{\lambda}\left[\frac{a_\pure^\ex(\eff)}{k_BT}+\ln(1-\eff)\right].
\label{17}
\eeq
In Equation~(\ref{17}), $\eta_{\text{eff}}$ and  $\eta$ are related through
\begin{equation}
\frac{\eta_{\text{eff}}}{1-\eta_{\text{eff}}}=\frac{1}{\lambda}\frac{\eta}{1-\eta},\quad \eff=\left[1+\lambda\left(\eta^{-1}-1\right)\right]^{-1},
\label{18a}
\end{equation}
while the parameters $\lambda$ and $\alpha$ are determined by imposing consistency with the (exact) second and third virial coefficients of the mixture, Equations \eqref{B2} and \eqref{B3exact}. More specifically~\cite{S16,SYHO17},
\begin{equation}
\lambda=\frac{\bar{B}_2-1}{b_2-1}\frac{b_3-2b_2+1}{\bar{B}_3-2\bar{B}_2+1},\quad
\alpha=\lambda^2\frac{\bar{B}_2-1}{b_2-1}.
\label{26}
\end{equation}
Note that the ratio $\eta/(1-\eta)$ represents a rescaled packing fraction, i.e.,~the ratio between
the volume occupied by the spheres and the remaining void
volume. Thus, according to Equation \eqref{18a}, the~effective monocomponent
fluid associated with a given mixture has a rescaled
packing fraction $\eff/(1-\eff)$ that is $\lambda$ times smaller
than that of the mixture.
Moreover, in~the case of three-dimensional hard-sphere mixtures, Equations \eqref{17}--\eqref{26} can be derived in the context of consistent fundamental-measure theories~\cite{S12,S12c,S16,SYHO17}.

Taking into account the thermodynamic relation
\beq
\label{Zfroma}
Z(\eta)=1+\eta\frac{\partial  a^\ex(\eta)/k_BT}{\partial \eta},
\eeq
the mapping between the compressibility factor of the $d$-dimensional monocomponent system ($Z_\pure$) and the approximate one of the  polydisperse mixture  that
is then obtained from Equation \eqref{17} may be expressed as
\beq
\eta Z_{\text{sp}}(\eta)-\frac{\eta}{1-\eta}=\alpha\left[\eff Z_\pure(\eff)-\frac{\eff}{1-\eff}\right],
\label{19}
\eeq
where a label ``sp'', motivated by the nomenclature already introduced in connection with the ``surplus'' pressure $\eta Z(\eta)-\eta/(1-\eta)$ \cite{S16}, has been added to distinguish this compressibility factor from the previous~approximations.

Equation \eqref{19} shares with Equations \eqref{Ze1'} and \eqref{Ze2'} the consistency with the exact second and third virial coefficients. On~the other hand, while $Z_{\text{\ep1}}(\eta)$ and $Z_{\text{\ep2}}(\eta)$ are related to the monocomponent compressibility factor $Z_\pure(\eta)$ evaluated at the same packing fraction $\eta$ as that of the mixture,  $Z_{\text{sp}}(\eta)$ is related to $Z_\pure(\eff)$ evaluated at a different (effective) packing fraction $\eff$.

\begin{figure}[H]%
\centering
\includegraphics[width=.45\columnwidth]{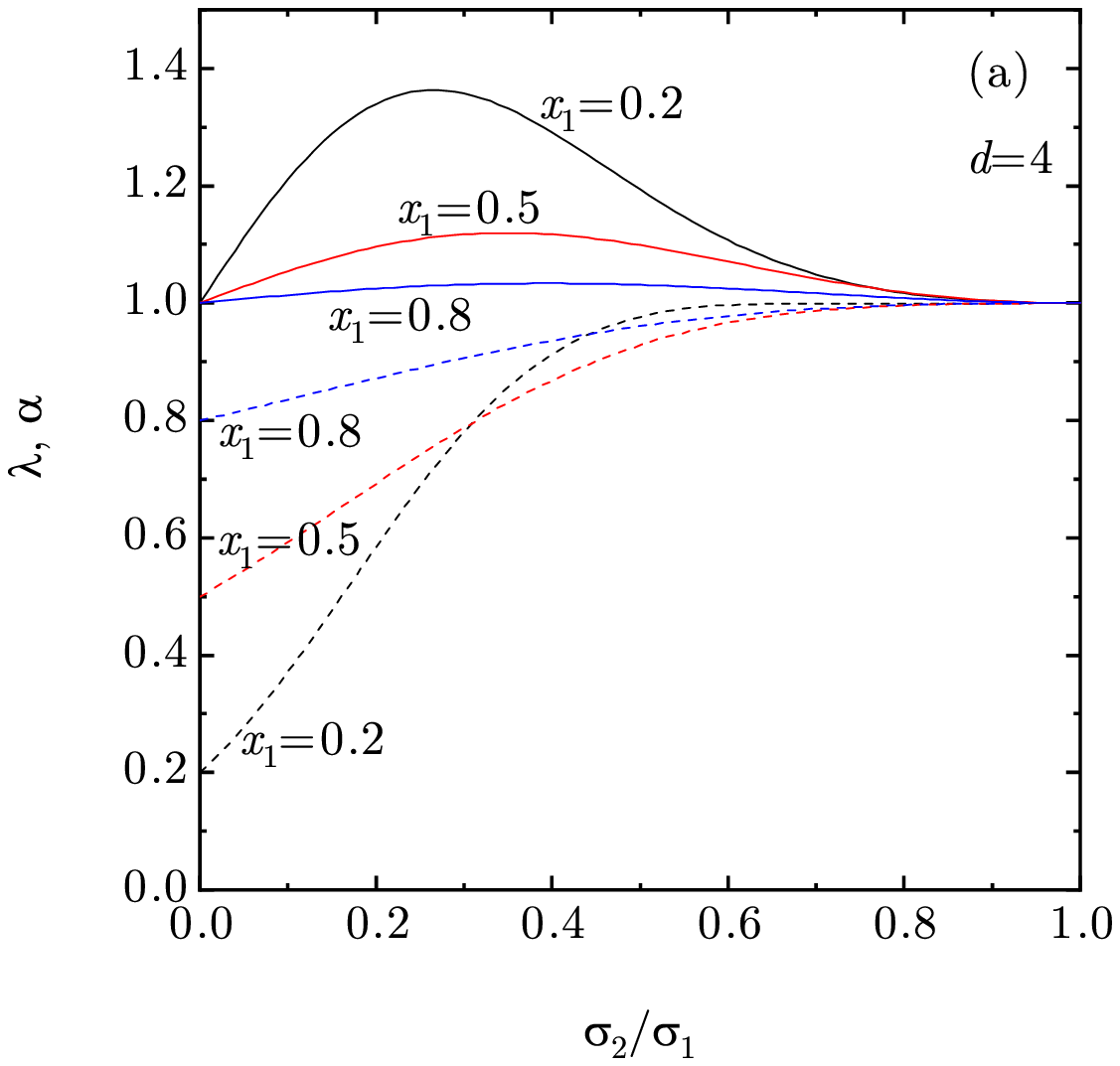}\hspace{1cm}\includegraphics[width=.45\columnwidth]{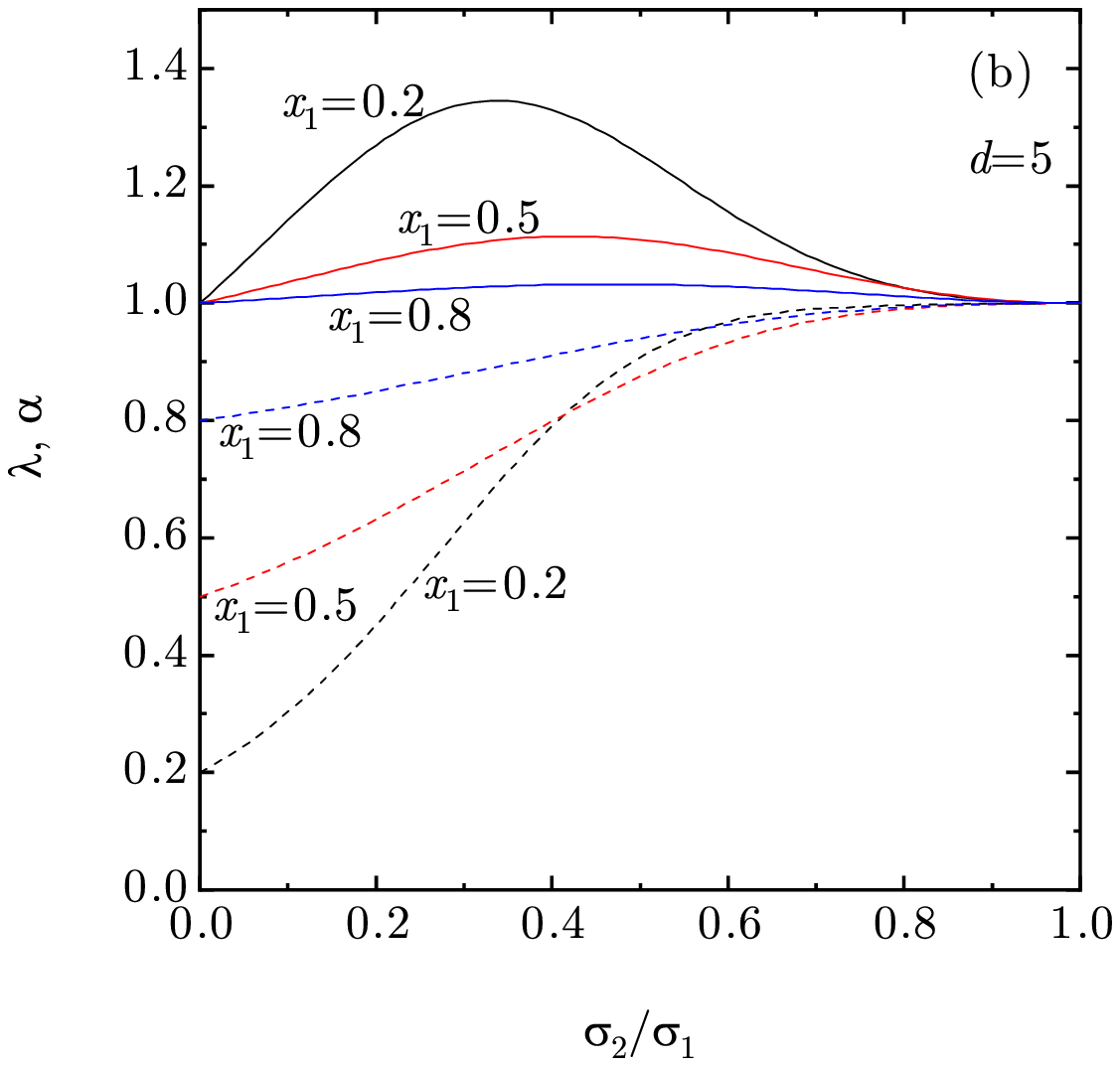}
\caption{Plot of the coefficients $\lambda$ (solid lines) and  $\alpha$ (dashed lines) [see Equation \eqref{26}] vs the size ratio $\sigma_2/\sigma_1$ for binary mixtures with mole fractions $x_1=0.2$, $0.5$, and~$0.8$. Panel (\textbf{a}) corresponds to $d=4$, while panel (\textbf{b}) corresponds to $d=5$.\label{fig_lambda_alpha}}
\end{figure}

Figure~\ref{fig_lambda_alpha} shows that $\lambda>1$, while  $\alpha<1$, except~if the mole fraction of the big spheres is large enough (not shown). According to Equations \eqref{18a} and \eqref{19}, this implies that (i) $\eff<\eta$ and (ii) the surplus pressure of the mixture at a packing fraction $\eta$ is generally smaller than that of the monocomponent fluid at the equivalent packing fraction $\eff$. It is also worthwhile noting that, in~contrast to what happens with $\bar{B}_2$ and $\bar{B}_3$ (see Figure~\ref{figB2B3}), $\lambda$ has a nonmonotonic dependence on the size ratio and $\alpha$ also exhibits a nonmonotonic behavior if  $x_1$ is small~enough.

While we have proved the sp approach to be successful for both hard-disk ($d=2$) \cite{SYHO17} and hard-sphere ($d=3$) \cite{SYHOO14} mixtures, one of our goals is to test it for $d=4$ and $d=5$ as~well.

\section{Comparison with Computer Simulation~Results}
\label{sec3}

In order to obtain explicit numerical results for the different approximations to the EOS of four- and five-dimensional hard-sphere mixtures, we require an expression for $Z_\pure(\eta)$. While other choices are available, we considered here the empirical proposal that works for both dimensionalities by Luban and Michels (LM) \cite{LM90}, which reads
\beq
\label{ZLM}
Z_\pure(\eta)=1 + b_2 \eta\frac{1+\left[b_3/b_2-\zeta(\eta)b_4/b_3\right]\eta}{1-\zeta(\eta)(b_4/b_3)\eta+\left[\zeta(\eta)-1\right](b_4/b_2)\eta^2},
\eeq
where $\zeta(\eta)=\zeta_0+\zeta_1\eta/\eta_{\text{cp}}$, $\eta_{\text{cp}}$ being the crystalline  close-packing value. The~values of $b_2$, $b_3$, $b_4$, $\zeta_0$, $\zeta_1$, and~$\eta_{\text{cp}}$ are given in Table~\ref{table0}.

\begin{table}[H]%
   \caption{Values of $b_2$--$b_4$, $\zeta_0$, $\zeta_1$, and~$\eta_{\text{cp}}$ for $d=4$ and $5$.\label{table0}}
\centering
\begin{tabular}{ccc}
\toprule
&\boldmath$d=4$&\boldmath$d=5$\\
\midrule
$b_2$&$8$&$16$\\
$b_3$&$2^6\left(\frac{4}{3}-\frac{3\sqrt{3}}{2\pi}\right)\simeq32.406$&$106$\\
$b_4$&$
2^9\left(2-\frac{27\sqrt{3}}{4\pi}+\frac{832}{45\pi^2}\right)
\simeq 77.7452
$
&$\frac{25\,315\,393}{8\,008}+\frac{3\,888\,425\sqrt{2}}{4\,004\pi}
-\frac{67\,183\,425\arccos(1/3)}{8\,008\pi}\simeq 311.183
$\\
$\zeta_0$&$1.2973(59)$&$1.074(16)$\\
$\zeta_1$&$-0.062(13)$&$0.163(45)$\\
$\eta_{\text{cp}}$&$\frac{\pi^2}{16}\simeq 0.617$&$\frac{\pi^2\sqrt{2}}{30}\simeq 0.465$\\
\bottomrule
   \end{tabular}
 \end{table}

In Table~\ref{table1} we list the systems whose compressibility factor has been obtained from simulation, either using MD~\cite{GAH01} or MC~\cite{BW12,BW16} methods. The~values of the corresponding coefficients $\bar{B}_2$ [see Equation \eqref{B2}], $\bar{B}_3$ [see Equations \eqref{B3exact}--\eqref{B111&B112}], $\lambda$, and~$\alpha$ [see Equation \eqref{26}] are also included.
We assigned a three-character label  to each system, where the first (capital) letter denotes the size ratio (A--F for $\sigma_2/\sigma_1=\frac{1}{4}$, $\frac{1}{3}$, $\frac{2}{5}$, $\frac{1}{2}$, $\frac{3}{5}$, and~$\frac{3}{4}$, respectively), the~second (lower-case) letter denotes the mole fraction (a, b, and~c for $x_1=0.25$, $0.50$, and~$0.75$, respectively), and~the digit ($4$ or $5$) denotes the~dimensionality.

\begin{table}[H]%
   \caption{Binary mixtures of four- and five-dimensional hard spheres studied through simulations (Monte Carlo---MC or molecular dynamics---MD) and the values of their coefficients $\bar{B}_2$ [see Equation \eqref{B2}], $\bar{B}_3$ [see Equations \eqref{B3exact}--\eqref{B111&B112}], $\lambda$, and~$\alpha$ [see Equation \eqref{26}].\label{table1}}
   \centering
\begin{tabular}{ccccccccc}
\toprule
$d$&Label&$\sigma_2/\sigma_1$&$x_1$&Simulation method&$\bar{B}_2$&$\bar{B}_3$&$\lambda$&$\alpha$\\
\midrule
$4$&Aa4&${1}/{4}$&$0.25$&MD \footnotemark[1]&$3.85618$&$12.2253$&$1.28824$&$0.677138$\\
&Ab4&${1}/{4}$&$0.50$&MD \footnotemark[1]&$5.21595$&$18.8828$&$1.10923$&$0.741033$\\
&Ac4&${1}/{4}$&$0.75$&MD \footnotemark[1]&$6.60436$&$25.6326$&$1.03810$&$0.862800$\\
&Ba4&${1}/{3}$&$0.25$&MD \footnotemark[1]&$4.42857$&$14.4931$&$1.28470$&$0.808392$\\
&Bb4&${1}/{3}$&$0.50$&MD \footnotemark[1]&$5.56098$&$20.2530$&$1.11943$&$0.816497$\\
&Bc4&${1}/{3}$&$0.75$&MD \footnotemark[1]&$6.77049$&$26.2935$&$1.04334$&$0.897356$\\
&Cb4&${2}/{5}$&$0.50$&MC \footnotemark[2]&$5.87285$&$21.5939$&$1.11692$&$0.868418$\\
&Da4&${1}/{2}$&$0.25$&MD \footnotemark[1]&$5.82895$&$20.8444$&$1.17876$&$0.958523$\\
&Db4&${1}/{2}$&$0.50$&MD \footnotemark[1] and MC \footnotemark[2]&$6.38235$&$23.9444$&$1.09883$&$0.928396$\\
&Dc4&${1}/{2}$&$0.75$&MD \footnotemark[1]&$7.15816$&$28.0333$&$1.04047$&$0.952376$\\
&Eb4&${3}/{5}$&$0.50$&MC \footnotemark[2]&$6.90085$&$26.5045$&$1.07078$&$0.966532$\\
&Fa4&${3}/{4}$&$0.25$&MD \footnotemark[1]&$7.55661$&$29.9061$&$1.03231$&$0.998173$\\
&Fb4&${3}/{4}$&$0.50$&MD \footnotemark[1]&$7.56231$&$29.9832$&$1.02894$&$0.992515$\\
&Fc4&${3}/{4}$&$0.75$&MD \footnotemark[1]&$7.73940$&$30.9790$&$1.01561$&$0.993060$\\\midrule
$5$&Aa5&${1}/{4}$&$0.25$&MD \footnotemark[1]&$6.30550$&$32.9426$&$1.24358$&$0.546995$\\
&Ab5&${1}/{4}$&$0.50$&MD \footnotemark[1]&$9.52439$&$57.2455$&$1.08739$&$0.671954$\\
&Ac5&${1}/{4}$&$0.75$&MD \footnotemark[1]&$12.7601$&$81.6145$&$1.02988$&$0.831562$\\
&Ba5&${1}/{3}$&$0.25$&MD \footnotemark[1]&$7.21951$&$37.7995$&$1.27656$&$0.675687$\\
&Bb5&${1}/{3}$&$0.50$&MD \footnotemark[1]&$10.0984$&$60.3097$&$1.10651$&$0.742645$\\
&Bc5&${1}/{3}$&$0.75$&MD \footnotemark[1]&$13.0411$&$83.1175$&$1.03739$&$0.863898$\\
&Cb5&${2}/{5}$&$0.50$&MC \footnotemark[3]$^{,4}$&$10.6565$&$63.6666$&$1.11369$&$0.798464$\\
&Da5&${1}/{2}$&$0.25$&MD \footnotemark[1]&$9.89286$&$55.1378$&$1.22316$&$0.886983$\\
&Db5&${1}/{2}$&$0.50$&MD \footnotemark[1] and MC \footnotemark[3]$^{,5}$&$11.6818$&$70.5615$&$1.10812$&$0.874437$\\
&Dc5&${1}/{2}$&$0.75$&MD \footnotemark[1]&$13.7964$&$88.0120$&$1.04172$&$0.925768$\\
&Fa5&${3}/{4}$&$0.25$&MD \footnotemark[1]&$14.5176$&$92.4875$&$1.04866$&$0.990981$\\
&Fb5&${3}/{4}$&$0.50$&MD \footnotemark[1]&$14.6327$&$93.8346$&$1.03957$&$0.982162$\\
&Fc5&${3}/{4}$&$0.75$&MD \footnotemark[1]&$15.2162$&$99.1168$&$1.02005$&$0.986104$\\
\bottomrule
\end{tabular}\\

   \begin{tabular}{ccc}
\multicolumn{1}{c}{\footnotesize $^1$ Ref. \cite{GAH01}, $^2$ Ref. \cite{BW12}, $^3 $ Ref. \cite{BW16}, $^4$ $x_1=\frac{971}{1944}=0.499486$, $^5$ $x_1=\frac{973}{1944}=0.500514$.}
\end{tabular}

 \end{table}

If, as~before, the~degree of bidispersity is measured by  $1-\bar{B}_2/b_2$ and $1-\bar{B}_3/b_3$, we can observe the following ordering of decreasing bidispersity in the four-dimensional systems:
Aa, Ba, Ab, Bb, Da, Cb, Db, Ac, Bc, Eb, Dc, Fa, Fb, and~Fc. The~same ordering applies in the case of the five-dimensional systems, except~that, apart from the absence of the system Eb, the~sequence \{Ab, Bb, Da\} is replaced by either \{Ab, Da, Bb\} or by \{Da, Ab, Bb\} if either $1-\bar{B}_2/b_2$ or $1-\bar{B}_3/b_3$ are used, respectively.

It should be stressed that the proposals implied by Equations \eqref{Ze1}, \eqref{Ze2}, \eqref{Ze1'}, \eqref{Ze2'}, and~\eqref{19} may be interpreted in two directions. On~the one hand, if~$Z_\pure$ is known as a function of the packing fraction, then one can readily compute the compressibility factor of the mixture for any packing fraction and composition [$\eff$ and $\eta$ being  related through Equation \eqref{18a} in the case of $Z_{\text{sp}}$]; this is the standard view. On~the other hand, if~simulation data for the EOS of the mixture are available for different densities, size ratios, and~mole fractions, Equations \eqref{Ze1}, \eqref{Ze2}, \eqref{Ze1'}, \eqref{Ze2'}, and~\eqref{19} can be used to \emph{infer} the compressibility factor of the monocomponent fluid. This is particularly important in the high-density region, where obtaining data from simulation may be accessible in the case of mixtures but either difficult or not feasible in the case of the monocomponent fluid, as~happens in the metastable fluid branch~\cite{SYHO17,SYHOO14}.

\begin{figure}[H]%
\centering
\includegraphics[width=.45\columnwidth]{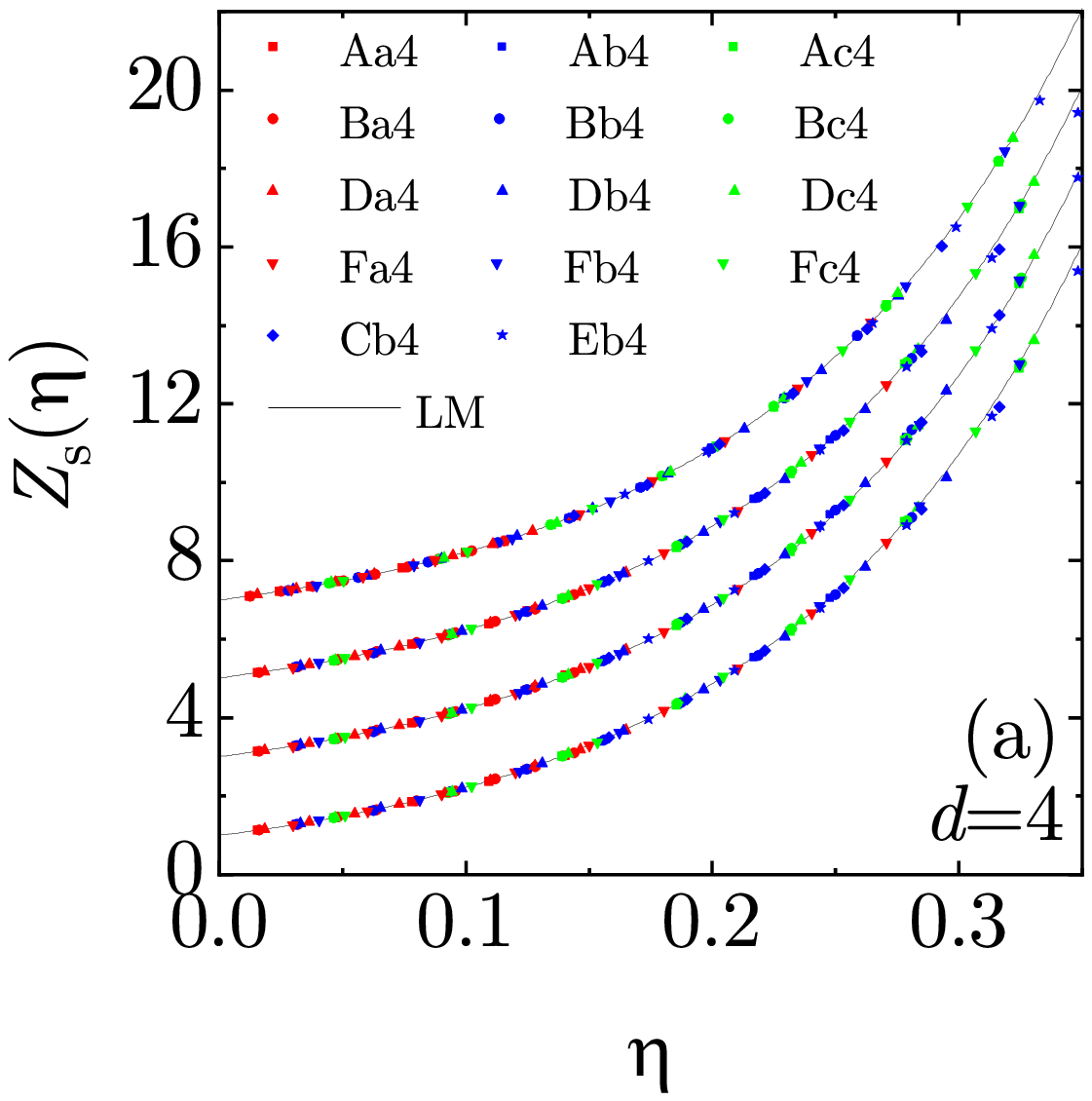}\hspace{1cm}\includegraphics[width=.45\columnwidth]{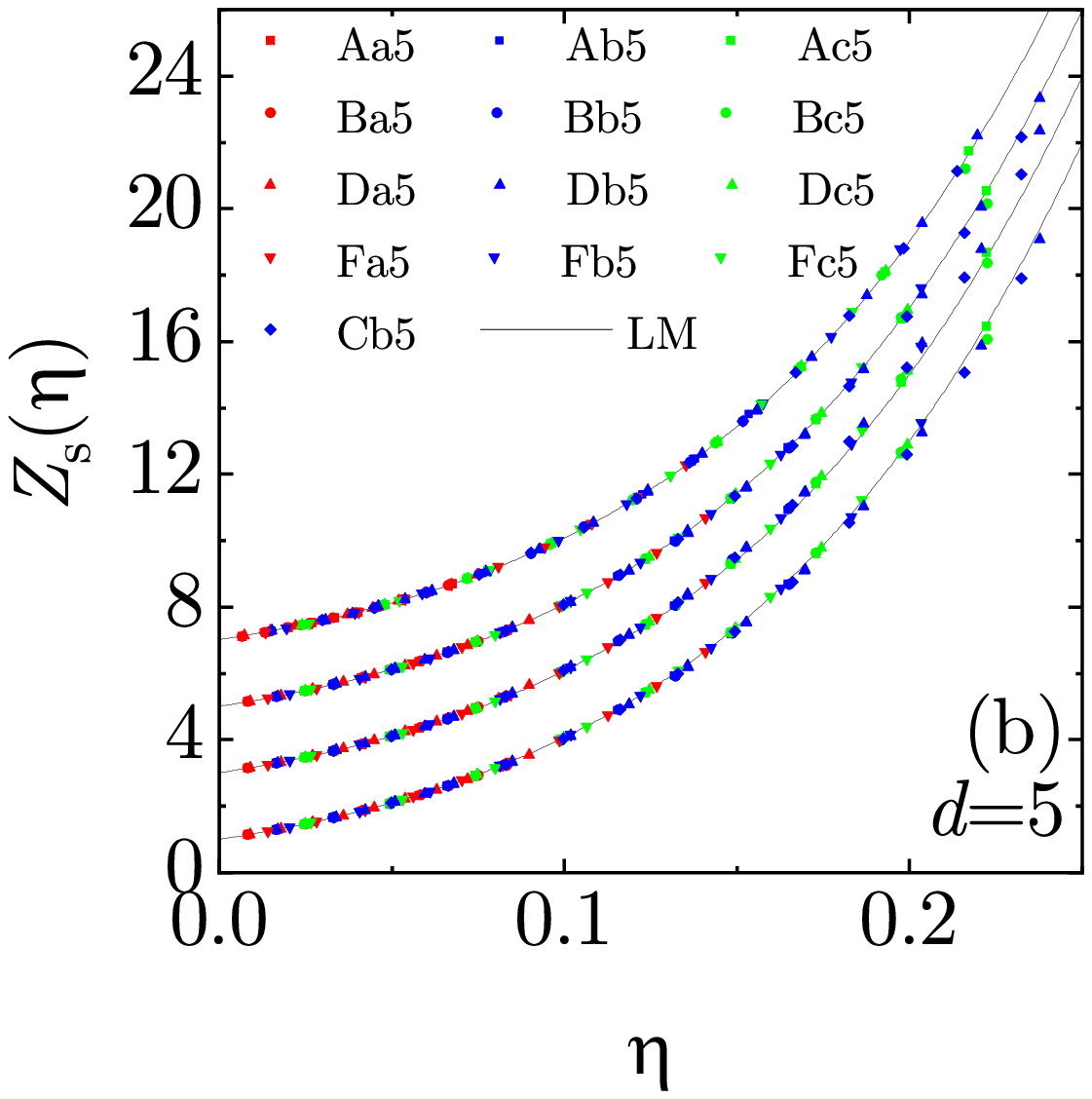}
\caption{Plot of the monocomponent  compressibility factor $Z_\pure(\eta)$, as~inferred from simulation data for the mixtures described in Table~\ref{table1}, according to the theories (from bottom to top) e1, e2, \ep1, and~sp (the three latter have been shifted vertically for better clarity). The~solid lines represent the Luban and Michels (LM) equation of state (EOS), Equation \eqref{ZLM}. Panel (\textbf{a}) corresponds to $d=4$, while panel (\textbf{b}) corresponds to $d=5$.\label{fig_infer}}
\end{figure}

In principle, simulation data for different mixtures would yield different inferred functions $Z_\pure(\eta)$. Thus, without~having to use an externally imposed monocomponent EOS, the~degree of collapse of the mapping from mixture compressibility factors onto a \emph{common} function $Z_\pure(\eta)$ is an efficient way of assessing the performance of Equations \eqref{Ze1}, \eqref{Ze2}, \eqref{Ze1'}, \eqref{Ze2'}, and~\eqref{19}. As~shown in Figure~\ref{fig_infer}, the~usefulness of those mappings is confirmed  by the nice collapse obtained for all the points corresponding to the mixtures described in Table~\ref{table1}. The~inferred data associated with $Z_{\text{\ep2}}$ are almost identical to those associated with $Z_{\text{e2}}$ and thus they are omitted in Figure~\ref{fig_infer}.
Figure~\ref{fig_infer} also shows that the inferred curves are very close to the  LM (monocomponent) EOS, Equation \eqref{ZLM}, what validates its choice as an accurate function $Z_\pure(\eta)$ in what follows. Notwithstanding this, one can observe in the high-density regime that the values inferred from simulation data via $Z_{\text{e1}}$ and $Z_{\text{\ep1}}$ tend to underestimate the LM curve for both $d=4$ and $d=5$, while the values inferred via  $Z_{\text{e2}}$ tend to overestimate it for $d=5$. Overall, one can say that the best agreement with the LM EOS is obtained by using $Z_{\text{e2}}$ and $Z_{\text{sp}}$ for $d=4$ and $d=5$, respectively.

Now we turn to a more a direct comparison between the simulation data and the approximate EOS for mixtures.
As expected from the indirect representation of Figure~\ref{fig_infer}, we observed  a very good agreement (not shown) between the simulation data for the systems displayed in Table~\ref{table1} and the theoretical predictions obtained from Equations \eqref{Ze1}, \eqref{Ze2}, \eqref{Ze1'}, \eqref{Ze2'}, and~\eqref{19}, supplemented by Equation \eqref{ZLM}.

In order to perform a more stringent assessment of the five theoretical EOS, we chose $Z_\text{e1}(\eta)$ as a \emph{reference} theory and focused on the percentage deviation $100[Z(\eta)/Z_\text{e1}(\eta)-1]$ from it. The~results are displayed in Figures~\ref{fig4D_1} and \ref{fig4D_2} for $d=4$ and Figures~\ref{fig5D_1} and \ref{fig5D_2} for $d=5$. Those figures { reinforce the view that all our theoretical proposals are rather accurate: the errors in $Z_{\text{e1}}$ were typically smaller than $1$\% and they are even smaller in the other approximate EOS. Note that we have not put error bars in the MD data since they were unfortunately not reported in Reference~\cite{GAH01}. We must also mention that the MD data were generally more scattered than the MC ones. Moreover, certain (small) discrepancies between MC and MD points can be observed in Figure~\ref{fig4D_2}c, MC data generally lying below MD data. The~same feature is also present (although somewhat less apparent) in Figure~\ref{fig5D_2}c. This~may be due to larger finite-size effects in the MD simulations than in the MC simulations: the MD simulations used $648$ hyperspheres for $d=4$ and $512$ or $1024$ hyperspheres for $d=5$, while the MC simulations used 10,000 hyperspheres for $d=4$ and 3888 or 7776 for $d=5$. In~any case, since the MC data were statistically precise, the~discrepancy might be eliminated by the inclusion of the (unknown) error bars in the MD results.
It is also worth pointing out that the representation of Figures~\ref{fig4D_1}--\ref{fig5D_2} is much more demanding than a conventional representation of $Z$ vs $\eta$ for each mixture or even the representation of Figure~\ref{fig_infer}.}

\unskip

\begin{figure}[H]%
\centering
\begin{tabular}{lll}
 \includegraphics[height=4.5cm]{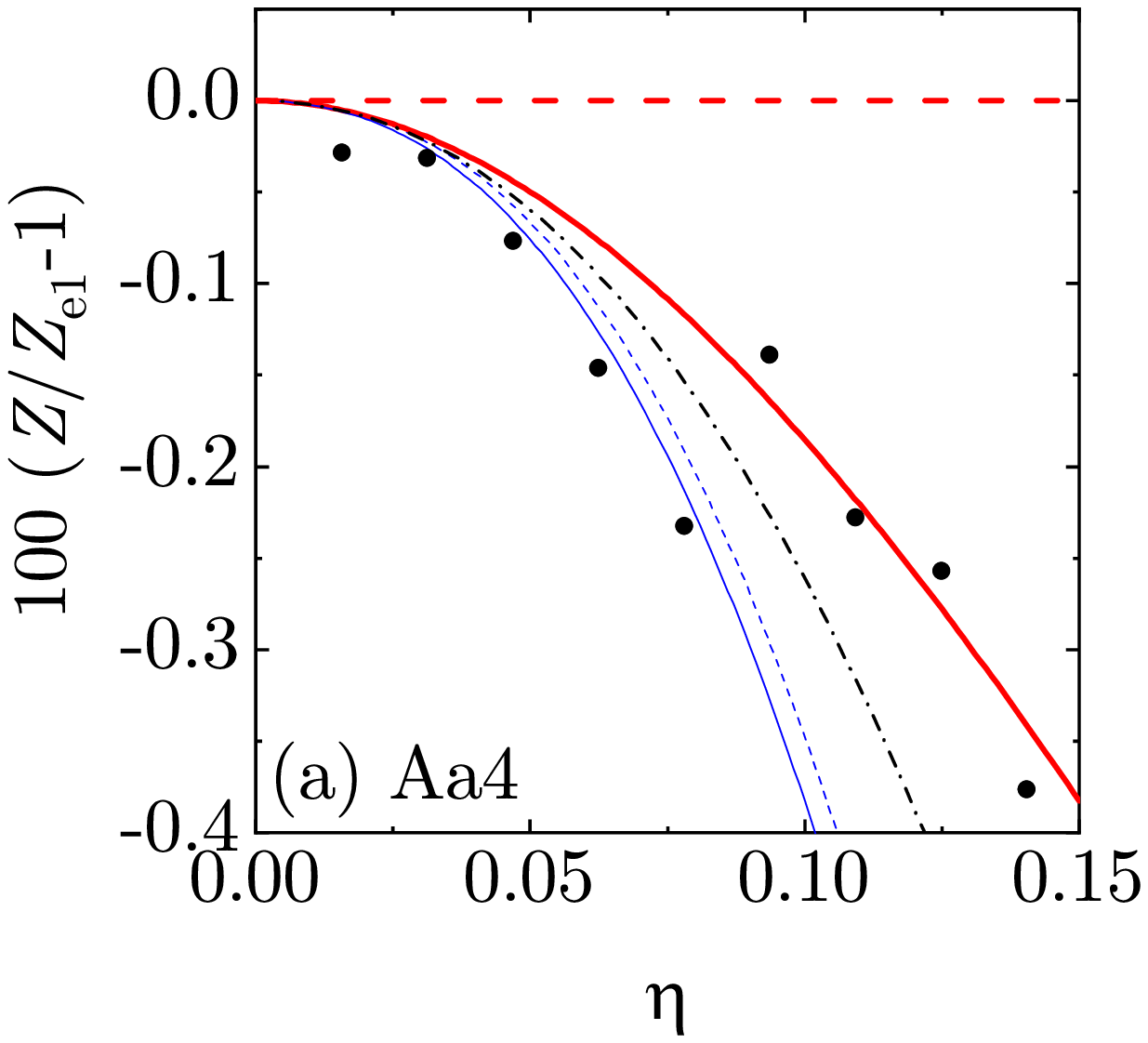}&\includegraphics[height=4.6cm]{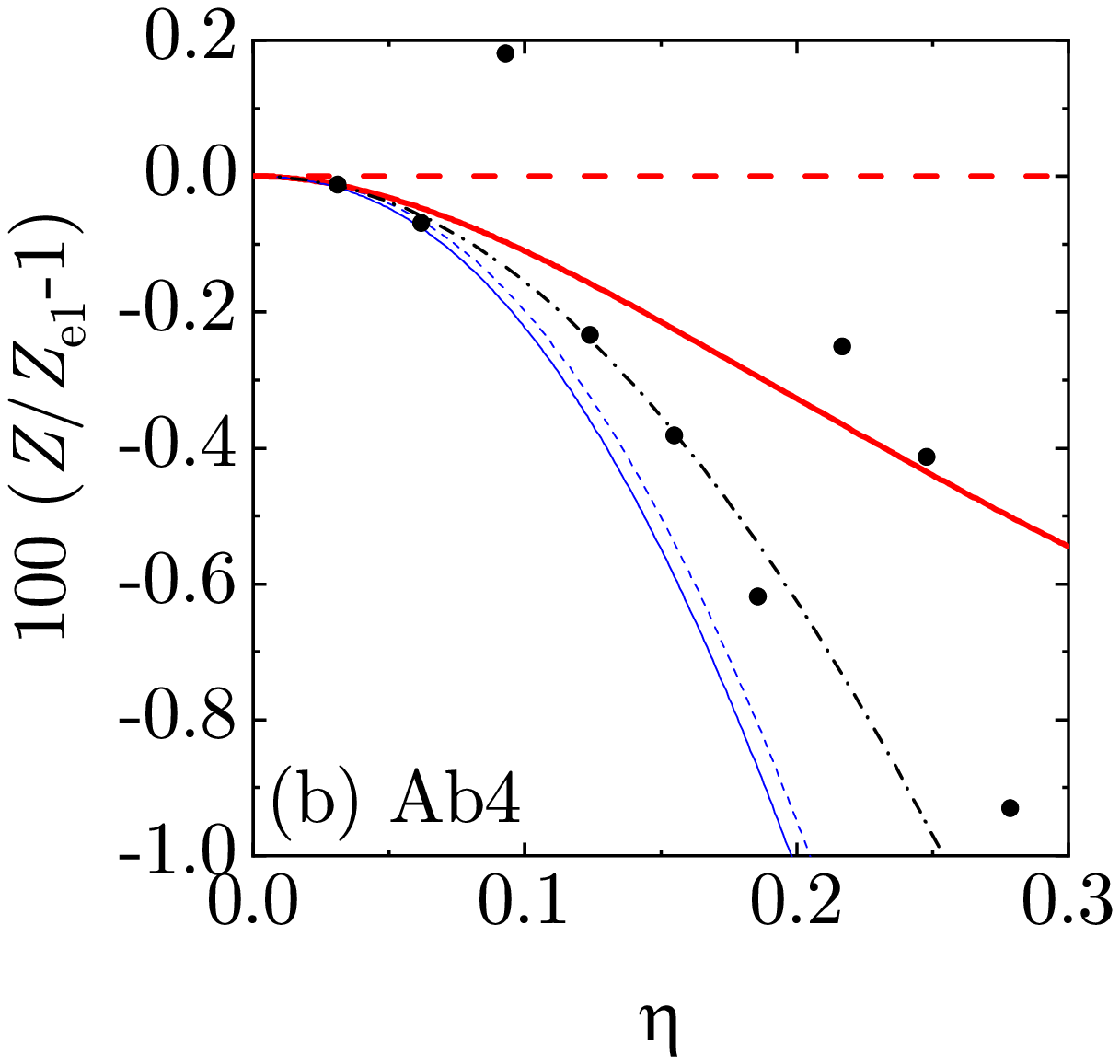}&\includegraphics[height=4.6cm]{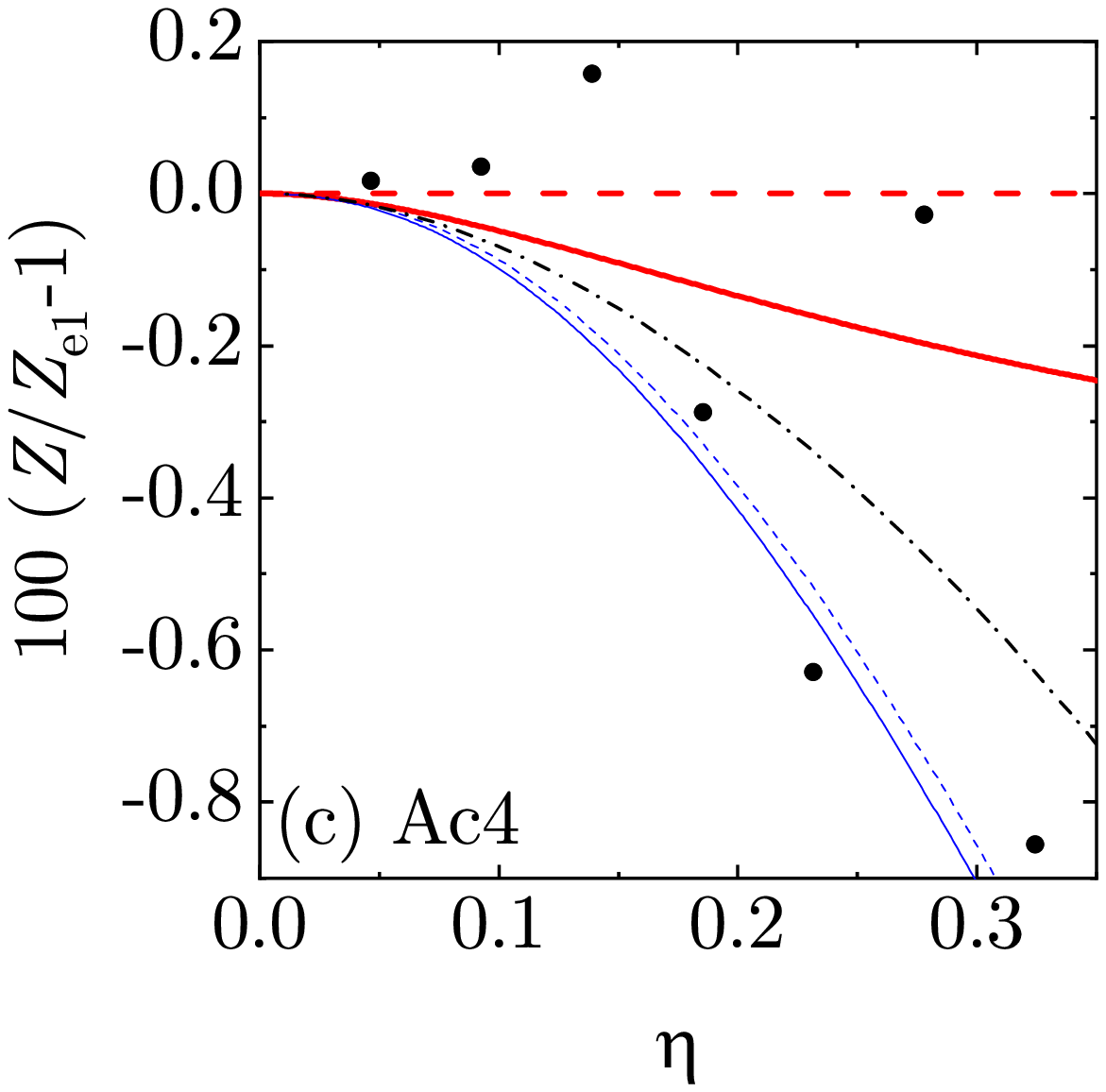}\\ \\
\includegraphics[height=4.5cm]{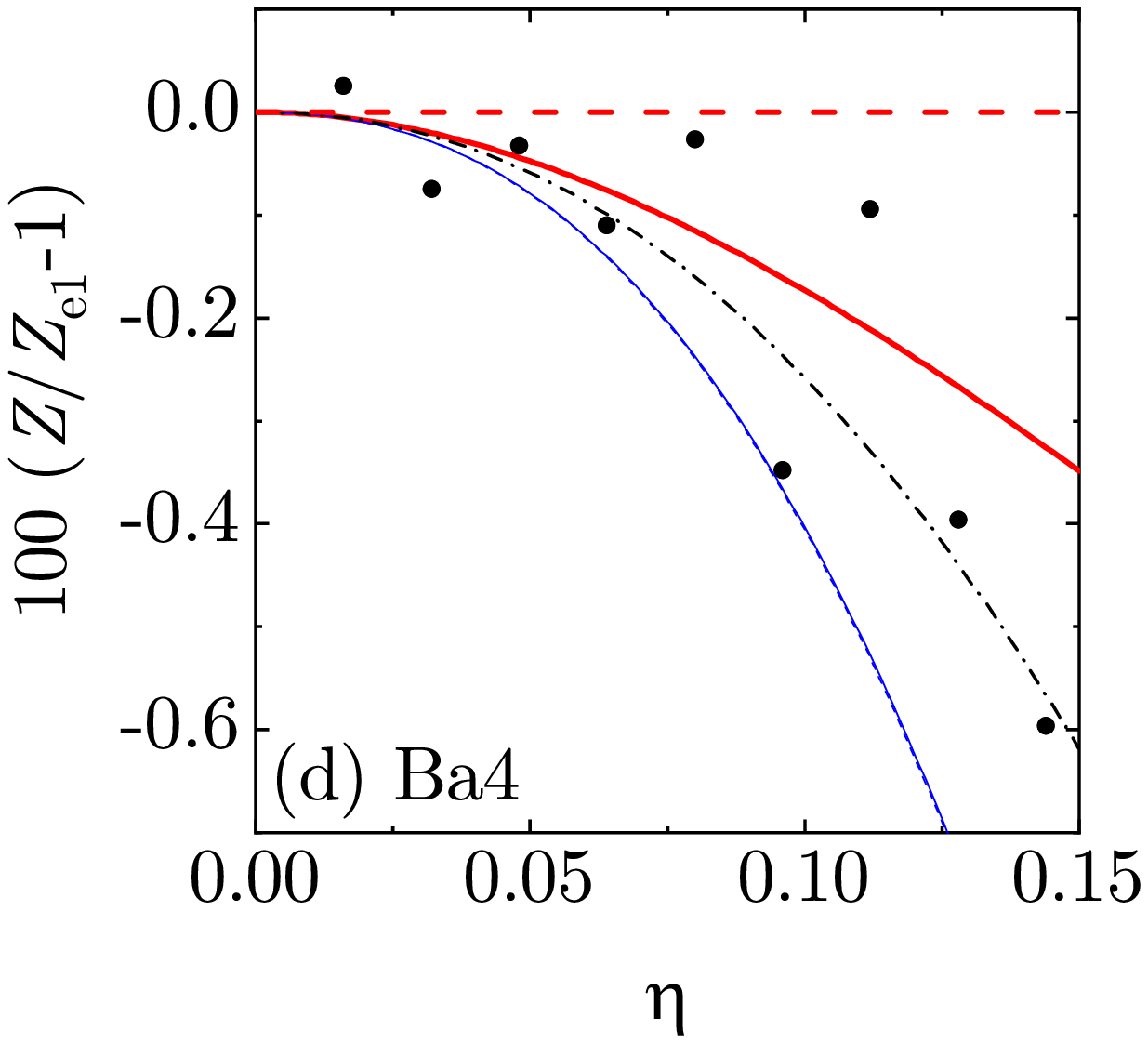}&\includegraphics[height=4.5cm]{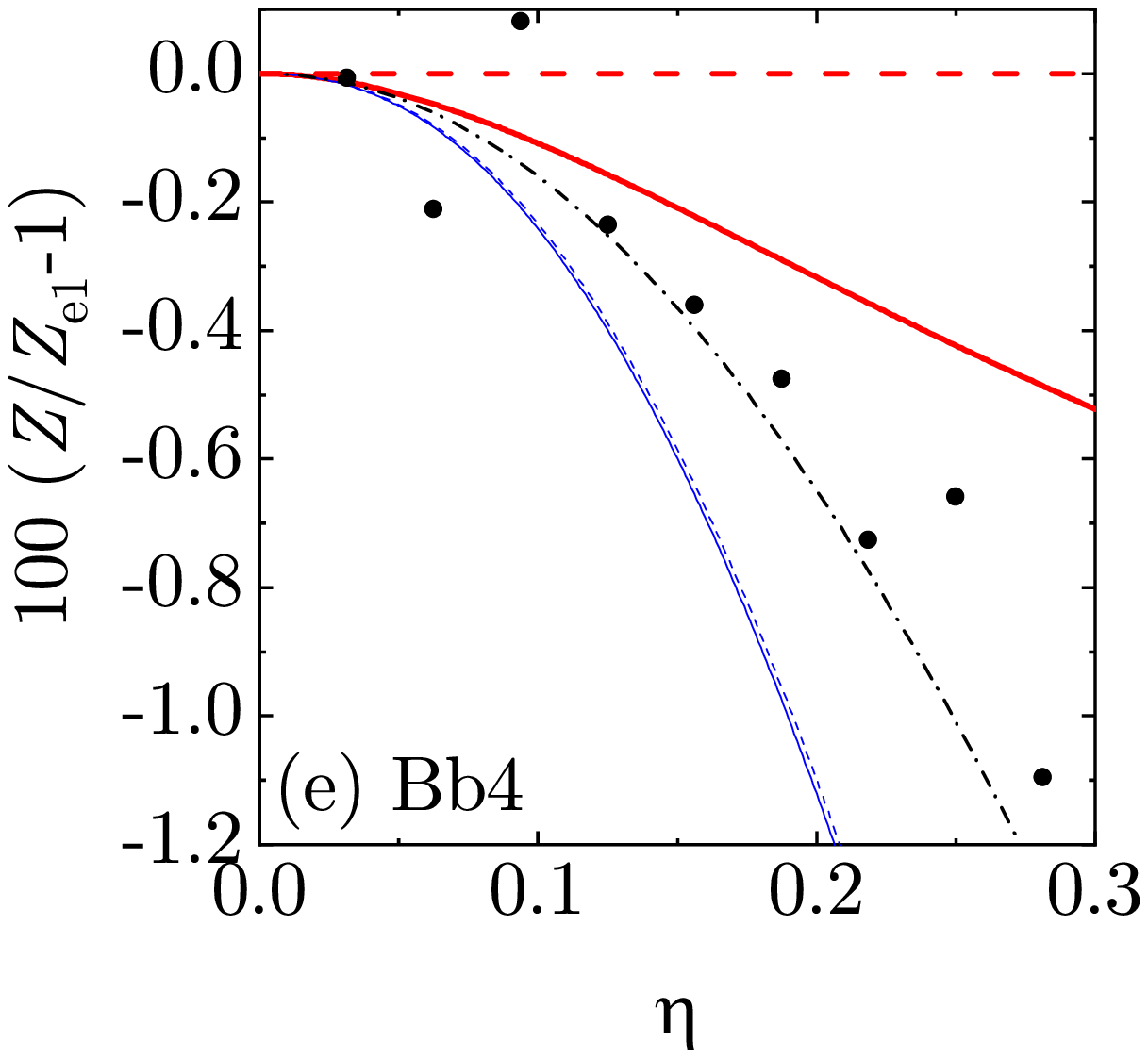}&\includegraphics[height=4.5cm]{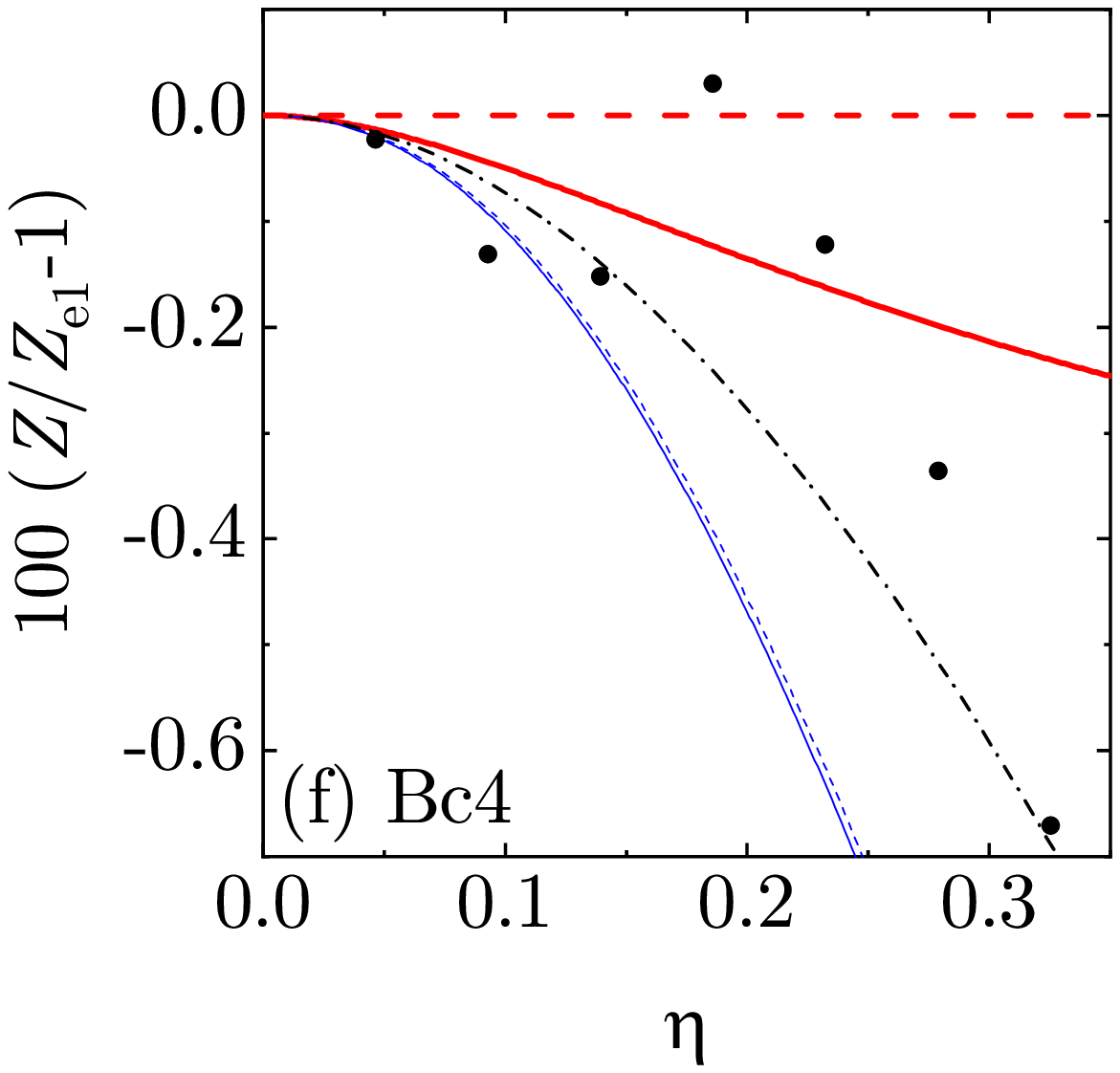}\\
\end{tabular}
\caption{Plot of the relative deviations $100[Z(\eta)/Z_\text{e1}(\eta)-1]$ from the theoretical EOS $Z_\text{e1}(\eta)$ for the four-dimensional mixtures Aa4--Bc4 (see Table~\ref{table1}). {Thick (red)} dashed lines: e1; {thick (red)} solid lines: \ep1; {thin (blue)} dashed lines: e2; {thin (blue)} solid lines: \ep2; {dash-dotted (black)} lines: sp; filled (black)  circles: MD.\label{fig4D_1}}
\end{figure}
\unskip

\begin{figure}[H]%
\centering
\begin{tabular}{lll}
\includegraphics[height=4.5cm]{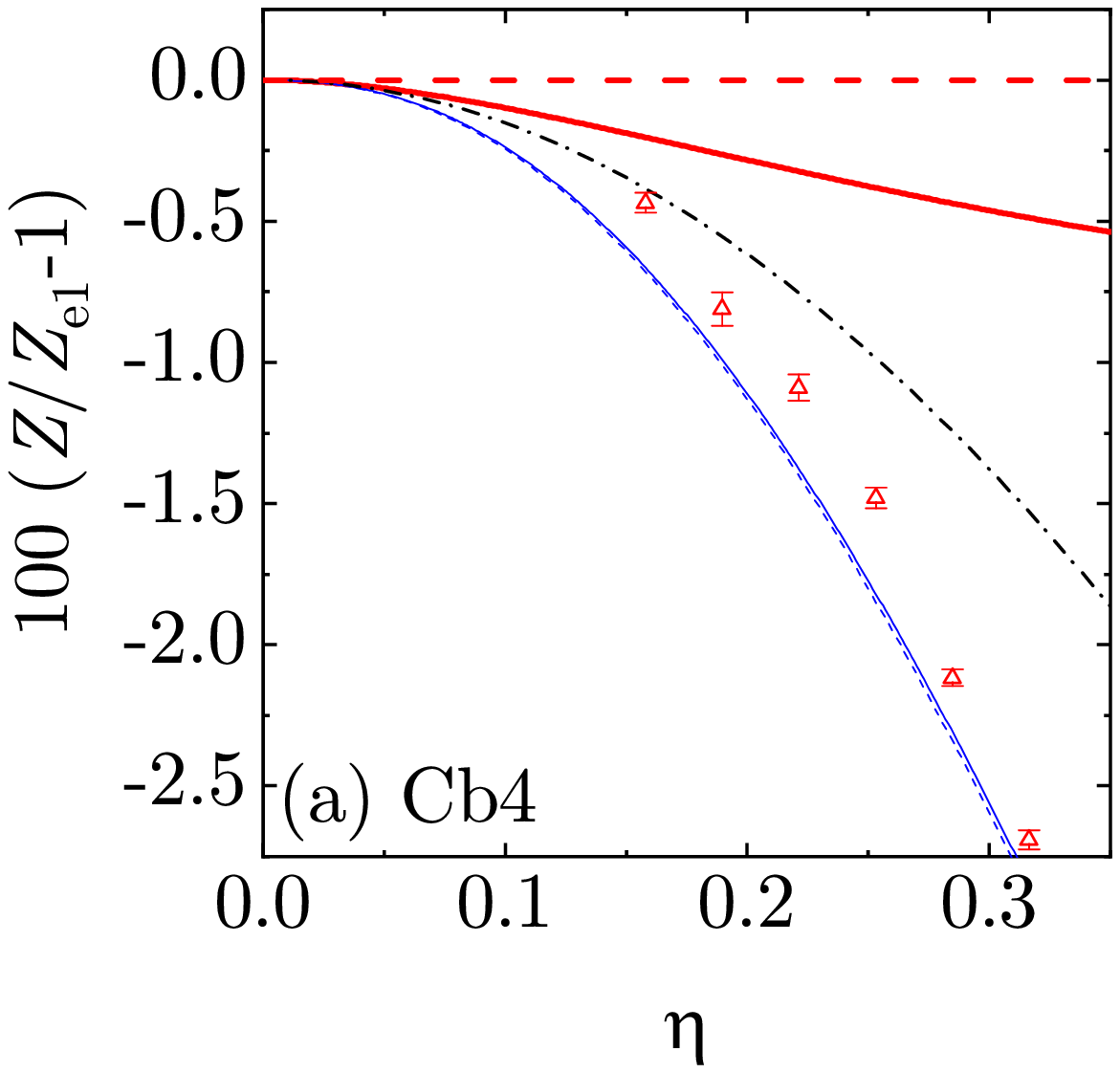}&\includegraphics[height=4.6cm]{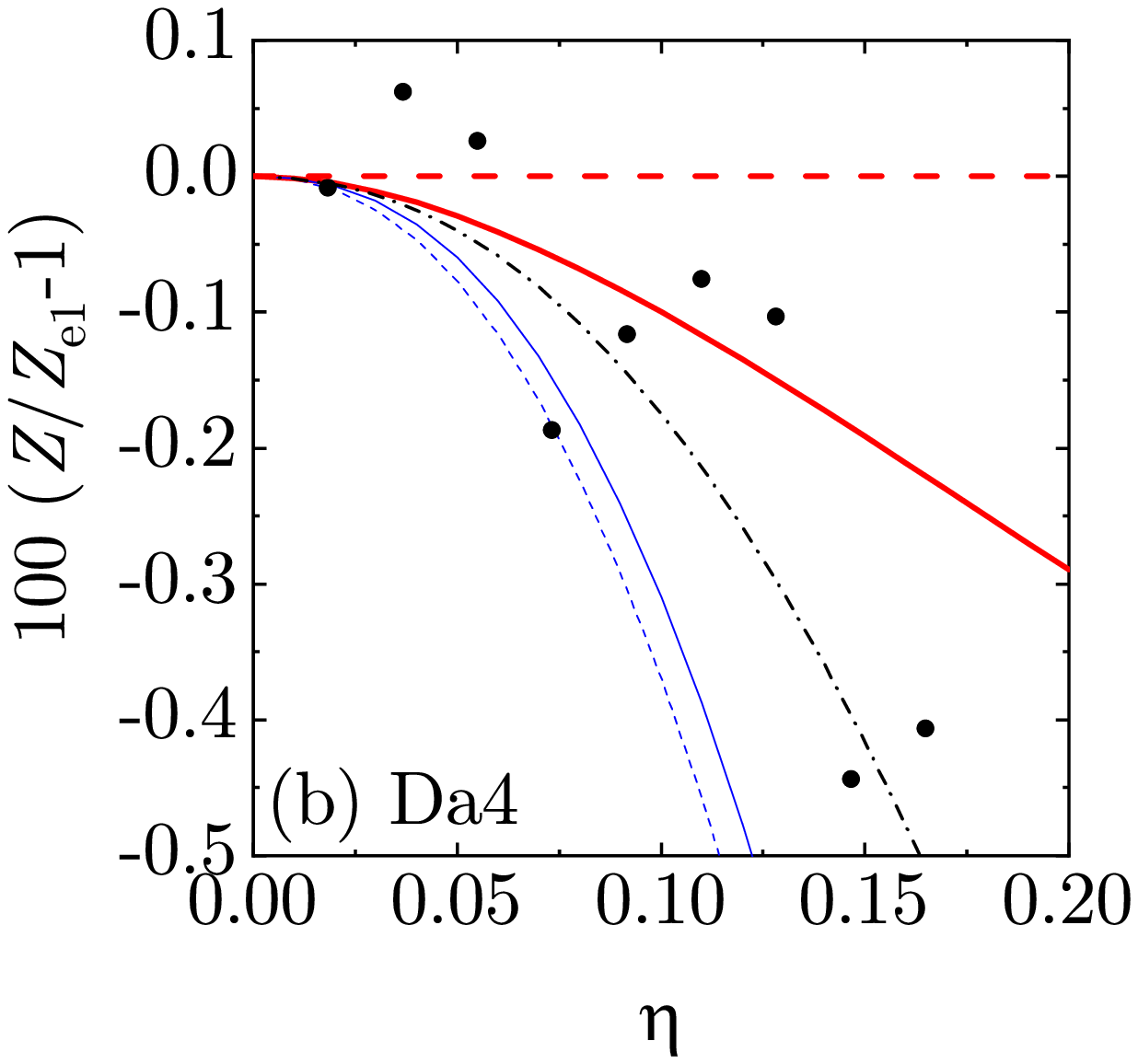}&\includegraphics[height=4.5cm]{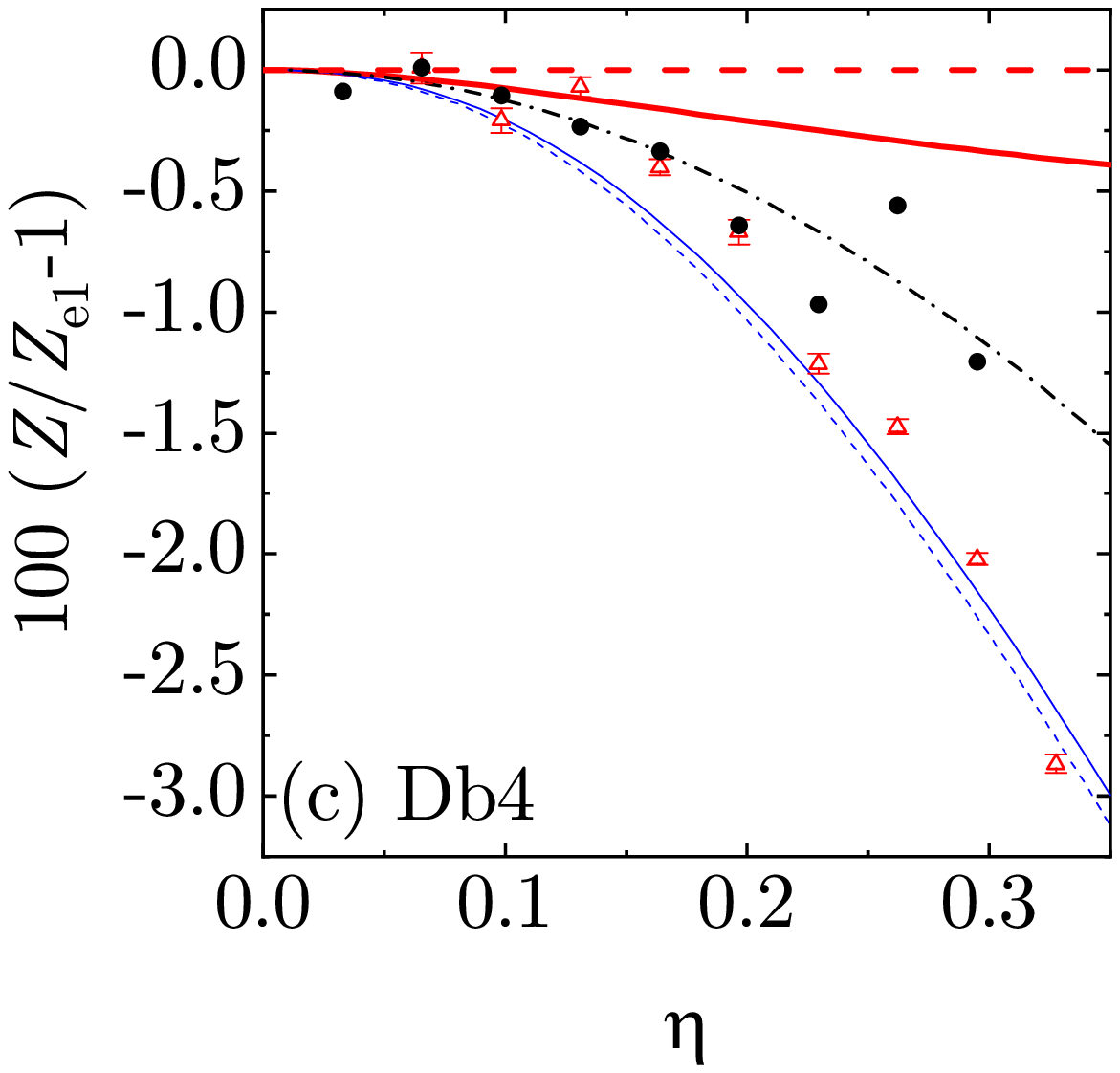}\\ \\
\includegraphics[height=4.6cm]{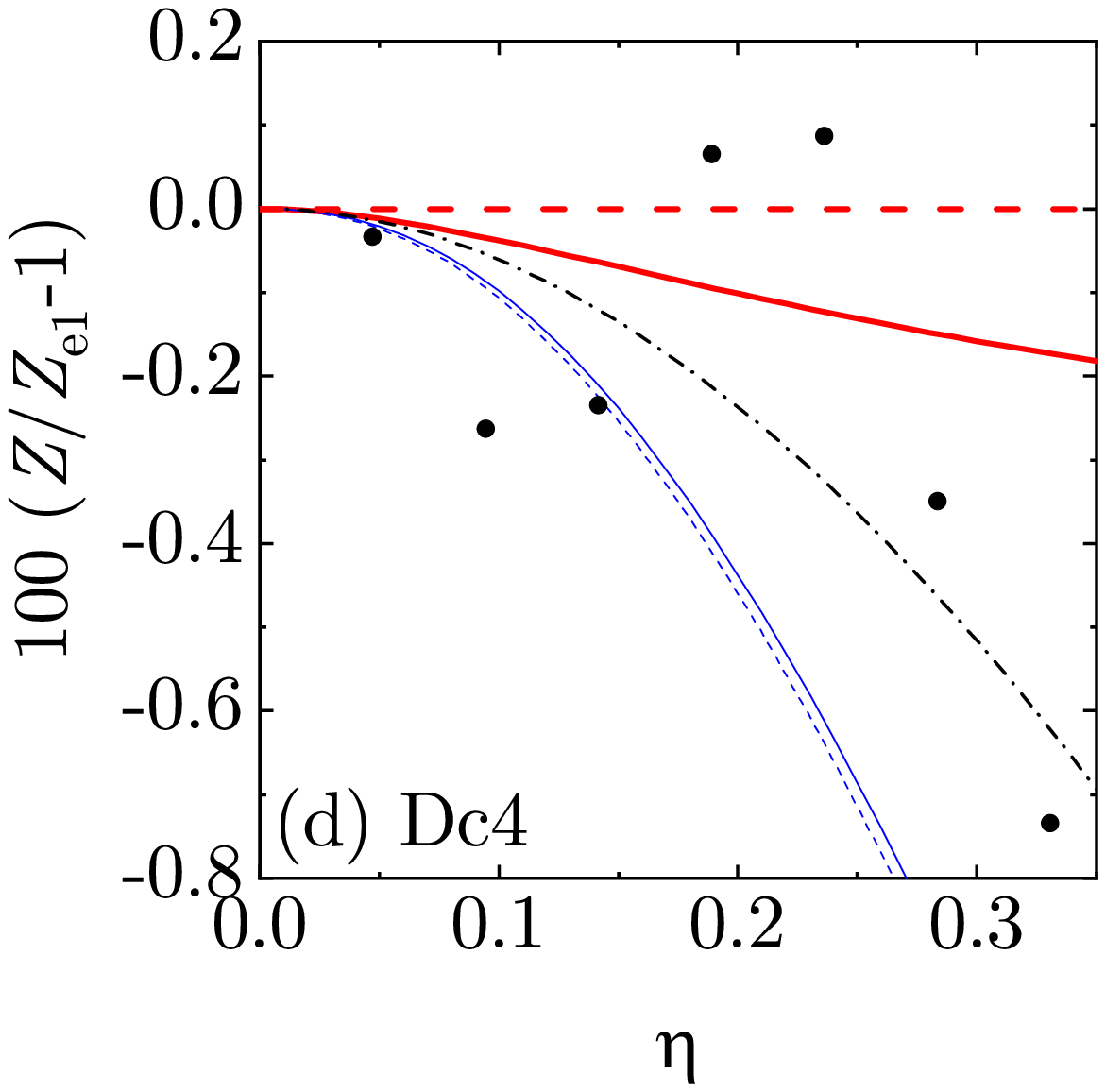}&\includegraphics[height=4.5cm]{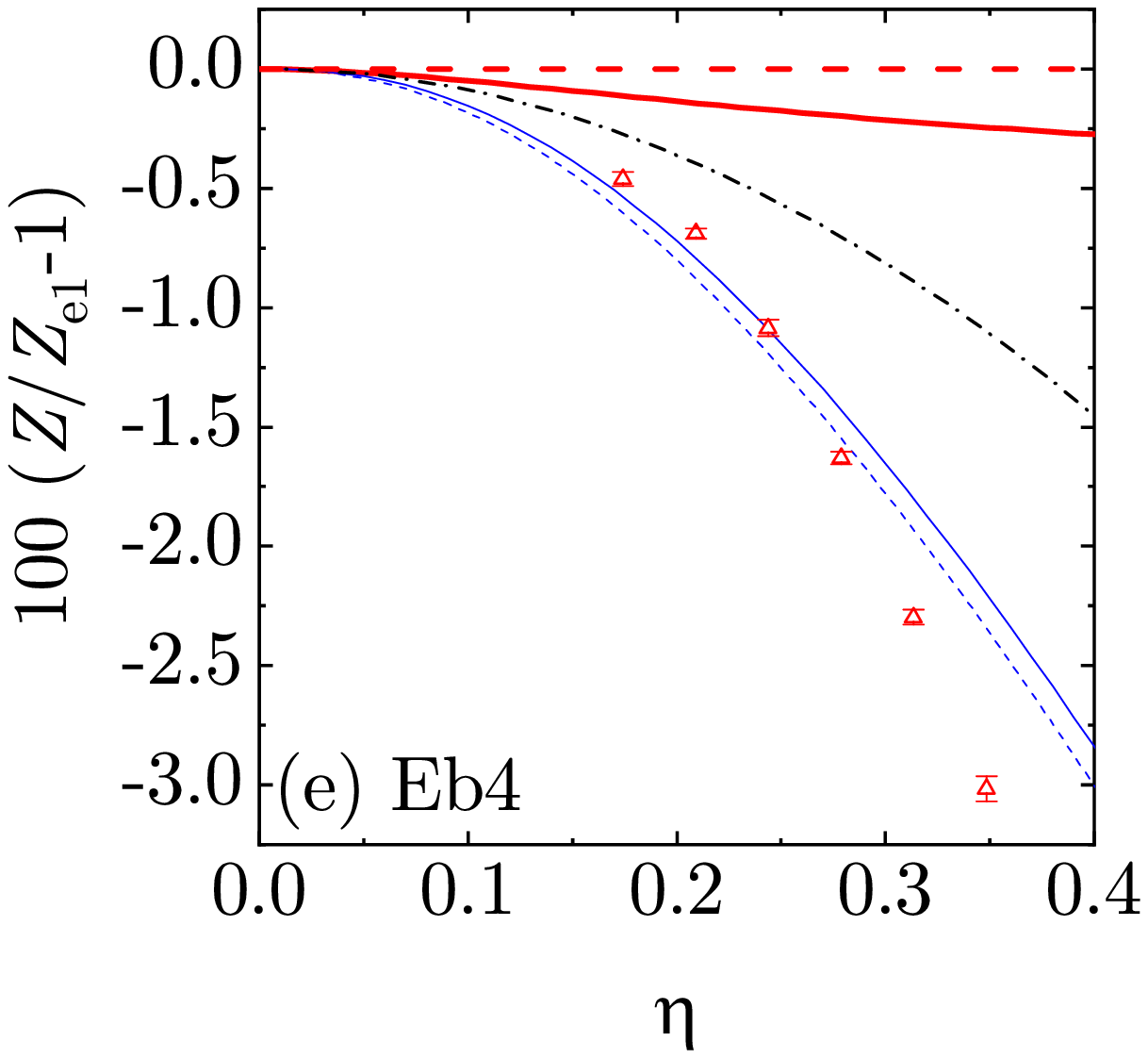}&\includegraphics[height=4.6cm]{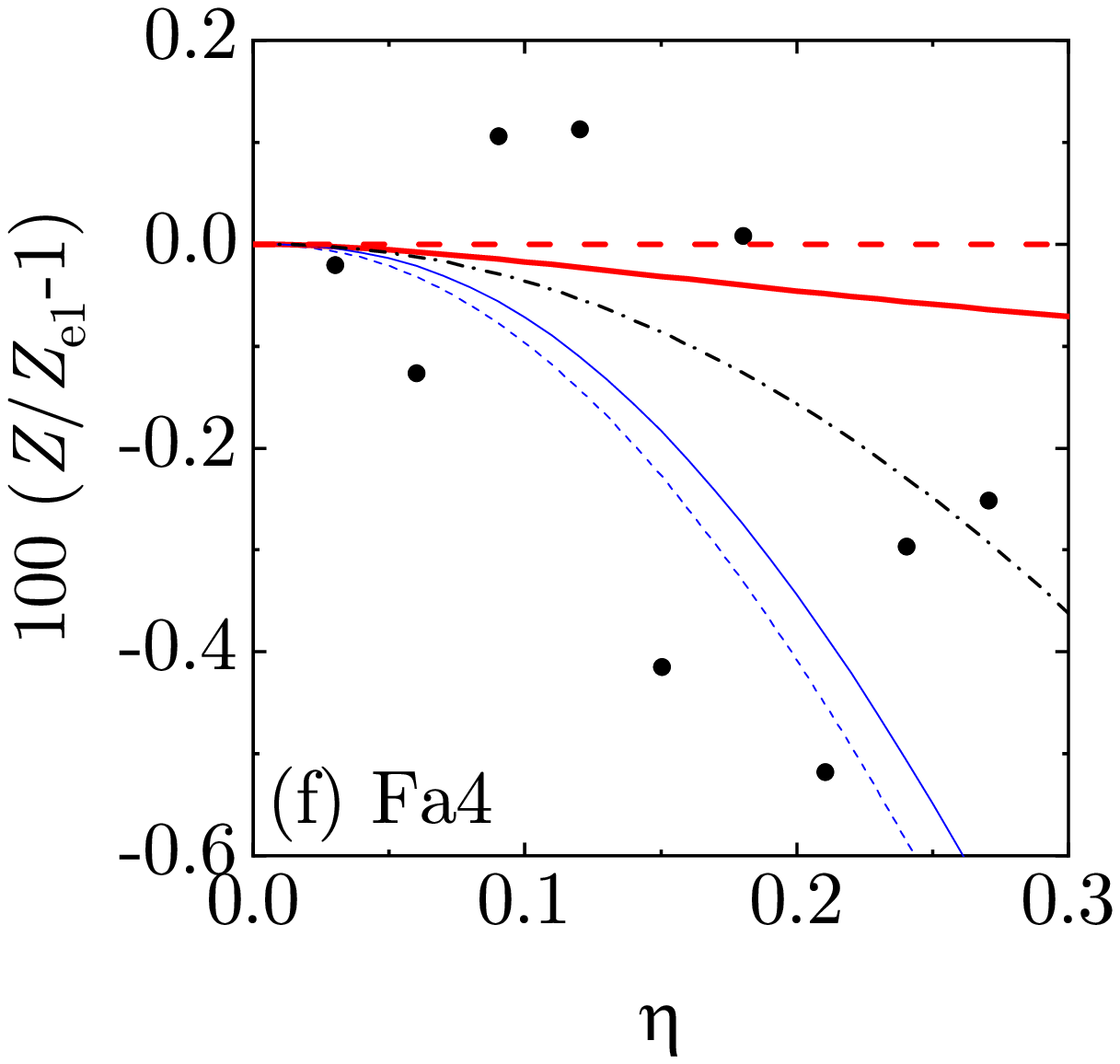}\\ \\
\end{tabular}
\caption{\textit{Cont.}}
\end{figure}

\begin{figure}[H]\ContinuedFloat
\centering	

\begin{tabular}{lll}

\includegraphics[height=4.5cm]{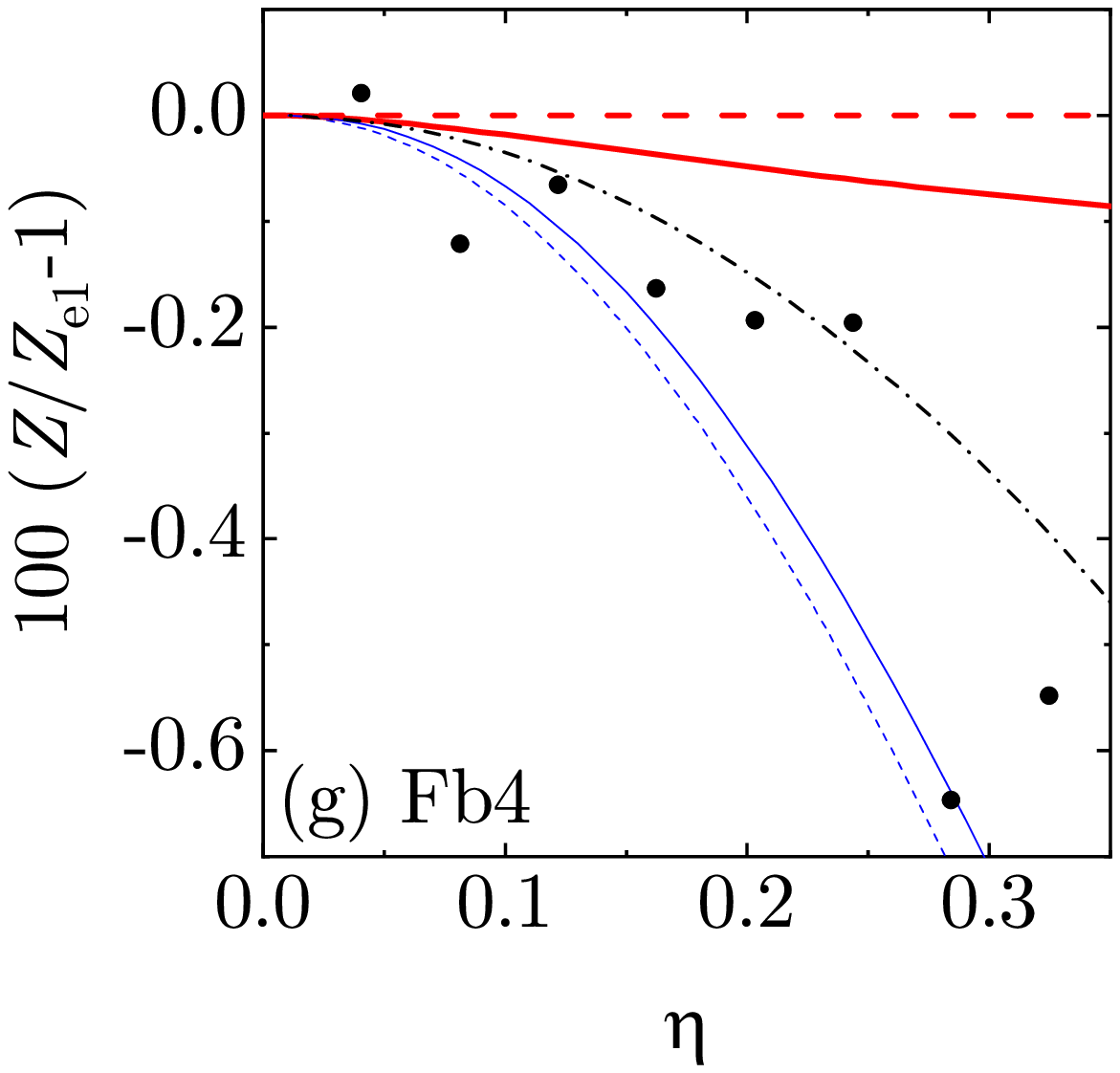}&\includegraphics[height=4.6cm]{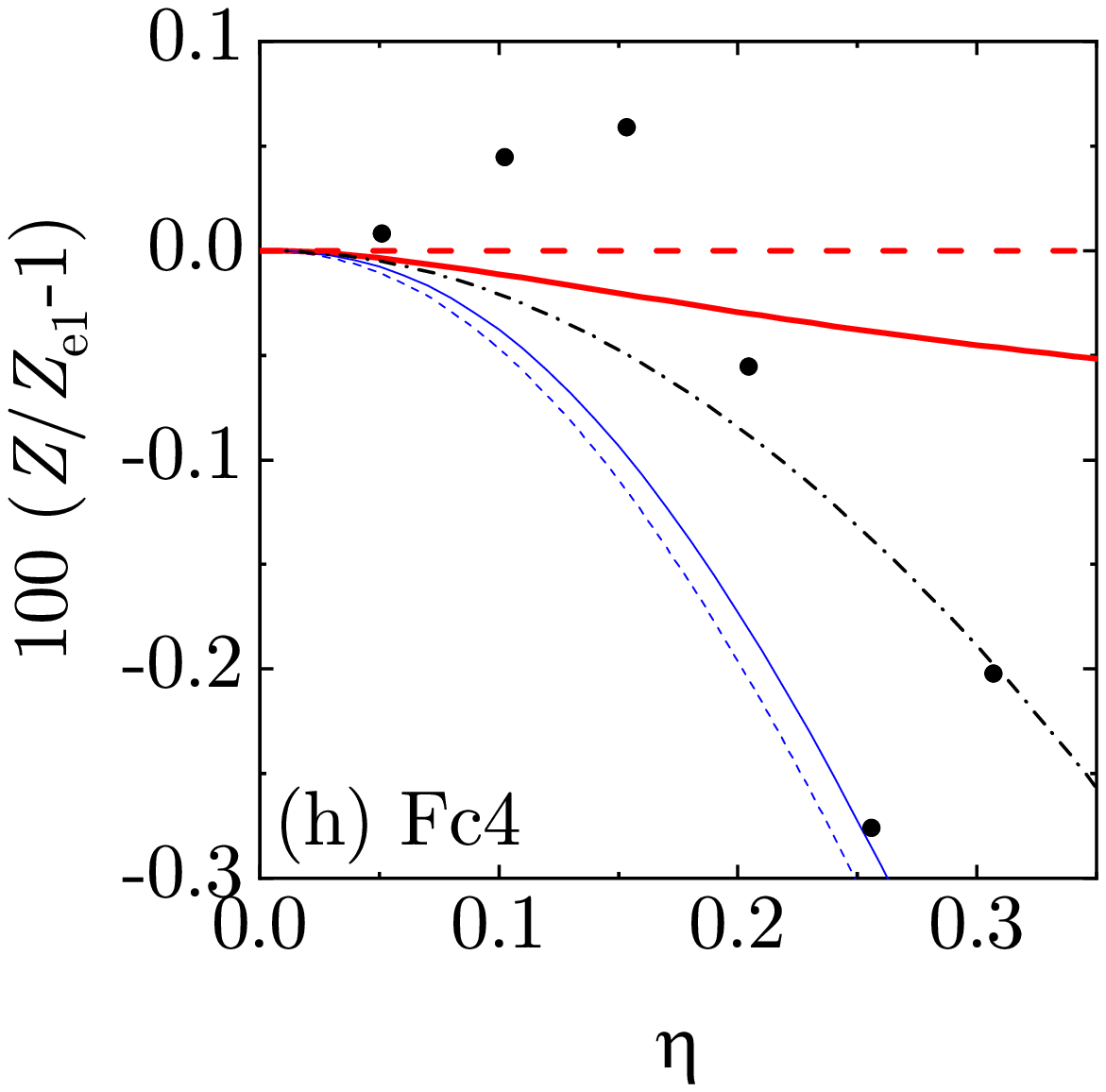}&\\
\end{tabular}
\caption{{Plot of the} relative deviations $100[Z(\eta)/Z_\text{e1}(\eta)-1]$ from the theoretical EOS $Z_\text{e1}(\eta)$ for the four-dimensional mixtures Cb4--Fc4 (see Table~\ref{table1}). {Thick (red)} dashed lines: e1; {thick (red)} solid lines: \ep1; {thin (blue)} dashed lines: e2; {thin (blue)} solid lines: \ep2; {dash-dotted (black)} lines: sp; filled (black) circles: MD; open (red) triangles with error bars in panels (\textbf{a}), (\textbf{c}), and~(\textbf{e}): MC.\label{fig4D_2}}
\end{figure}
\unskip

\begin{figure}[H]%
\centering
\begin{tabular}{lll}
\includegraphics[height=4.5cm]{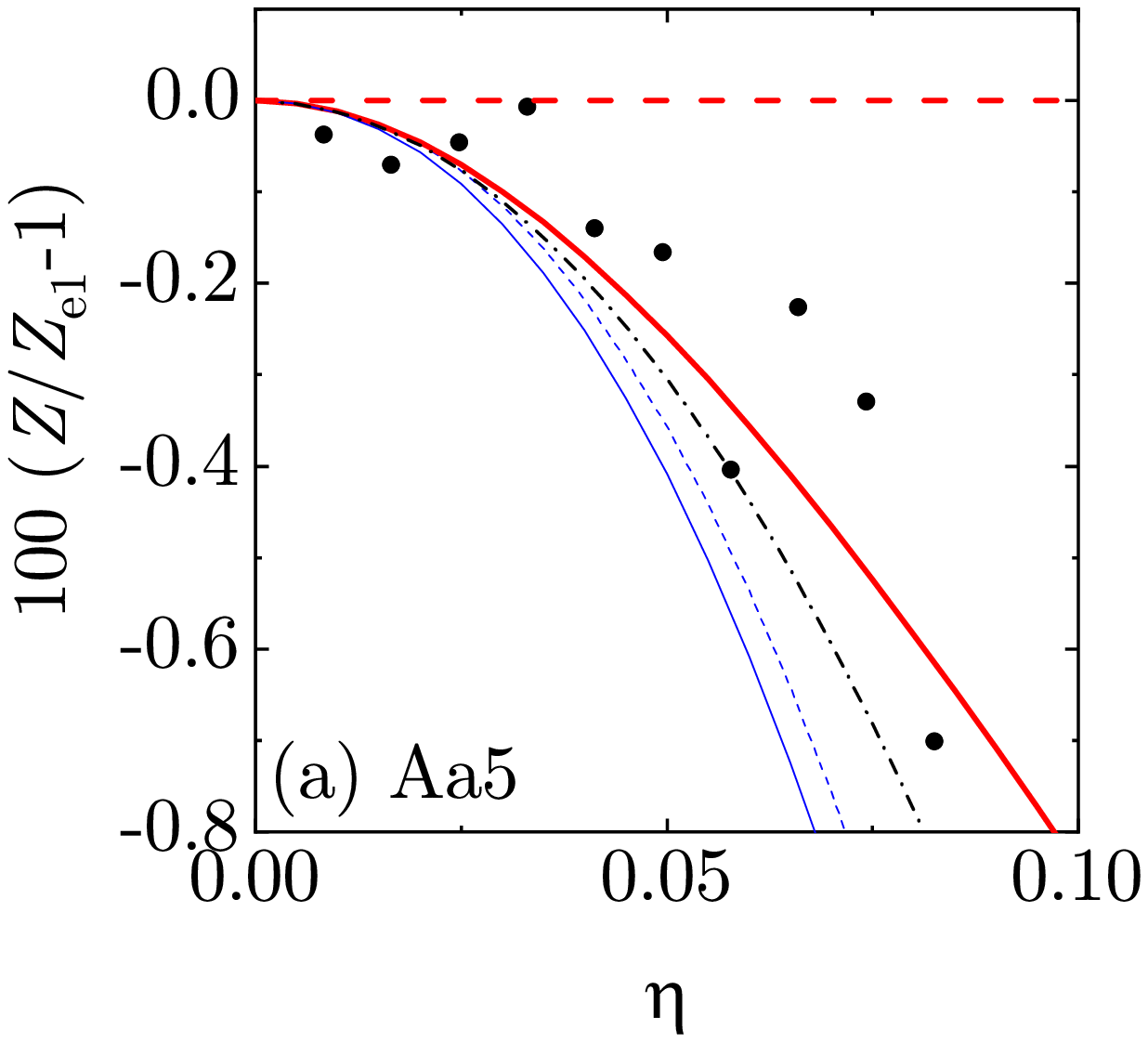}&\includegraphics[height=4.5cm]{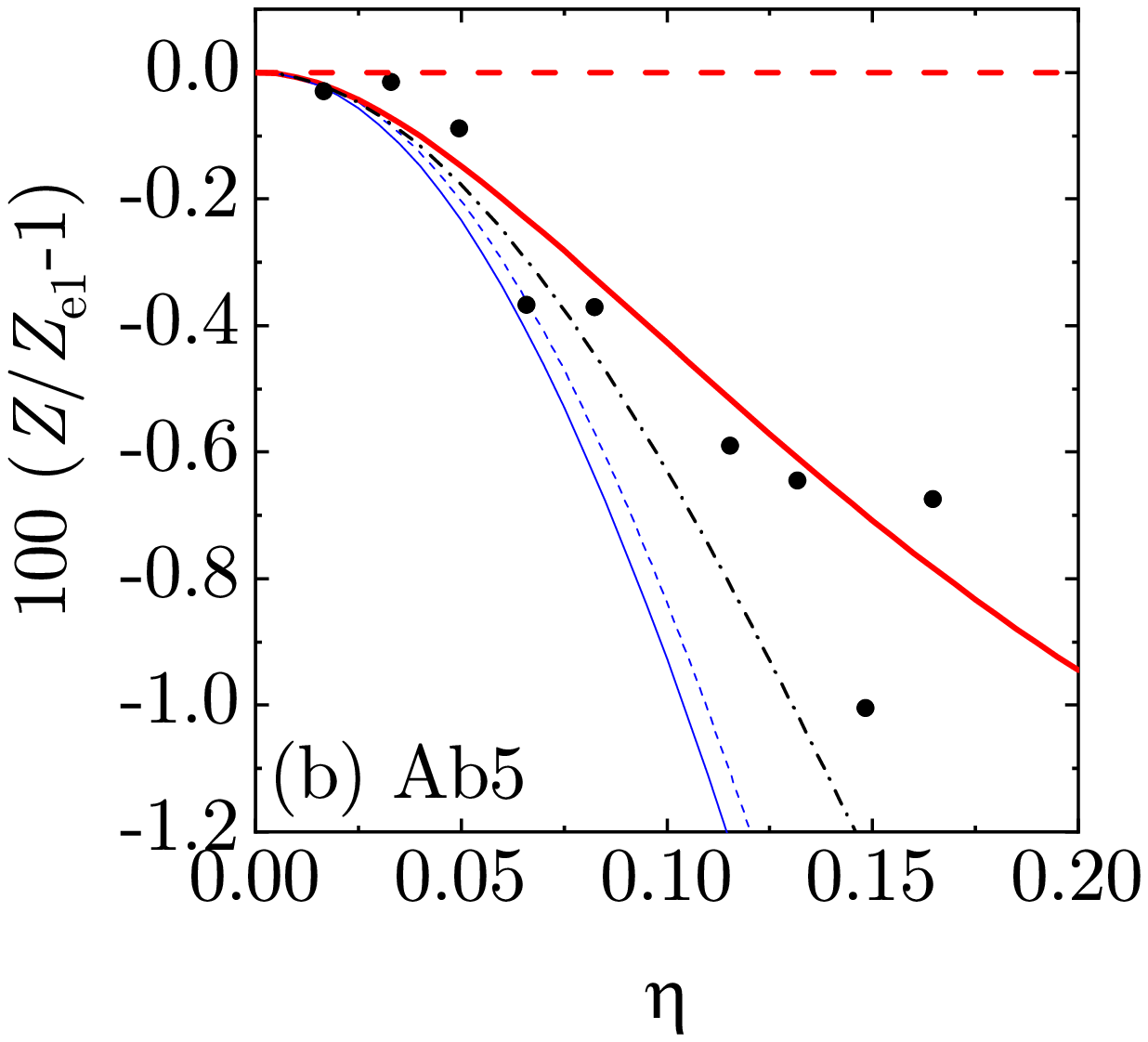}&\includegraphics[height=4.5cm]{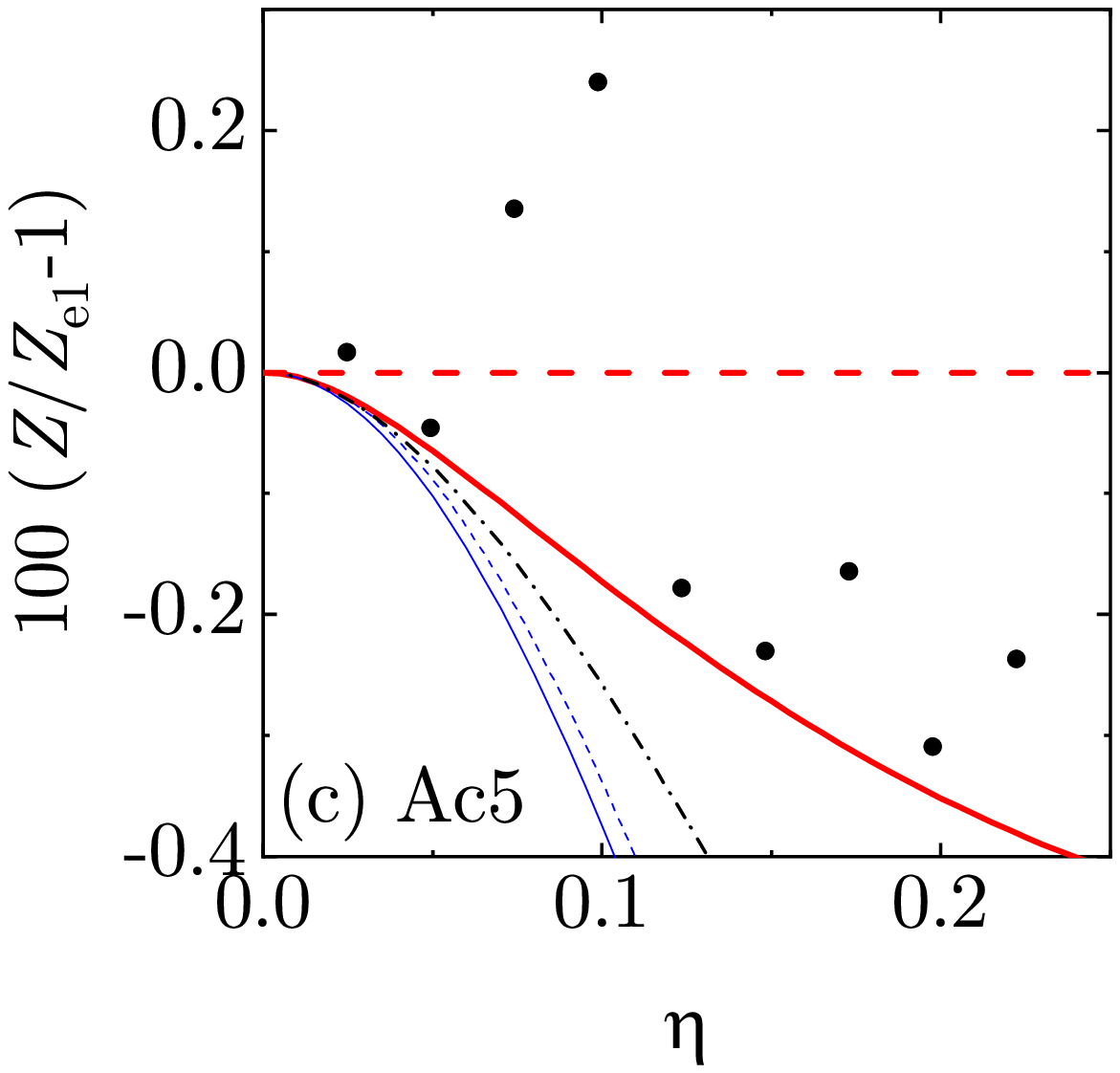}\\ \\
\includegraphics[height=4.5cm]{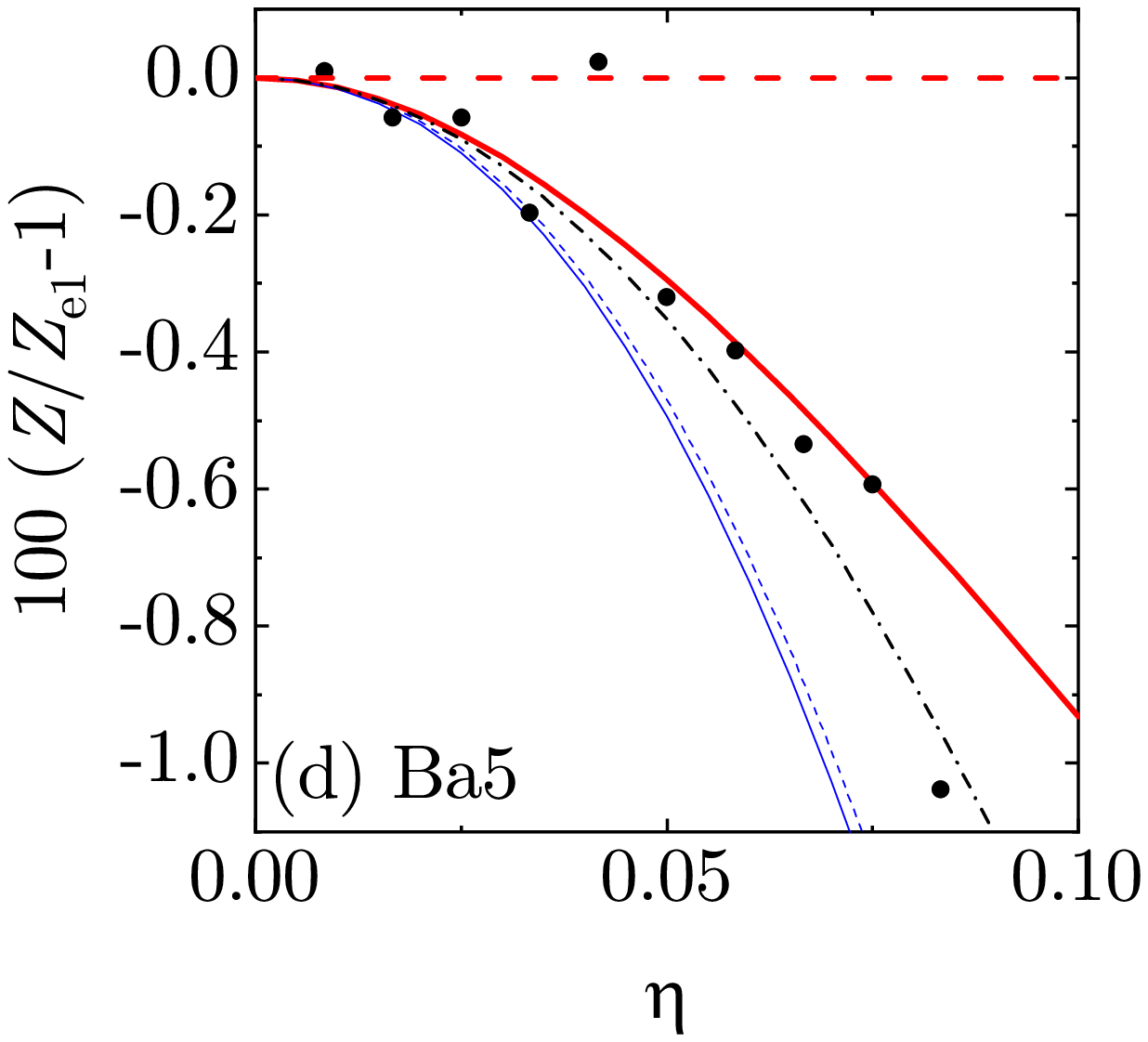}&\includegraphics[height=4.5cm]{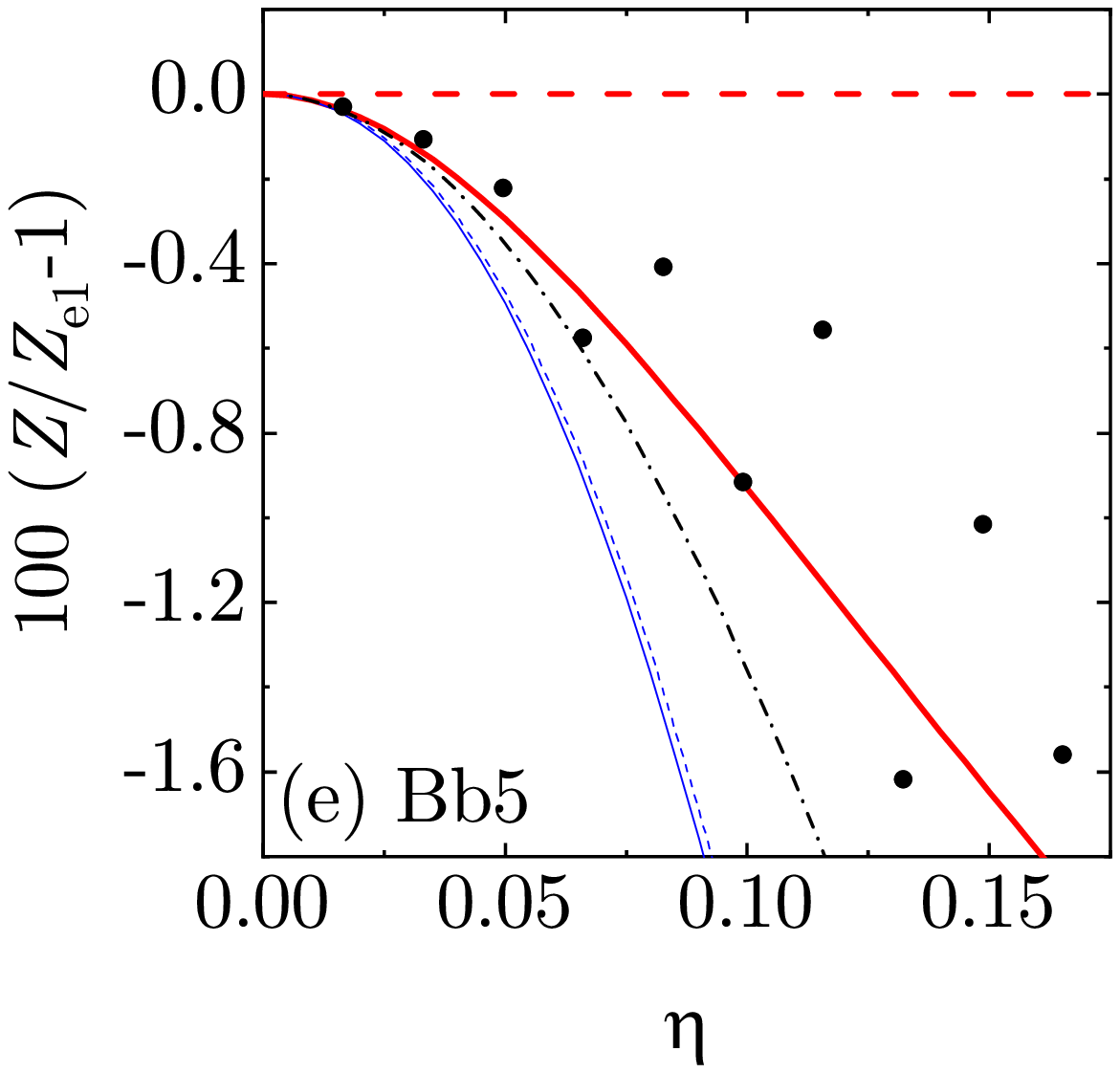}&\includegraphics[height=4.5cm]{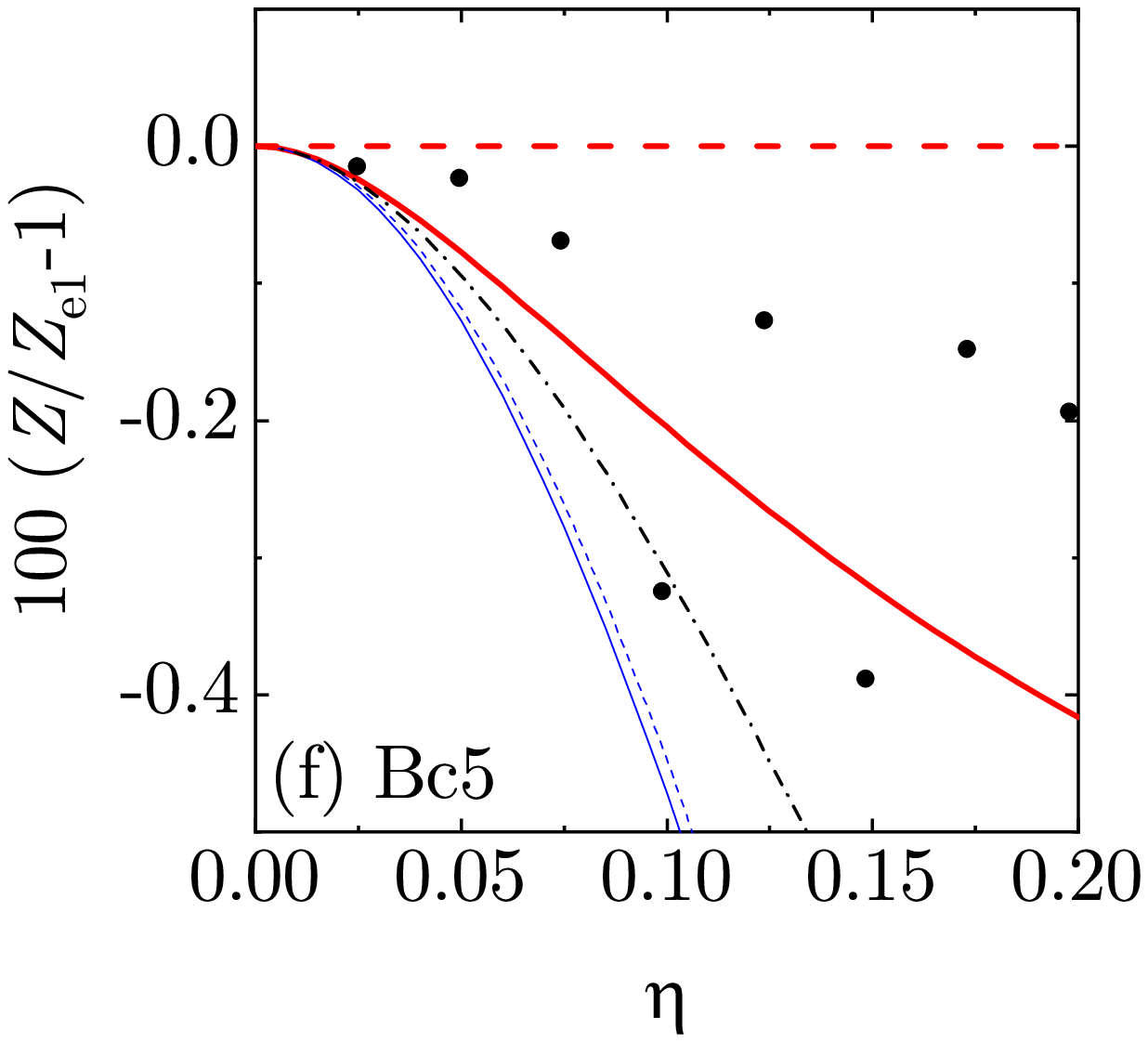}\\
\end{tabular}
\caption{Plot of the relative deviations $100[Z(\eta)/Z_\text{e1}(\eta)-1]$ from the theoretical EOS $Z_\text{e1}(\eta)$ for the five-dimensional mixtures Aa5--Bc5 (see Table~\ref{table1}). {Thick (red)} dashed lines: e1; {thick (red)} solid lines: \ep1; {thin (blue)} dashed lines: e2; {thin (blue)} solid lines: \ep2; {dash-dotted (black)} lines: sp; filled (black)  circles: MD.\label{fig5D_1}}
\end{figure}
\unskip

\begin{figure}[H]%
\centering
\begin{tabular}{lll}
\includegraphics[height=4.5cm]{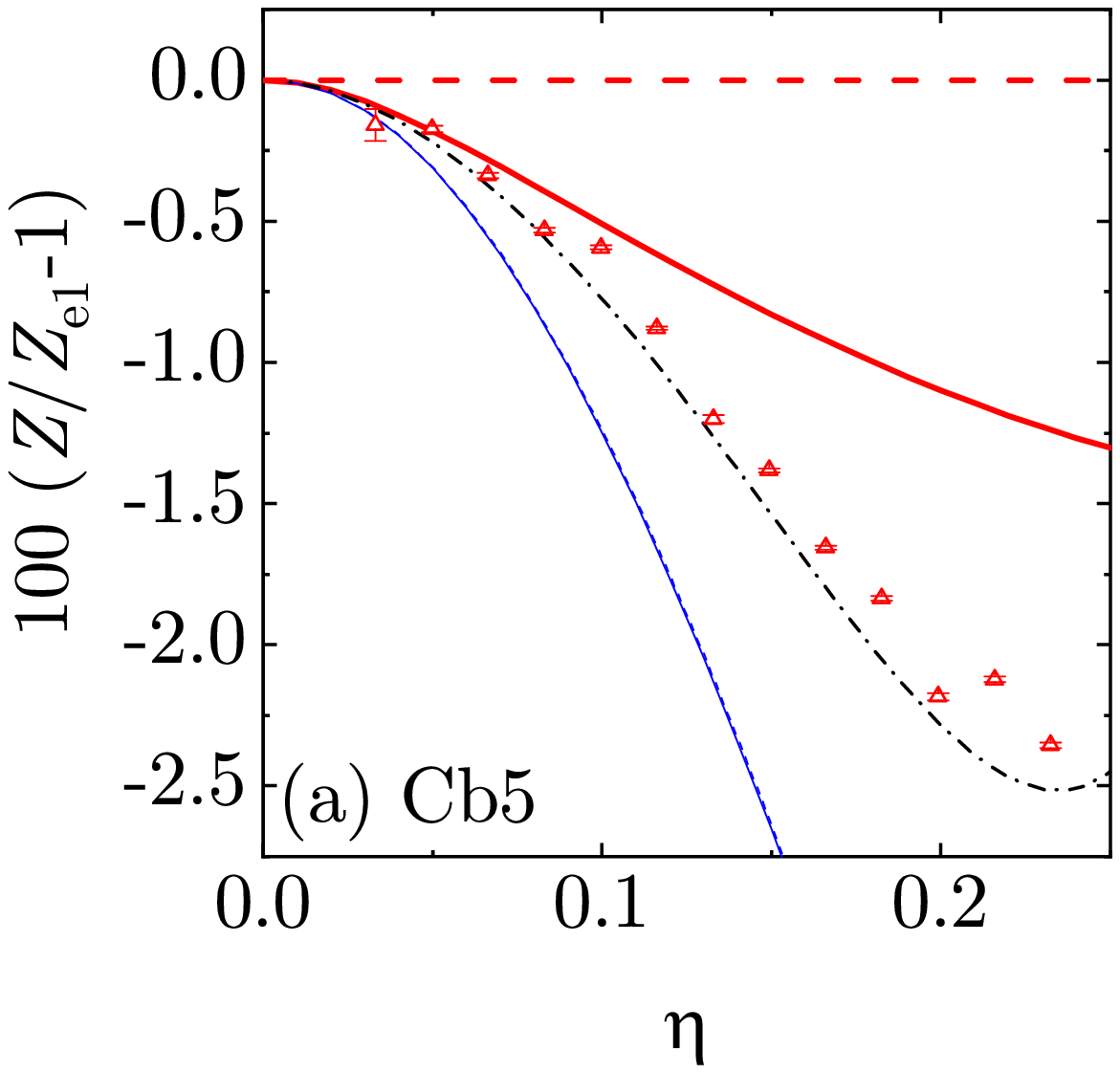}&\includegraphics[height=4.5cm]{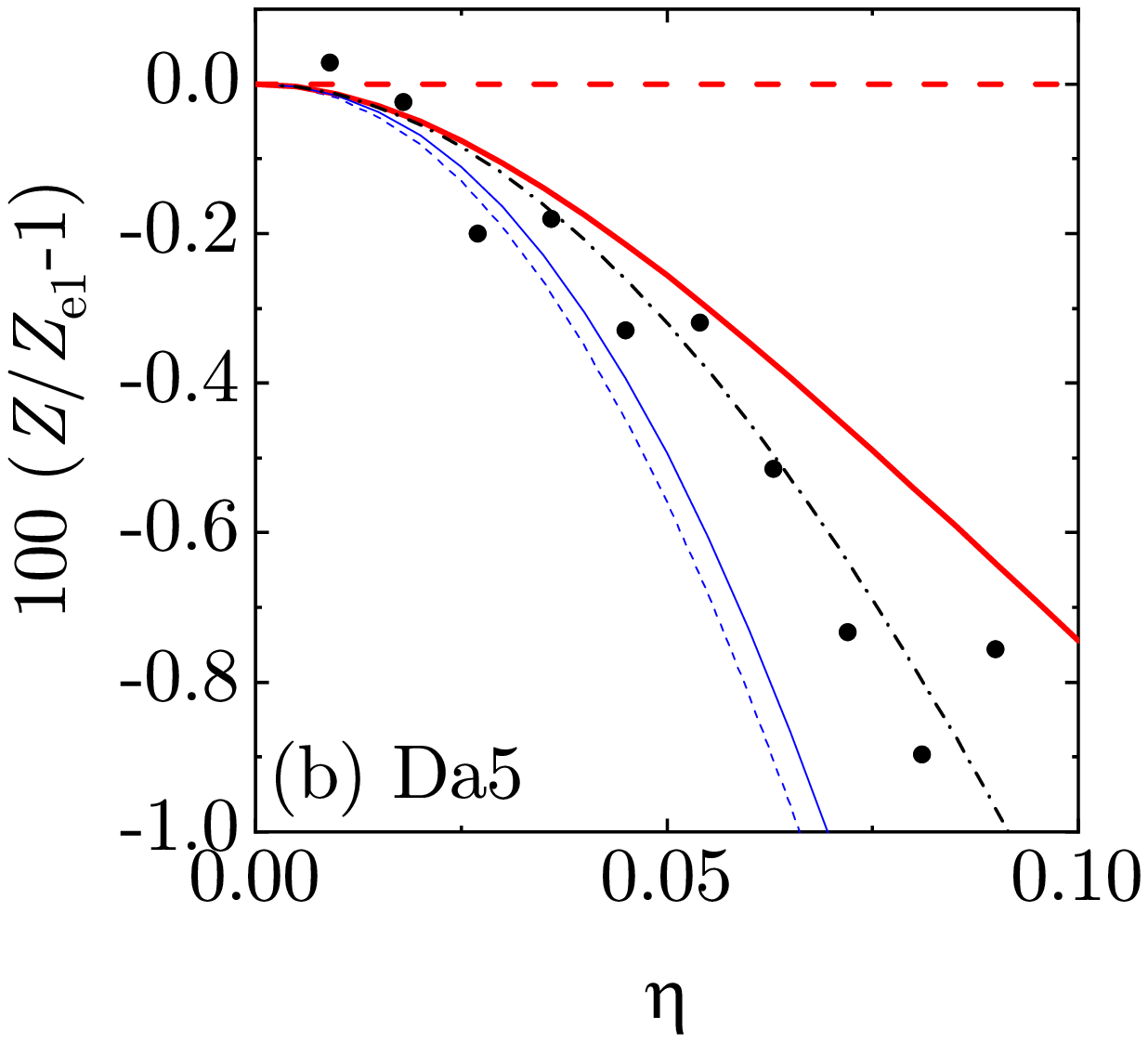}&\includegraphics[height=4.5cm]{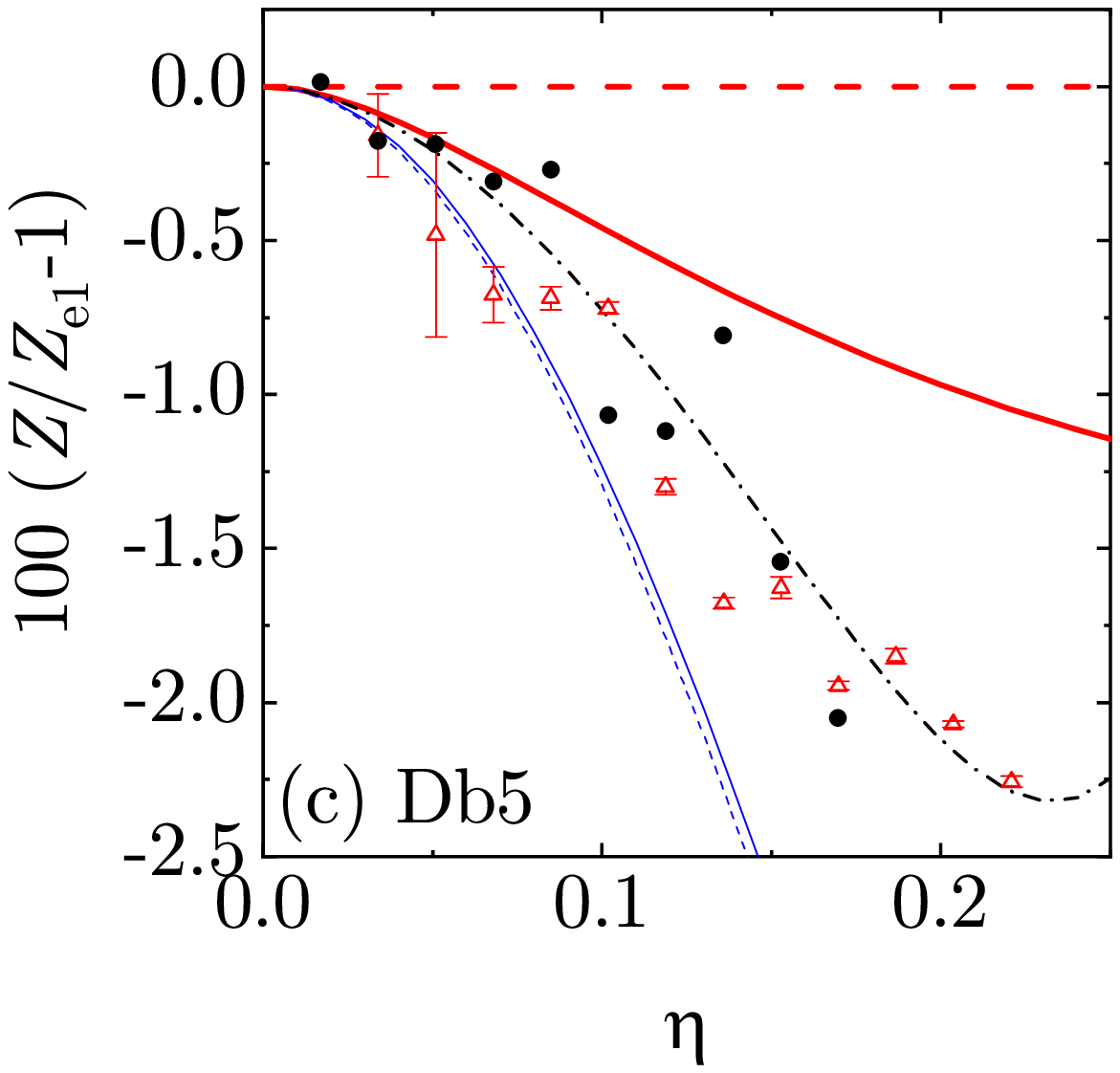}\\ \\
\includegraphics[height=4.6cm]{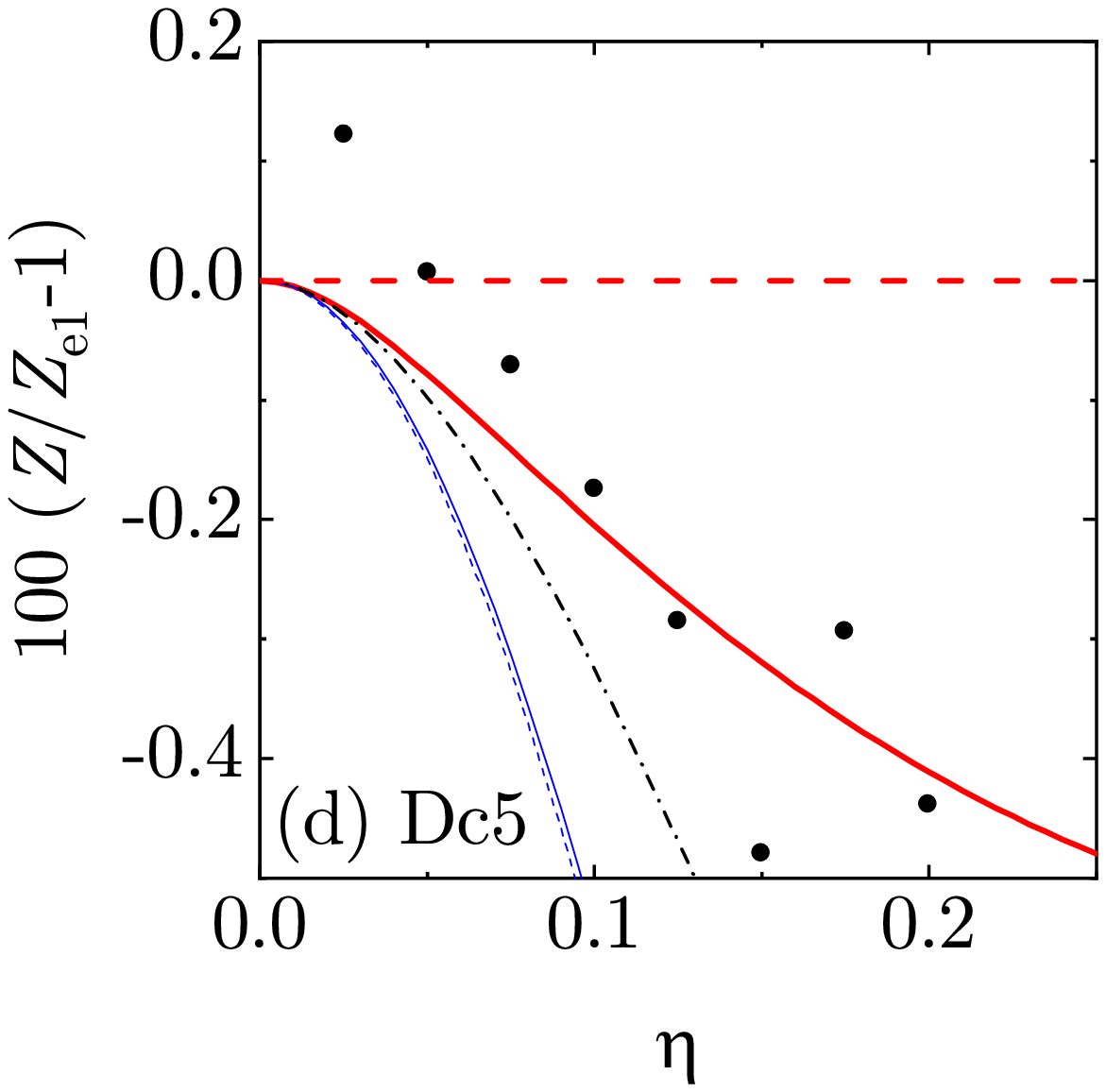}&\includegraphics[height=4.6cm]{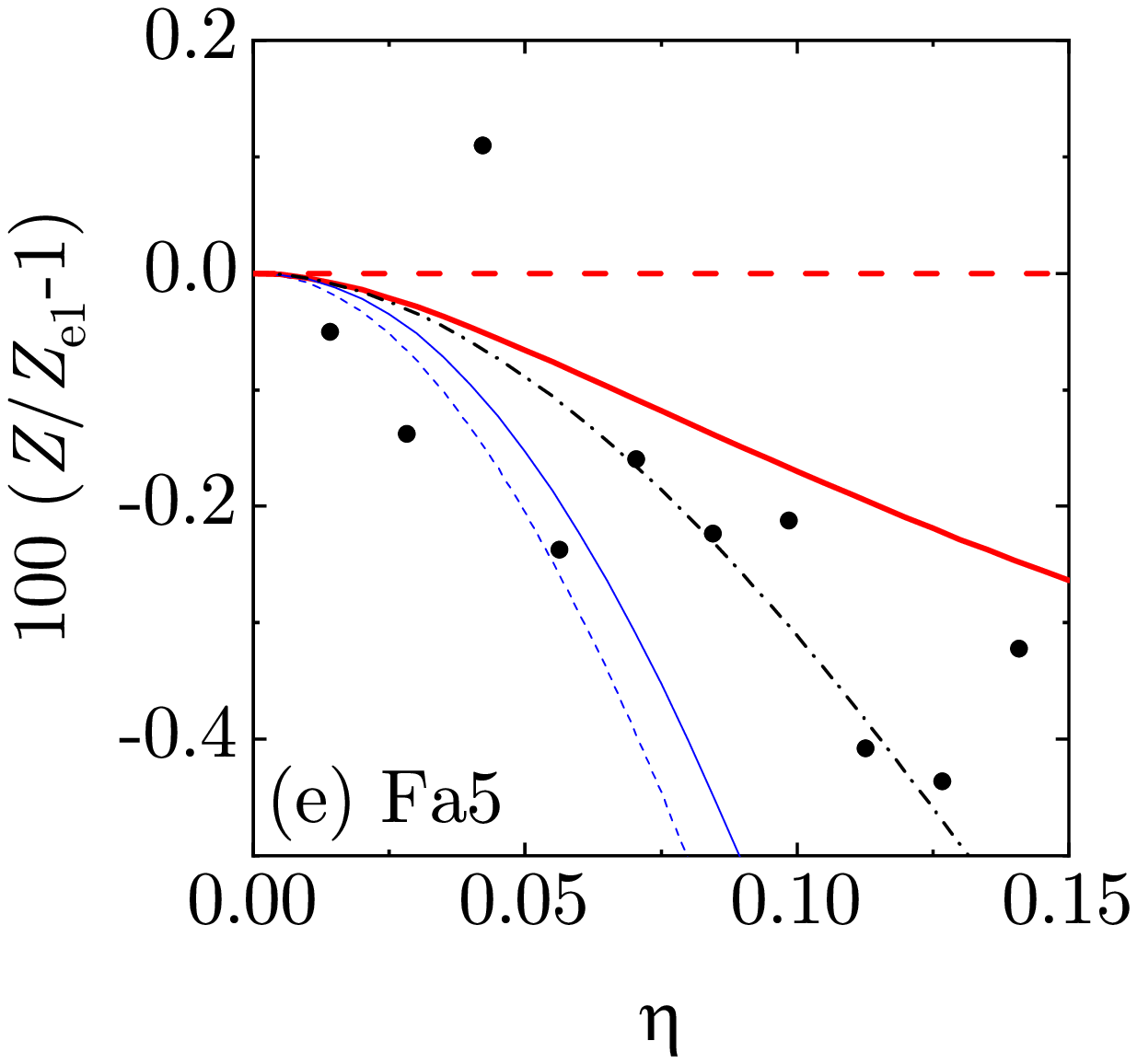}&\includegraphics[height=4.5cm]{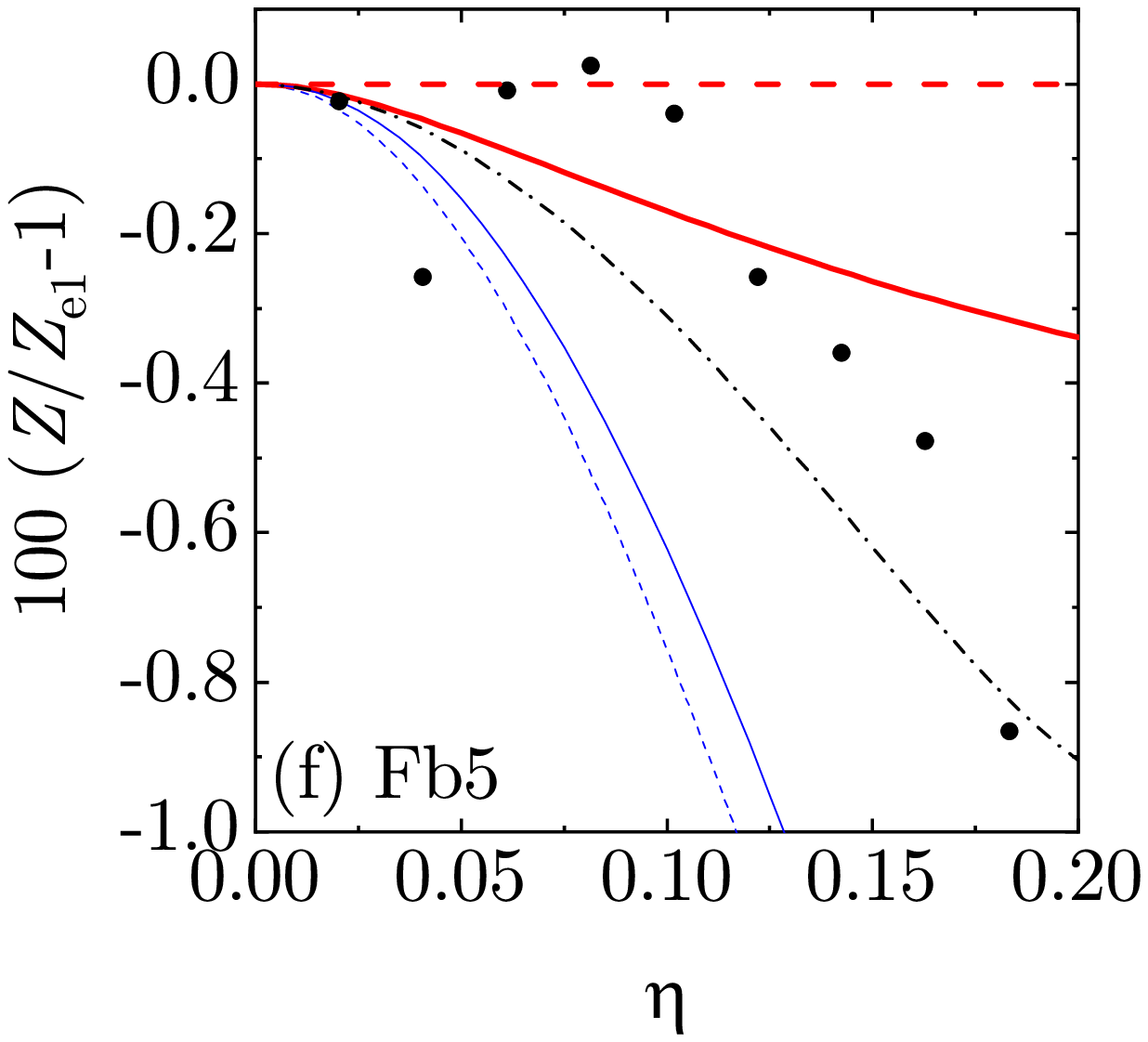}\\ \\
\includegraphics[height=4.6cm]{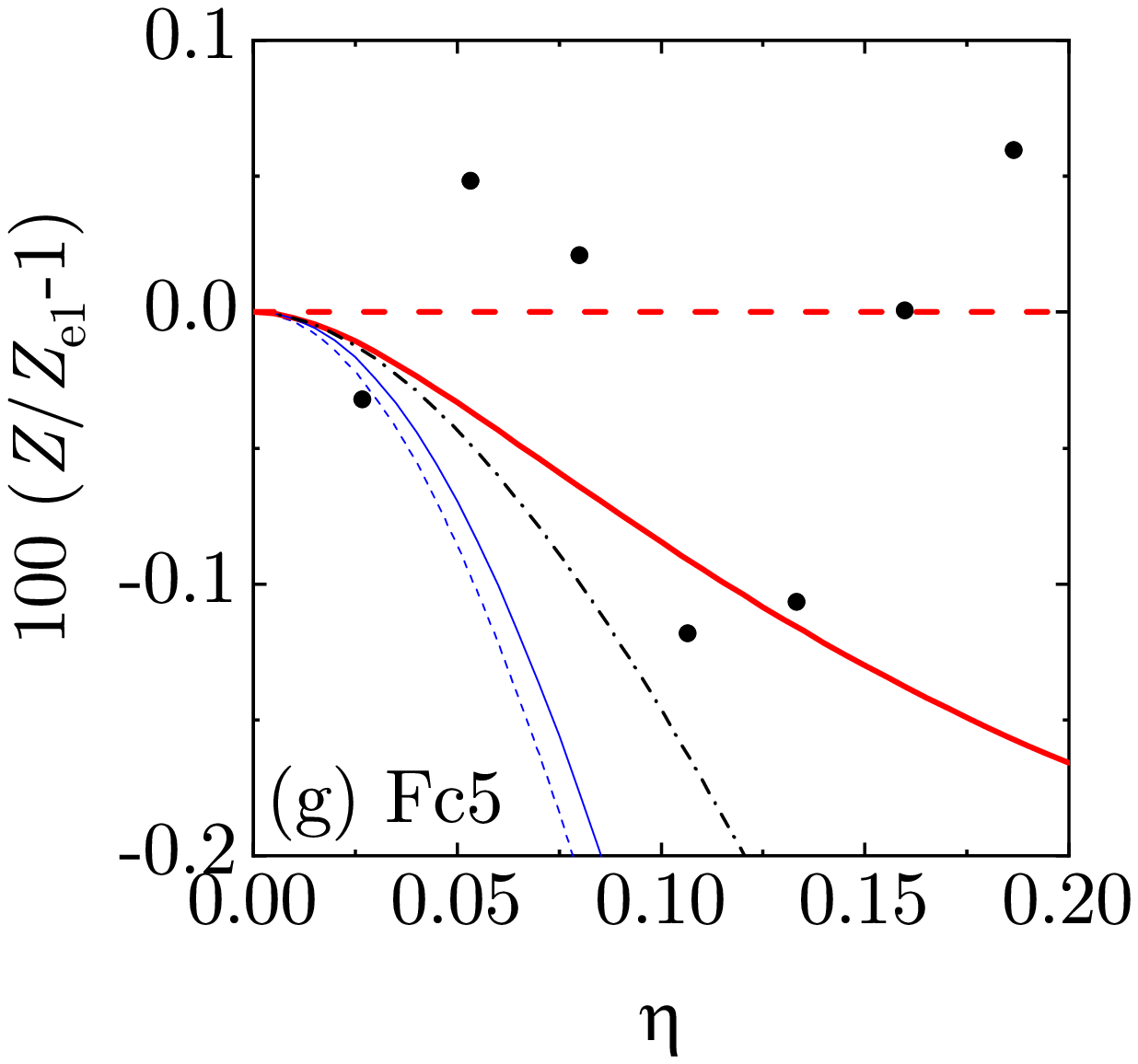}&&\\
\end{tabular}
\caption{Plot of the relative deviations $100[Z(\eta)/Z_\text{e1}(\eta)-1]$ from the theoretical EOS $Z_\text{e1}(\eta)$ for the five-dimensional mixtures Cb5--Fc5 (see Table~\ref{table1}). {Thick (red)} dashed lines: e1; {thick (red)} solid lines: \ep1; {thin (blue)} dashed lines: e2; {thin (blue)} solid lines: \ep2; {dash-dotted (black)} lines: sp; filled (black) circles: MD; open (red) triangles  with error bars  in panels (\textbf{a}) and (\textbf{c}): MC.\label{fig5D_2}}
\end{figure}
\section{Discussion and Concluding~Remarks}
\label{sec4}

In this paper we have carried out a thorough comparison between our theoretical proposals for the EOS of a multicomponent $d$-dimensional mixture of hard hyperspheres and the available simulation results for binary mixtures of both four- and five-dimensional hard hyperspheres. It should be stressed that in this comparison we have restricted ourselves to the liquid
branch. Let us now summarize the outcome of the different theories for the compressibility~factor.

First, we note that $Z_{\text{\ep2}}(\eta)\approx Z_{\text{e2}}(\eta)<Z_{\text{sp}}(\eta)<Z_{\text{\ep1}}(\eta)<Z_{\text{e1}}(\eta)$. The~fact that $Z_{\text{\ep2}}(\eta)\approx Z_{\text{e2}}(\eta)$ is a consequence of the small deviations of $B_3^{\text{e2}}$ from the exact third virial coefficient (see Figure~\ref{fig0}). Thus, there does not seem to be any practical advantage in choosing $Z_{\text{\ep2}}$ instead of $Z_{\text{e2}}$, especially if $d=4$ [where the exact $B_3$ has a rather involved expression, see Equations \eqref{B111&B112}]. If~one restricts oneself to the comparison between those approximate EOS that do not yield the exact $B_3$, namely $Z_{\text{e1}}$ and $Z_{\text{e2}}$, we find that $Z_{\text{e2}}$ performs generally better. On~the other hand, if~approximations requiring the exact $B_3$ as input are considered, namely $Z_{\text{\ep1}}$, $Z_{\text{\ep2}}$, and~$Z_{\text{sp}}$, the~conclusion is that  $Z_{\text{sp}}$ generally outperforms the other~two.

The comparison with the simulation data confirms that the good agreement between the results of $Z_\text{e1}(\eta)$  that had been found earlier in connection with both MD~\cite{GAH01} and MC~\cite{BW12,BW16} simulation data are even improved by the other approximate theories. In~fact, in~both the four- and five-dimensional cases, the~best agreement with the MD results is generally obtained from $Z_{\text{\ep1}}$ and $Z_{\text{sp}}$. On~the other hand, for~the four-dimensional case, the~best agreement with the MC results corresponds to  $Z_{\text{\ep2}}\approx Z_{\text{e2}}$,  while that for the five-dimensional case corresponds to  $Z_{\text{sp}}$.

{Finally, it must be pointed out that it seems that overall $Z_{\text{sp}}$ exhibits the best global behavior. However, more accurate simulation data would be needed to confirm this conclusion. It should also be stressed that the performance of the analyzed approximate {EOS for fluid mixtures}  might be affected by the reliability of the (monocomponent) LM EOS. In~any event, one may reasonably argue that the mapping between the compressibility factor of the mixture and the one of the monocomponent system with an effective packing fraction [see Equations \eqref{18a} and \eqref{19}] that had already been tested in two- \cite{SYHO17} and three-dimensional~\cite{SYHOO14} mixtures is confirmed as an excellent approach also for higher dimensions.}


\vspace{6pt}


\authorcontributions{
A.S. proposed the idea and the three authors performed the calculations.
The three authors also participated in the analysis and discussion of the results and worked on the revision and writing of the final~manuscript. All authors have read and agreed to the published version of the manuscript.}

\funding{A.S. and S.B.Y. acknowledge financial support from the Spanish Agencia Estatal de Investigaci\'on through Grant No.\ FIS2016-76359-P and the Junta de Extremadura
(Spain) through Grant No.\ GR18079, both partially financed by Fondo Europeo de Desarrollo Regional funds.}


\conflictsofinterest{The authors declare no conflict of interest. 
}

\abbreviations{The following abbreviations are used in this manuscript:\\

\noindent
\begin{tabular}{@{}ll}
EOS & Equation of state\\
LM & Luban--Michels\\
MC  & Monte Carlo\\
MD  & Molecular dynamics\\
\end{tabular}}

\reftitle{References}

\begin{thebibliography}{999}

\bibitem[Frisch \em{et~al.}(1985)Frisch, Rivier, and Wyler]{FRW85}
Frisch, H.L.; Rivier, N.; Wyler, D.
\newblock Classical Hard-Sphere Fluid in Infinitely Many Dimensions.
\newblock {\em Phys. Rev. Lett.} {\bf 1985}, {\em 54},~2061--2063, doi:10.1103/PhysRevLett.54.2061.

\bibitem[Luban(1986)]{L86}
Luban, M.
\newblock Comment on {``Classical Hard-Sphere Fluid in Infinitely Many
  Dimensions''}.
\newblock {\em Phys. Rev. Lett.} {\bf 1986}, {\em 56},~2330--2330, doi:10.1103/PhysRevLett.56.2330.

\bibitem[Frisch \em{et~al.}(1986)Frisch, Rivier, and Wyler]{FRW86}
Frisch, H.L.; Rivier, N.; Wyler, D.
\newblock {F}risch, {R}ivier, and {W}yler Respond.
\newblock {\em Phys. Rev. Lett.} {\bf 1986}, {\em 56},~2331--2331, doi:10.1103/PhysRevLett.56.2331.

\bibitem[Klein and Frisch(1986)]{KF86}
Klein, W.; Frisch, H.L.
\newblock Instability in the infinite dimensional hard-sphere fluid.
\newblock {\em J. Chem. Phys.} {\bf 1986}, {\em 84},~968--970, doi:10.1063/1.450544.

\bibitem[Wyler \em{et~al.}(1987)Wyler, Rivier, and Frisch]{WRF87}
Wyler, D.; Rivier, N.; Frisch, H.L.
\newblock Hard-sphere fluid in infinite dimensions.
\newblock {\em Phys. Rev. A} {\bf 1987}, {\em 36},~2422--2431, doi:10.1103/PhysRevA.36.2422.

\bibitem[Bagchi and Rice(1988)]{BR88}
Bagchi, B.; Rice, S.A.
\newblock On the stability of the infinite dimensional fluid of hard
  hyperspheres: {A} statistical mechanical estimate of the density of closest
  packing of simple hypercubic lattices in spaces of large dimensionality.
\newblock {\em J. Chem. Phys.} {\bf 1988}, {\em 88},~1177--1184, doi:10.1063/1.454237.

\bibitem[Elskens and Frisch(1988)]{EF88}
Elskens, Y.; Frisch, H.L.
\newblock Kinetic theory of hard spheres in infinite dimensions.
\newblock {\em Phys. Rev. A} {\bf 1988}, {\em 37},~4351--4353, doi:10.1103/PhysRevA.37.4351.

\bibitem[Carmesin \em{et~al.}(1991)Carmesin, Frisch, and Percus]{CFP91}
Carmesin, H.O.; Frisch, H.; Percus, J.
\newblock Binary nonadditive hard-sphere mixtures at high dimension.
\newblock {\em J. Stat. Phys.} {\bf 1991}, {\em 63},~791--795, doi:10.1007/BF01029212.

\bibitem[Frisch and Percus(1999)]{FP99}
Frisch, H.L.; Percus, J.K.
\newblock {High dimensionality as an organizing device for classical fluids}.
\newblock {\em Phys. Rev. E} {\bf 1999}, {\em 60},~2942--2948, doi:10.1103/PhysRevE.60.2942.

\bibitem[Parisi and Slanina(2000)]{PS00}
Parisi, G.; Slanina, F.
\newblock {Toy model for the mean-field theory of hard-sphere liquids}.
\newblock {\em Phys. Rev. E} {\bf 2000}, {\em 62},~6554--6559, doi:10.1103/PhysRevE.62.6554.

\bibitem[Yukhimets \em{et~al.}(2000)Yukhimets, Frisch, and Percus]{YFP00}
Yukhimets, A.; Frisch, H.L.; Percus, J.K.
\newblock Molecular Fluids at High Dimensionality.
\newblock {\em J. Stat. Phys.} {\bf 2000}, {\em 100},~135--151, doi:10.1023/A:1018635527522.

\bibitem[Charbonneau \em{et~al.}(2017)Charbonneau, Kurchan, Parisi, Urbani, and
  Zamponi]{CKPUZ16}
Charbonneau, P.; Kurchan, J.; Parisi, G.; Urbani, P.; Zamponi, F.
\newblock Glass and Jamming Transitions: {F}rom Exact Results to
  Finite-Dimensional Descriptions.
\newblock {\em Annu. Rev. Cond. Matter Phys.} {\bf 2017}, {\em 8},~265--288, doi:10.1146/annurev-conmatphys-031016-025334.

\bibitem[Santos and {L\'opez de Haro}(2005)]{SH05}
Santos, A.; {L\'opez de Haro}, M.
\newblock Demixing can occur in binary hard-sphere mixtures with negative
  non-additivity.
\newblock {\em Phys. Rev. E} {\bf 2005}, {\em 72},~010501(R), doi:10.1103/PhysRevE.72.010501.

\bibitem[Freasier and Isbister(1981)]{FI81}
Freasier, C.; Isbister, D.J.
\newblock {A remark on the Percus--Yevick approximation in high dimensions.
  Hard core systems}.
\newblock {\em Mol. Phys.} {\bf 1981}, {\em 42},~927--936, doi:10.1080/00268978100100711.

\bibitem[Leutheusser(1984)]{L84}
Leutheusser, E.
\newblock Exact solution of the {Percus}--{Yevick} equation for a hard-core
  fluid in odd dimensions.
\newblock {\em Physica~A} {\bf 1984}, {\em 127},~667--676, doi:10.1016/0378-4371(84)90050-5.

\bibitem[Michels and Trappeniers(1984)]{MT84}
Michels, J.P.J.; Trappeniers, N.J.
\newblock {Dynamical computer simulations on hard hyperspheres in four- and
  five-dimensional space}.
\newblock {\em Phys. Lett. A} {\bf 1984}, {\em 104},~425--429, doi:10.1016/0375-9601(84)90749-7.

\bibitem[Baus and Colot(1986)]{BC86}
Baus, M.; Colot, J.L.
\newblock Theoretical structure factors for hard-core fluids.
\newblock {\em J. Phys. C} {\bf 1986}, {\em 19},~L643--L648, doi:10.1088/0022-3719/19/28/002.

\bibitem[Baus and Colot(1987)]{BC87}
Baus, M.; Colot, J.L.
\newblock {Thermodynamics and structure of a fluid of hard rods, disks,
  spheres, or hyperspheres from rescaled virial expansions}.
\newblock {\em Phys. Rev. A} {\bf 1987}, {\em 36},~3912--3925, doi:10.1103/PhysRevA.36.3912.

\bibitem[Rosenfeld(1987)]{R87}
Rosenfeld, Y.
\newblock Distribution function of two cavities and {P}ercus--{Y}evick direct
  correlation functions for a hard sphere fluid in $D$ dimensions: {O}verlap
  volume function representation.
\newblock {\em J. Chem. Phys.} {\bf 1987}, {\em 87},~4865--4869, doi:10.1063/1.452797.

\bibitem[Rosenfeld(1988)]{R88}
Rosenfeld, Y.
\newblock Scaled field particle theory of the structure and thermodynamics of
  isotropic hard particle fluids.
\newblock {\em J. Chem. Phys.} {\bf 1988}, {\em 89},~4272--4287, doi:10.1063/1.454810.

\bibitem[Amor\'os \em{et~al.}(1989)Amor\'os, Solana, and Villar]{ASV89}
Amor\'os, J.; Solana, J.R.; Villar, E.
\newblock {Equations of state for four- and five-dimensional hard hypersphere
  fluids}.
\newblock {\em Phys. Chem. Liq.} {\bf 1989}, {\em 19},~119--124, doi:10.1080/00319108908028443.

\bibitem[Song \em{et~al.}(1989)Song, Mason, and Stratt]{SMS89}
Song, Y.; Mason, E.A.; Stratt, R.M.
\newblock {Why does the Carnahan-Starling equation work so well?}
\newblock {\em J. Phys. Chem.} {\bf 1989}, {\em 93},~6916--6919, doi:10.1021/j100356a008.

\bibitem[Song and Mason(1990)]{SM90}
Song, Y.; Mason, E.A.
\newblock {Equation of state for fluids of spherical particles in $d$
  dimensions}.
\newblock {\em J. Chem. Phys.} {\bf 1990}, {\em 93},~686--688, doi:10.1063/1.459517.

\bibitem[Gonz\'alez \em{et~al.}(1990)Gonz\'alez, Gonz\'alez, and
  Silbert]{GGS90}
Gonz\'alez, D.J.; Gonz\'alez, L.E.; Silbert, M.
\newblock {Thermodynamics of a fluid of hard $D$-dimensional spheres:
  Percus-Yevick and Carnahan-Starling-like results for $D=4$ and $5$}.
\newblock {\em Phys. Chem. Liq.} {\bf 1990}, {\em 22},~95--102, doi:10.1080/00319109008036415.

\bibitem[Luban and Michels(1990)]{LM90}
Luban, M.; Michels, J.P.J.
\newblock {Equation of state of hard $D$-dimensional hyperspheres}.
\newblock {\em Phys. Rev. A} {\bf 1990}, {\em 41},~6796--6804, doi:10.1103/PhysRevA.41.6796.

\bibitem[Maeso \em{et~al.}(1991)Maeso, Solana, Amor\'os, and Villar]{MSAV91}
Maeso, M.J.; Solana, J.R.; Amor\'os, J.; Villar, E.
\newblock {Equations of state for D-dimensional hard sphere fluids}.
\newblock {\em Mater. Chem. Phys.} {\bf 1991}, {\em 30},~39--42, doi:10.1016/0254-0584(91)90151-J.

\bibitem[Gonz\'alez \em{et~al.}(1991)Gonz\'alez, Gonz\'alez, and
  Silbert]{GGS91}
Gonz\'alez, D.J.; Gonz\'alez, L.E.; Silbert, M.
\newblock {Structure and thermodynamics of hard $D$-dimensional spheres:
  overlap volume function approach}.
\newblock {\em Mol. Phys.} {\bf 1991}, {\em 74},~613--627, doi:10.1080/00268979100102461.

\bibitem[Gonz\'alez \em{et~al.}(1992)Gonz\'alez, Gonz\'alez, and
  Silbert]{GGS92}
Gonz\'alez, L.E.; Gonz\'alez, D.J.; Silbert, M.
\newblock {Structure and thermodynamics of mixtures of hard $D$-dimensional
  spheres: Overlap volume function approach}.
\newblock {\em J. Chem. Phys.} {\bf 1992}, {\em 97},~5132--5141, doi:10.1063/1.463810.

\bibitem[Velasco \em{et~al.}(1999)Velasco, Mederos, and Navascu\'es]{VMN99}
Velasco, E.; Mederos, L.; Navascu\'es, G.
\newblock Analytical approach to the thermodynamics and density distribution of
  crystalline phases of hard spheres spheres.
\newblock {\em Mol. Phys.} {\bf 1999}, {\em 97},~1273--1277, doi:10.1080/00268979909482929.

\bibitem[Bishop \em{et~al.}(1999)Bishop, Masters, and Clarke]{BMC99}
Bishop, M.; Masters, A.; Clarke, J.H.R.
\newblock {Equation of state of hard and Weeks--Chandle--Anderson hyperspheres
  in four and five dimensions}.
\newblock {\em J. Chem. Phys.} {\bf 1999}, {\em 110},~11449--11453, doi:10.1063/1.479086.

\bibitem[Finken \em{et~al.}(2001)Finken, Schmidt, and L{\"o}wen]{FSL01}
Finken, R.; Schmidt, M.; L{\"o}wen, H.
\newblock {Freezing transition of hard hyperspheres}.
\newblock {\em Phys. Rev. E} {\bf 2001}, {\em 65},~016108, doi:10.1103/PhysRevE.65.016108.

\bibitem[Santos \em{et~al.}(1999)Santos, Yuste, and {L\'opez de Haro}]{SYH99}
Santos, A.; Yuste, S.B.; {L\'opez de Haro}, M.
\newblock {Equation of state of a multicomponent $d$-dimensional hard-sphere
  fluid}.
\newblock {\em Mol. Phys.} {\bf 1999}, {\em 96},~1--5, doi:10.1080/00268979909482932.

\bibitem[Mon and Percus(1999)]{MP99}
Mon, K.K.; Percus, J.K.
\newblock Virial expansion and liquid-vapor critical points of high dimension
  classical fluids.
\newblock {\em J.~Chem. Phys.} {\bf 1999}, {\em 110},~2734--2735, doi:10.1063/1.477998.

\bibitem[Santos(2000)]{S00}
Santos, A.
\newblock An equation of state \textit{\`a La} {C}arnahan-{S}tarling A Five-Dimens. Fluid Hard Hyperspheres.\newblock {\em J. Chem. Phys.} {\bf 2000}, {\em 112},~10680--10681, doi:10.1063/1.481701.

\bibitem[Yuste \em{et~al.}(2000)Yuste, Santos, and {L\'opez de Haro}]{YSH00}
Yuste, S.B.; Santos, A.; {L\'opez de Haro}, M.
\newblock {Demixing in binary mixtures of hard hyperspheres}.
\newblock {\em Europhys. Lett.} {\bf 2000}, {\em 52},~158--164, doi:10.1209/epl/i2000-00411-9.

\bibitem[Gonz\'alez-Melchor \em{et~al.}(2001)Gonz\'alez-Melchor, Alejandre, and
  {L\'opez de Haro}]{GAH01}
Gonz\'alez-Melchor, M.; Alejandre, J.; {L\'opez de Haro}, M.
\newblock {Equation of state and structure of binary mixtures of hard
  $d$-dimensional hyperspheres}.
\newblock {\em J. Chem. Phys.} {\bf 2001}, {\em 114},~4905--4911, doi:10.1063/1.1349094.

\bibitem[Santos \em{et~al.}(2002)Santos, Yuste, and {L\'opez de Haro}]{SYH02}
Santos, A.; Yuste, S.B.; {L\'opez de Haro}, M.
\newblock Contact values of the radial distribution functions of additive
  hard-sphere mixtures in $d$ dimensions: A new proposal.
\newblock {\em J. Chem. Phys.} {\bf 2002}, {\em 117},~5785--5793, doi:10.1063/1.1502247.

\bibitem[Robles \em{et~al.}(2004)Robles, {L\'opez de Haro}, and Santos]{RHS04}
Robles, M.; {L\'opez de Haro}, M.; Santos, A.
\newblock Equation of state of a seven-dimensional hard-sphere fluid.
  {Percus--Yevick} theory and molecular-dynamics simulations.
\newblock {\em J. Chem. Phys.} {\bf 2004}, {\em 120},~9113--9122, doi:10.1063/1.1701840.

\bibitem[Santos \em{et~al.}(2005)Santos, {L\'opez de Haro}, and Yuste]{SHY05}
Santos, A.; {L\'opez de Haro}, M.; Yuste, S.B.
\newblock Equation of state of nonadditive $d$-dimensional hard-sphere
  mixtures.
\newblock {\em J. Chem. Phys.} {\bf 2005}, {\em 122},~{024}{514}, doi:10.1063/1.1832591.

\bibitem[Bishop \em{et~al.}(2005)Bishop, Whitlock, and Klein]{BWK05}
Bishop, M.; Whitlock, P.A.; Klein, D.
\newblock The structure of hyperspherical fluids in various dimensions.
\newblock {\em J. Chem. Phys.} {\bf 2005}, {\em 122},~{074}{508}, doi:10.1063/1.1848091.

\bibitem[Bishop and Whitlock(2005)]{BW05}
Bishop, M.; Whitlock, P.A.
\newblock The equation of state of hard hyperspheres in four and five
  dimensions.
\newblock {\em J. Chem. Phys.} {\bf 2005}, {\em 123},~{014}{507}, doi:10.1063/1.1874793.

\bibitem[Lue and Bishop(2006)]{LB06}
Lue, L.; Bishop, M.
\newblock Molecular dynamics study of the thermodynamics and transport
  coefficients of hard hyperspheres in six and seven dimensions.
\newblock {\em Phys. Rev. E} {\bf 2006}, {\em 74},~{021}{201}, doi:10.1103/PhysRevE.74.021201.

\bibitem[{L\'opez de Haro} \em{et~al.}(2006){L\'opez de Haro}, Yuste, and
  Santos]{HYS06}
{L\'opez de Haro}, M.; Yuste, S.B.; Santos, A.
\newblock Test of a universality ansatz for the contact values of the radial
  distribution functions of hard-sphere mixtures near a hard wall.
\newblock {\em Mol. Phys.} {\bf 2006}, {\em 104},~3461--3467, doi:10.1080/00268970601028963.

\bibitem[Bishop and Whitlock(2007)]{BW07}
Bishop, M.; Whitlock, P.A.
\newblock Monte {C}arlo Simulation of Hard Hyperspheres in Six, Seven and Eight
  Dimensions for Low to Moderate Densities.
\newblock {\em J. Stat. Phys.} {\bf 2007}, {\em 126},~299--314, doi:10.1007/s10955-006-9266-9.

\bibitem[Robles \em{et~al.}(2007)Robles, {L\'opez de Haro}, and Santos]{RHS07}
Robles, M.; {L\'opez de Haro}, M.; Santos, A.
\newblock {Percus--Yevick} theory for the structural properties of the
  seven-dimensional hard-sphere fluid.
\newblock {\em J. Chem. Phys.} {\bf 2007}, {\em 126},~{016}{101}, doi:10.1063/1.2424459.

\bibitem[Whitlock \em{et~al.}(2007)Whitlock, Bishop, and Tiglias]{WBT07}
Whitlock, P.A.; Bishop, M.; Tiglias, J.L.
\newblock Structure factor for hard hyperspheres in higher dimensions.
\newblock {\em J. Chem. Phys.} {\bf 2007}, {\em 126},~{224}{505}, doi:10.1063/1.2743031.

\bibitem[Rohrmann and Santos(2007)]{RS07}
Rohrmann, R.D.; Santos, A.
\newblock Structure of hard-hypersphere fluids in odd dimensions.
\newblock {\em Phys. Rev. E} {\bf 2007}, {\em 76},~{051}{202}, doi:10.1103/PhysRevE.76.051202.

\bibitem[{L\'opez de Haro} \em{et~al.}(2008){L\'opez de Haro}, Yuste, and
  Santos]{HYS08}
{L\'opez de Haro}, M.; Yuste, S.B.; Santos, A.
\newblock {Alternative Approaches to the Equilibrium Properties of Hard-Sphere
  Liquids}.
\newblock  In {\em Theory and Simulation of Hard-Sphere Fluids and Related Systems};
  Mulero, A., Ed.; Lecture Notes in
  Physics; Springer: Berlin, Germany, 2008; Volume 753,  pp. 183--245.

\bibitem[Bishop \em{et~al.}(2008)Bishop, Clisby, and Whitlock]{BCW08}
Bishop, M.; Clisby, N.; Whitlock, P.A.
\newblock The equation of state of hard hyperspheres in nine dimensions for low
  to moderate densities.
\newblock {\em J. Chem. Phys.} {\bf 2008}, {\em 128},~{034}{506}, doi:10.1063/1.2821962.

\bibitem[Adda-Bedia \em{et~al.}(2008)Adda-Bedia, Katzav, and Vella]{AKV08}
Adda-Bedia, M.; Katzav, E.; Vella, D.
\newblock Solution of the Percus--Yevick equation for hard hyperspheres in even
  dimensions.
\newblock {\em J. Chem. Phys.} {\bf 2008}, {\em 129},~{144}{506}, doi:10.1063/1.2991338.

\bibitem[Rohrmann \em{et~al.}(2008)Rohrmann, Robles, {L\'opez de Haro}, and
  Santos]{RRHS08}
Rohrmann, R.D.; Robles, M.; {L\'opez de Haro}, M.; Santos, A.
\newblock Virial series for fluids of hard hyperspheres in odd dimensions.
\newblock {\em J. Chem. Phys.} {\bf 2008}, {\em 129},~{014}{510}, doi:10.1063/1.2951456.

\bibitem[van Meel \em{et~al.}(2009)van Meel, Charbonneau, Fortini, and
  Charbonneau]{vMCFC09}
van Meel, J.A.; Charbonneau, B.; Fortini, A.; Charbonneau, P.
\newblock Hard-sphere crystallization gets rarer with increasing dimension.
\newblock {\em Phys. Rev. E} {\bf 2009}, {\em 80},~{061}{110}, doi:10.1103/PhysRevE.80.061110.

\bibitem[Lue \em{et~al.}(2010)Lue, Bishop, and Whitlock]{LBW10}
Lue, L.; Bishop, M.; Whitlock, P.A.
\newblock The fluid to solid phase transition of hard hyperspheres in four and
  five dimensions.
\newblock {\em J. Chem. Phys.} {\bf 2010}, {\em 132},~{104}{509}, doi:10.1063/1.3354115.

\bibitem[Rohrmann and Santos(2011)]{RS11}
Rohrmann, R.D.; Santos, A.
\newblock Multicomponent fluids of hard hyperspheres in odd dimensions.
\newblock {\em Phys. Rev. E} {\bf 2011}, {\em 83},~{011}{201}, doi:10.1103/PhysRevE.83.011201.

\bibitem[Leithall and Schmidt(2011)]{LS11}
Leithall, G.; Schmidt, M.
\newblock Density functional for hard hyperspheres from a
  tensorial-diagrammatic series.
\newblock {\em Phys. Rev. E} {\bf 2011}, {\em 83},~{021}{201}, doi:10.1103/PhysRevE.83.021201.

\bibitem[Estrada and Robles(2011)]{ER11}
Estrada, C.D.; Robles, M.
\newblock Fluid--solid transition in hard hypersphere systems.
\newblock {\em J. Chem. Phys.} {\bf 2011}, {\em 134},~{044}{115}, doi:10.1063/1.3530780.

\bibitem[Bishop and Whitlock(2012)]{BW12}
Bishop, M.; Whitlock, P.A.
\newblock Monte {C}arlo study of four dimensional binary hard hypersphere
  mixtures.
\newblock {\em J.~Chem. Phys.} {\bf 2012}, {\em 136},~{014}{506}, doi:10.1063/1.3671651.

\bibitem[Bishop and Whitlock(2013)]{BW13}
Bishop, M.; Whitlock, P.A.
\newblock Phase transitions in four-dimensional binary hard hypersphere
  mixtures.
\newblock {\em J. Chem. Phys.} {\bf 2013}, {\em 138},~{084}{502}, doi:10.1063/1.4789953.

\bibitem[Bishop and Whitlock(2016)]{BW16}
Bishop, M.; Whitlock, P.A.
\newblock Five dimensional binary hard hypersphere mixtures: {A Monte Carlo}
  study.
\newblock {\em J.~Chem. Phys.} {\bf 2016}, {\em 145},~{154}{502}, doi:10.1063/1.4964614.

\bibitem[Amor\'os and Ravi(2013)]{AR13}
Amor\'os, J.; Ravi, S.
\newblock On the application of the {C}arnahan--{S}tarling method for hard
  hyperspheres in several dimensions.
\newblock {\em Phys. Lett. A} {\bf 2013}, {\em 377},~2089--2092, doi:10.1016/j.physleta.2013.06.004.

\bibitem[Amor\'os(2014)]{A14}
Amor\'os, J.
\newblock Equations of state for tetra-dimensional hard-sphere fluids.
\newblock {\em Phys. Chem. Liq.} {\bf 2014}, {\em 52},~287--290, doi:10.1080/00319104.2013.820301.

\bibitem[Heinen \em{et~al.}(2015)Heinen, Horbach, and L\"owen]{HHL15}
Heinen, M.; Horbach, J.; L\"owen, H.
\newblock Liquid pair correlations in four spatial dimensions: Theory versus
  simulation.
\newblock {\em Mol. Phys.} {\bf 2015}, {\em 113},~1164--1169, doi:10.1080/00268976.2014.993736.

\bibitem[Santos(2016)]{S16}
Santos, A.
\newblock {\em {A Concise Course on the Theory of Classical Liquids. Basics and
  Selected Topics}}; {Lecture Notes in Physics}; Springer: New
  York, NY, USA, 2016; Volume 923.

\bibitem[Santos \em{et~al.}(2017)Santos, Yuste, {L\'opez de Haro}, and
  Ogarko]{SYHO17}
Santos, A.; Yuste, S.B.; {L\'opez de Haro}, M.; Ogarko, V.
\newblock Equation of state of polydisperse hard-disk mixtures in the
  high-density regime.
\newblock {\em Phys. Rev. E} {\bf 2017}, {\em 93},~{062}{603}, doi:10.1103/PhysRevE.062603.

\bibitem[Akhouri(2017)]{A17}
Akhouri, B.P.
\newblock Equations of state for hard hypersphere fluids in high dimensional
  spaces.
\newblock {\em Int. J. Chem. Stud.} {\bf 2017}, {\em 5},~39--45, doi:10.22271/chemi.

\bibitem[Ivanizki(2018)]{I18}
Ivanizki, D.
\newblock A generalization of the {C}arnahan--{S}tarling approach with
  applications to four- and five-dimensional hard spheres.
\newblock {\em Phys. Lett. A} {\bf 2018}, {\em 382},~1745--1751, doi:10.1016/j.physleta.2018.04.036.

\bibitem[Santos \em{et~al.}(2001)Santos, Yuste, and {L\'opez de Haro}]{SYH01}
Santos, A.; Yuste, S.B.; {L\'opez de Haro}, M.
\newblock Virial coefficients and equations of state for mixtures of hard
  discs, hard spheres, and hard hyperspheres.
\newblock {\em Mol. Phys.} {\bf 2001}, {\em 99},~1959--1972, doi:10.1080/00268970110063890.

\bibitem[Ree and Hoover(1964)]{RH64a}
Ree, F.H.; Hoover, W.G.
\newblock On the Signs of the Hard Sphere Virial Coefficients.
\newblock {\em J. Chem. Phys.} {\bf 1964}, {\em 40},~2048--2049, doi:10.1063/1.1725456.

\bibitem[Luban and Baram(1982)]{LB82}
Luban, M.; Baram, A.
\newblock {{Third} and fourth virial coefficients of hard hyperspheres of
  arbitrary dimensionality}.  
\newblock {\em J. Chem. Phys.} {\bf 1982}, {\em 76},~3233--3241, doi:10.1063/1.443316.

\bibitem[Joslin(1982)]{J82}
Joslin, C.G.
\newblock Third and fourth virial coefficients of hard hyperspheres of
  arbitrary dimensionality.
\newblock {\em J. Chem. Phys.} {\bf 1982}, {\em 77},~2701--2702, doi:10.1063/1.444104.

\bibitem[Loeser \em{et~al.}(1991)Loeser, Zhen, Kais, and Herschbach]{LZKH91}
Loeser, J.G.; Zhen, Z.; Kais, S.; Herschbach, D.R.
\newblock Dimensional interpolation of hard sphere virial coefficients.
\newblock {\em J. Chem. Phys.} {\bf 1991}, {\em 95},~4525--4544, doi:10.1063/1.461776.

\bibitem[Enciso \em{et~al.}(2002)Enciso, Almarza, Gonz\'alez, and
  Bermejo]{EAGB02}
Enciso, E.; Almarza, N.G.; Gonz\'alez, M.A.; Bermejo, F.J.
\newblock The virial coefficients of hard hypersphere binary mixtures.
\newblock {\em Mol. Phys.} {\bf 2002}, {\em 100},~1941--1944, doi:10.1080/00268970110108322.

\bibitem[Bishop \em{et~al.}(2004)Bishop, Masters, and Vlasov]{BMV04}
Bishop, M.; Masters, A.; Vlasov, A.Y.
\newblock Higher virial coefficients of four and five dimensional hard
  hyperspheres.
\newblock {\em J. Chem. Phys.} {\bf 2004}, {\em 121},~6884--6886, doi:10.1063/1.1777574.

\bibitem[Clisby and McCoy(2004{\natexlab{a}})]{CM04a}
Clisby, N.; McCoy, B.M.
\newblock Analytic Calculation of ${B_4}$ for Hard Spheres in Even Dimensions.
\newblock {\em J. Stat. Phys.} {\bf 2004}, {\em 114},~1343--1360, doi:10.1023/B:JOSS.0000013959.30878.d2.

\bibitem[Clisby and McCoy(2004{\natexlab{b}})]{CM04b}
Clisby, N.; McCoy, B.
\newblock Negative Virial Coefficients and the Dominance of Loose Packed
  Diagrams for $D$-Dimensional Hard Spheres.
\newblock {\em J. Stat. Phys.} {\bf 2004}, {\em 114},~1361--1392, doi:10.1023/B:JOSS.0000013960.83555.7d.

\bibitem[Bishop \em{et~al.}(2005)Bishop, Masters, and Vlasov]{BMV05}
Bishop, M.; Masters, A.; Vlasov, A.Y.
\newblock The eighth virial coefficient of four- and five-dimensional hard
  hyperspheres.
\newblock {\em J. Chem. Phys.} {\bf 2005}, {\em 122},~{154}{502}, doi:10.1063/1.1882273.

\bibitem[Clisby and McCoy(2005)]{CM05}
Clisby, N.; McCoy, B.M.
\newblock New results for virial coeffcients of hard spheres in ${D}$
  dimensions.
\newblock {\em Pramana} {\bf 2005}, {\em 64},~775--783, doi:10.1007/BF02704582.

\bibitem[Lyberg(2005)]{L05}
Lyberg, I.
\newblock The fourth virial coefficient of a fluid of hard spheres in odd
  dimensions.
\newblock {\em J. Stat. Phys.} {\bf 2005}, {\em 119},~747--764, doi:10.1007/s10955-005-3020-6.

\bibitem[Clisby and McCoy(2006)]{CM06}
Clisby, N.; McCoy, B.M.
\newblock Ninth and Tenth Order Virial Coefficients for Hard Spheres in ${D}$
  Dimensions.
\newblock {\em J.~Stat. Phys.} {\bf 2006}, {\em 122},~15--57, doi:10.1007/s10955-005-8080-0.

\bibitem[Zhang and Pettitt(2016)]{ZM16b}
Zhang, C.; Pettitt, B.M.
\newblock Computation of high-order virial coefficients in high-dimensional
  hard-sphere fluids by {M}ayer sampling.
\newblock {\em Mol. Phys.} {\bf 2016}, {\em 112},~1427--1447, doi:10.1080/00268976.2014.904945.

\bibitem[Skoge \em{et~al.}(2006)Skoge, Donev, Stillinger, and Torquato]{SDST06}
Skoge, M.; Donev, A.; Stillinger, F.H.; Torquato, S.
\newblock Packing Hyperspheres in high-dimensional Euclidean spaces.
\newblock {\em Phys. Rev. E} {\bf 2006}, {\em 74},~{041}{127}, doi:10.1103/PhysRevE.74.041127.

\bibitem[Torquato and Stillinger(2006{\natexlab{a}})]{TS06}
Torquato, S.; Stillinger, F.H.
\newblock New Conjectural Lower Bounds on the Optimal Density of Sphere
  Packings.
\newblock {\em Exp. Math.} {\bf 2006}, {\em 15},~307--331, doi:10.1080/10586458.2006.10128964.

\bibitem[Torquato and Stillinger(2006{\natexlab{b}})]{TS06a}
Torquato, S.; Stillinger, F.H.
\newblock Exactly Solvable Disordered Hard-Sphere Packing Model in
  Arbitrary-Dimensional {E}uclidean Spaces.
\newblock {\em Phys. Rev. E} {\bf 2006}, {\em 73},~{031}{106}, doi:10.1103/PhysRevE.73.031106.

\bibitem[Torquato \em{et~al.}(2006)Torquato, Uche, and Stillinger]{TUS06}
Torquato, S.; Uche, O.U.; Stillinger, F.H.
\newblock Random sequential addition of hard spheres in high {E}uclidean
  dimensions.
\newblock {\em Phys. Rev. E} {\bf 2006}, {\em 74},~{061}{308}, doi:10.1103/PhysRevE.74.061308.

\bibitem[Parisi and Zamponi(2006)]{PZ06}
Parisi, G.; Zamponi, F.
\newblock Amorphous packings of hard spheres for large space dimension.
\newblock {\em J. Stat. Mech.} {\bf 2006},  P03017, doi:10.1088/1742-5468/2006/03/p03017.

\bibitem[Scardicchio \em{et~al.}(2008)Scardicchio, Stillinger, and
  Torquato]{SST08}
Scardicchio, A.; Stillinger, F.H.; Torquato, S.
\newblock Estimates of the optimal density of sphere packings in high
  dimensions.
\newblock {\em J. Math. Phys.} {\bf 2008}, {\em 49},~{043}{301}, doi:10.1063/1.2897027.

\bibitem[van Meel \em{et~al.}(2009)van Meel, Frenkel, and Charbonneau]{vMFC09}
van Meel, J.A.; Frenkel, D.; Charbonneau, P.
\newblock Geometrical frustration: A study of four-dimensional hard spheres.
\newblock {\em Phys. Rev. E} {\bf 2009}, {\em 79},~{030}{201}(R), doi:10.1103/PhysRevE.79.030201.

\bibitem[Agapie and Whitlock(2010)]{AW10}
Agapie, S.C.; Whitlock, P.A.
\newblock Random packing of hyperspheres and {M}arsaglia's parking lot test.
\newblock {\em Monte Carlo Methods Appl.} {\bf 2010}, {\em
  16},~197--209, doi:10.1515/mcma.2010.019.

\bibitem[Torquato and Stillinger(2010)]{TS10}
Torquato, S.; Stillinger, F.H.
\newblock Jammed hard-particle packings: {F}rom {K}epler to {B}ernal and
  beyond.
\newblock {\em Rev. Mod. Phys.} {\bf 2010}, {\em 82},~2633--2672, doi:10.1103/RevModPhys.82.2633.

\bibitem[Zhang and Torquato(2013)]{ZT13}
Zhang, G.; Torquato, S.
\newblock Precise algorithm to generate random sequential addition of hard
  hyperspheres at saturation.
\newblock {\em Phys. Rev. E} {\bf 2013}, {\em 88},~{053}{312}, doi:10.1103/PhysRevE.88.053312.

\bibitem[Kazav \em{et~al.}(2019)Kazav, Berdichevsky, and Schwartz]{KBS19}
Kazav, E.; Berdichevsky, R.; Schwartz, M.
\newblock Random close packing from hard-sphere {P}ercus-{Y}evick theory.
\newblock {\em Phys. Rev. E} {\bf 2019}, {\em 99},~{012}{146}, doi:10.1103/PhysRevE.99.012146.

\bibitem[Berthier \em{et~al.}(2019)Berthier, Charbonneau, and Kundu]{BCK19}
Berthier, L.; Charbonneau, P.; Kundu, J.
\newblock Bypassing sluggishness: {SWAP} algorithm and glassiness in high
  dimensions.
\newblock {\em Phys. Rev. E} {\bf 2019}, {\em 99},~{031}{301}(R), doi:10.1103/PhysRevE.99.031301.

\bibitem[Santos \em{et~al.}(2014)Santos, Yuste, {L\'opez de Haro}, Odriozola,
  and Ogarko]{SYHOO14}
Santos, A.; Yuste, S.B.; {L\'opez de Haro}, M.; Odriozola, G.; Ogarko, V.
\newblock Simple effective rule to estimate the jamming packing fraction of
  polydisperse hard spheres.
\newblock {\em Phys. Rev. E} {\bf 2014}, {\em 89},~{040}{302}(R), doi:10.1103/PhysRevE.89.040302.

\bibitem[Bishop \em{et~al.}(1985)Bishop, Michels, and de~Schepper]{BMS85}
Bishop, M.; Michels, J.P.J.; de~Schepper, I.M.
\newblock The short-time behavior of the velocity autocorrelation function of
  smooth, hard hyperspheres in three, four and five dimensions.
\newblock {\em Phys. Lett. A} {\bf 1985}, {\em 111},~169--170, doi:10.1016/0375-9601(85)90568-7.

\bibitem[Colot and Baus(1986)]{CB86}
Colot, J.L.; Baus, M.
\newblock {The freezing of hard disks and hyperspheres}.
\newblock {\em Phys. Lett. A} {\bf 1986}, {\em 119},~135--139, doi:10.1016/0375-9601(86)90432-9.

\bibitem[Lue(2005)]{L05b}
Lue, L.
\newblock Collision statistics, thermodynamics, and transport coefficients of
  hard hyperspheres in three, four, and five dimensions.
\newblock {\em J. Chem. Phys.} {\bf 2005}, {\em 122},~{044}{513}, doi:10.1063/1.1834498.

\bibitem[Santos(2012{\natexlab{a}})]{S12}
Santos, A.
\newblock Note: {A}n exact scaling relation for truncatable free energies of
  polydisperse hard-sphere mixtures.
\newblock {\em J. Chem. Phys.} {\bf 2012}, {\em 136},~{136}{102}, doi:10.1063/1.3702439.

\bibitem[Santos(2012{\natexlab{b}})]{S12c}
Santos, A.
\newblock Class of consistent fundamental-measure free energies for hard-sphere
  mixtures.
\newblock {\em Phys. Rev. E} {\bf 2012}, {\em 86},~{040}{102}(R), doi:10.1103/PhysRevE.86.040102.

\end{thebibliography}

\end{document}